\documentclass[Journal,twocolumn]{IEEEtran}
\IEEEoverridecommandlockouts

\usepackage{color}
\usepackage{graphicx}
\usepackage{amsmath}
\usepackage{amssymb}
\usepackage{algorithm}
\usepackage{algorithmic}
\usepackage{multirow}
\usepackage{booktabs}
\usepackage{array}
\usepackage{amsthm}
\usepackage{stfloats}
\usepackage{caption}
\usepackage{bm}
\usepackage{booktabs}
\usepackage{setspace}
\usepackage{subcaption}
\usepackage{url}
\usepackage{arydshln}
\usepackage{tikz}
\usepackage{accents}
\usepackage{comment}

\theoremstyle{remark}
\newtheorem{remark}{Remark} 
\usetikzlibrary{arrows.meta,positioning,fit,shapes,calc}

\renewcommand{\thesubsubsection}{\thesubsection.\arabic{subsubsection}}

\allowdisplaybreaks[4]

\title{Clutter-Aware Integrated Sensing and Communication: Models, Methods, and Future Directions
\thanks{R. Liu and A. Lee Swindlehurst are with the Nhu Department of Electrical Engineering and Computer Science, University of California, Irvine, CA 92697, USA (e-mail: rangl2@uci.edu; swindle@uci.edu).}
\thanks{P. Li and M. Li are with the School of Information and Communication Engineering, Dalian University of Technology, Dalian 116024, China (e-mail: lipeishi@mail.dlut.edu.cn; mli@dlut.edu.cn).}
}

\author{Rang Liu,~\IEEEmembership{Member,~IEEE,}
        Peishi Li,
        Ming Li,~\IEEEmembership{Senior Member,~IEEE,}
        and A. Lee Swindlehurst,~\IEEEmembership{Life Fellow,~IEEE}}

\begin{document}

\maketitle

\begin{abstract}
Integrated sensing and communication (ISAC) can substantially improve spectral, hardware, and energy efficiency by unifying radar sensing and data communications. In wideband and scattering-rich environments, clutter often dominates weak target reflections and becomes a fundamental bottleneck for reliable sensing. Practical ISAC clutter includes ``cold'' clutter arising from environmental backscatter of the probing waveform, and ``hot'' clutter induced by external interference and reflections from the environment whose statistics can vary rapidly over time.
In this article, we develop a unified wideband multiple-input multiple-output orthogonal frequency-division multiplexing (MIMO-OFDM) signal model that captures both clutter types across the space, time, and frequency domains. Building on this model, we review clutter characterization at multiple levels, including amplitude statistics, robust spherically invariant random vector (SIRV) modeling, and structured covariance representations suitable for limited-snapshot regimes. We then summarize receiver-side suppression methods in the temporal and spatial domains, together with extensions to space-time adaptive processing (STAP) and space-frequency-time adaptive processing (SFTAP), and we provide guidance on selecting techniques under different waveform and interference conditions. To move beyond reactive suppression, we discuss clutter-aware transceiver co-design that couples beamforming and waveform optimization with practical communication quality-of-service (QoS) constraints to enable proactive clutter avoidance. We conclude with open challenges and research directions toward environment-adaptive and clutter-resilient ISAC for next-generation networks. 
\end{abstract}


\section{Introduction}  

Integrated sensing and communication (ISAC) refers to the unified use of wireless resources for radar sensing and data communication within the same system, facilitating a mutual trade-off between these two functionalities \cite{Fan-Liu-TCOM-2020}--\cite{Lu-2024}. By integrating sensing and communication into a single infrastructure, ISAC significantly improves spectrum, energy, and hardware efficiency, while concurrently reducing signaling overhead and deployment costs. The dedicated spectrum and hardware resources conventionally assigned separately to radar and communication are shared, enabling dual-functional ``green'' networks and cost-effective hardware platforms \cite{Fan Liu SPM 2023}. Beyond resource efficiency, ISAC also targets a deeper integration in which the sensing and communication subsystems are jointly designed. Such integration supports communication-assisted sensing and sensing-assisted communication, leading to gains beyond what either system can achieve alone \cite{Wei-CM-2022}.

The ISAC concept has attracted considerable attention as an enabling technology for next-generation 6G wireless networks. Recent research advances have substantially expanded the range of ISAC application scenarios to include automotive sensing for intelligent transportation, vehicle-to-everything (V2X) communications, uncrewed aerial vehicle (UAV) networks, smart homes, smart factories, and human-centered sensing such as gesture or activity recognition. These developments have stimulated standardization activities in bodies such as the 3rd Generation Partnership Project (3GPP) and IEEE toward integrating dual-functional capabilities into future wireless network specifications. International organizations such as the International Telecommunication Union (ITU) have also identified ISAC as an integral part of their 6G vision, highlighting sensing as a native capability of future intelligent and perceptive networks.

A fundamental challenge for ISAC systems is clutter, which refers to undesired echoes and interference that obscure the target of interest \cite{Zhang-JSTSP-2021}--\cite{Luo-TWC-2024}. The distinction between targets and clutter is task-dependent. Scatterers that are relevant to the sensing objective are treated as targets, while other undesired returns, including those from irrelevant moving objects, are regarded as clutter. Following radar terminology, we distinguish between {\em cold} and {\em hot} clutter. While both arise from environmental scattering, they differ in whether or not the illuminating waveform is controlled and known at the sensing receiver. Cold clutter, also known as self-echo or ``slow-time'' clutter, represents undesired backscatter from the environment of the ISAC transmitter's own waveform. Since this waveform is known at the sensing receiver, cold clutter is coherent with the transmitted signal and (after ``de-randomization'') is well approximated as quasi-stationary across multiple coherent processing intervals (CPIs). It can nevertheless exhibit Doppler spread when the receiver and scatterers are in relative motion. Hot clutter, also referred to as scattered external interference, arises when signals emitted by external transmitters illuminate the environment and reach the sensing receiver both directly and after scattering \cite{Kogon1996}--\cite{Abramovich1998}. Since the interference may not originate from a cooperative source, its unknown randomness will result in rapid (``fast-time'') variations across space, time, and frequency.

Clutter effects are especially pronounced in ISAC deployments, particularly indoors and in dense urban environments, where multiple transmitters and receivers operate in the presence of static structures such as buildings and dynamic scatterers such as vehicles and people. Moreover, interference may originate from malicious emitters or from the network itself. Even in upward-looking sensing scenarios where a base station (BS) attempts to detect airborne targets, electromagnetic waves guided by atmospheric ducts can create interference at the sensing receiver. As a result, clutter can dominate the intended target echoes, and accurate clutter modeling, effective clutter estimation, and robust clutter suppression are critical for reliable ISAC operations \cite{Li-TAES-2017}.

Clutter mitigation in ISAC also differs from interference management in radar- and communication-only systems. Conventional radar processing exploits Doppler-domain filtering under quasi-stationary clutter assumptions, while communication systems primarily rely on scheduling, coding, and resource allocation. In ISAC, sensing must cope with both radar-like environmental reflections and communication-induced interference whose statistics may violate the stationarity and narrowband assumptions underlying traditional radar processing \cite{Li-TAES-2017}. This motivates clutter models and mitigation algorithms that operate over wide bandwidths and jointly exploit spatial, temporal, and spectral structure with limited sample support.

\begin{figure}
    \centering
    \includegraphics[width=\linewidth]{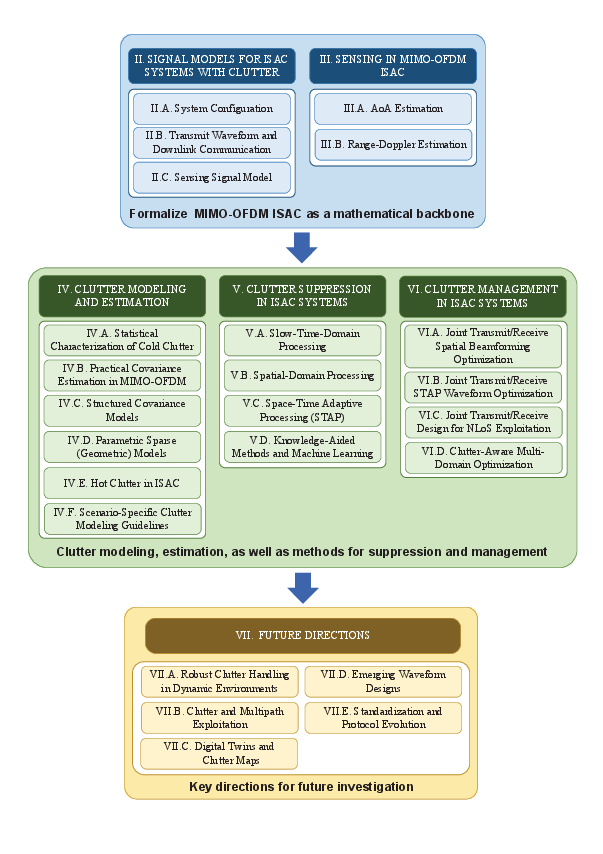}
    \vspace{-10mm}
    \caption{Organization of the paper.}
    \vspace{-2mm}
    \label{fig:outline}
\end{figure}

To address these issues, various techniques have been explored including enhanced space-time adaptive processing (STAP) tailored to ISAC \cite{RLiu-JSAC-2022}, clutter‑aware precoding/beamforming \cite{Liao-TCOM-2023}, and knowledge‑assisted or map‑based approaches that exploit prior environmental information \cite{Xu-ICCC-2024}. Building on these efforts, this paper provides a unified treatment of signal and clutter models, receiver‑side suppression techniques, and proactive transmitter‑side resource allocation for clutter‑aware ISAC systems. 
While we adopt a wideband MIMO-OFDM baseline to align with cellular standards and to establish a consistent notation across space, time, and frequency, most of the clutter models and mitigation techniques discussed in the paper are applicable to other waveforms and can be applied given an appropriate angle-range-Doppler representation. The OFDM-specific portions of the paper focus on range-Doppler formation through the subcarrier and symbol grid and the de-randomization required under data-embedded transmissions. For deterministic probing waveforms such as frequency-modulated continuous-wave (FMCW) signals, range discrimination is obtained through conventional matched filtering or dechirp processing. Emerging waveform designs beyond OFDM are briefly discussed in Sec. VII-D. The structure of the paper is outlined in Fig.~\ref{fig:outline}, and the primary contributions of the paper are  summarized below:
\begin{itemize}
  \item We provide a unified wideband MIMO-OFDM baseline ISAC signal model that accounts for both cold and hot clutter. We also outline an end-to-end receiver processing pipeline across space, time, and frequency, establishing a common framework for the rest of the paper.

  \item We discuss clutter models that connect amplitude distributions, Gaussian or spherically invariant random vector (SIRV) statistics, and structured space-frequency covariance matrices estimated with limited sample support. We also provide scenario-specific guidelines for selecting suitable models.

  \item We review clutter suppression methods over time, frequency, and space, including extensions to joint STAP and space-frequency-time adaptive processing (SFTAP). We provide concise guidance for practical ISAC implementations.

  \item We extend clutter mitigation from reactive suppression to proactive transceiver co‑design, and we formulate clutter-aware resource allocation under communication quality-of-service (QoS) constraints.

  \item We outline future research directions that address multi-domain joint optimization, multi-node cooperative clutter management, antenna-based electromagnetic environment reshaping, and cross-layer MAC and protocol integration toward deployable clutter-aware ISAC architectures.
\end{itemize}
To support further research on this topic, the source code for simulations presented in this work is publicly available at \url{https://github.com/LS-Wireless/Clutter-Aware-ISAC-Tutorial}.

\emph{Notation}: Scalars, column vectors, and matrices are denoted by
plain (e.g., $a$), boldface lower-case (e.g., $\mathbf{a}$), and boldface upper-case variables (e.g., $\mathbf{A}$), respectively. The sets of real and complex numbers are
$\mathbb{R}$ and $\mathbb{C}$. For a matrix or vector, $(\cdot)^*$, $(\cdot)^T$, $(\cdot)^H$, $(\cdot)^{-1}$, and $(\cdot)^\dagger$ denote conjugate, transpose, Hermitian (conjugate transpose), inverse, and Moore-Penrose pseudoinverse. For a scalar $a$, $|a|$ is its magnitude. For a vector $\mathbf{a}$, $\|\mathbf{a}\|$ is the Euclidean norm. For a matrix $\mathbf{A}$, $\|\mathbf{A}\|_F$ and $\|\mathbf{A}\|_2$ denote the Frobenius and spectral norms, respectively. The trace, determinant, rank, and $i$-th eigenvalue (singular value) of $\mathbf{A}$ are $\operatorname{tr}(\mathbf{A})$, $\det(\mathbf{A})$, $\operatorname{rank}(\mathbf{A})$, $\lambda_i(\mathbf{A})$ ($\sigma_i(\mathbf{A})$). The real and imaginary parts are $\Re\{\cdot\}$ and $\Im\{\cdot\}$, and $\arg(\cdot)$ denotes the argument (phase) of a complex quantity. The identity, all-zero, and all-ones matrices/vectors are $\mathbf{I}_N$, $\mathbf{0}_{m\times n}$, and $\mathbf{1}_N$. The operators $\operatorname{diag}\{\mathbf{a}\}$ and $\operatorname{diag}(\mathbf{A})$ denote, respectively, a diagonal matrix whose diagonal is $\mathbf{a}$ and the vector collecting the diagonal of $\mathbf{A}$; $\operatorname{blkdiag}(\cdot)$ forms a block-diagonal matrix; $\operatorname{vec}(\mathbf{A})$ stacks the columns of $\mathbf{A}$. For a matrix $\mathbf{A}$, $[\mathbf{A}]_{i,j}$ denotes its $(i,j)$-th entry. The Kronecker product, Hadamard (elementwise) product, and elementwise division are denoted by $\otimes$, $\odot$, and $\oslash$. The set of $N\times N$ Hermitian positive semidefinite matrices is $\mathbb{S}_N^+$; $\mathbf{A}\succeq\mathbf{0}$ indicates $\mathbf{A}\in\mathbb{S}_N^+$. $\mathbb{Z}_+$ denotes the set of positive integers. 
Statistical expectation is represented by $\mathbb{E}\{\cdot\}$.
$\mathcal{N}(\boldsymbol{\mu},\mathbf{\Sigma})$ and $\mathcal{CN} (\boldsymbol{\mu},\mathbf{\Sigma})$ denote real and circularly symmetric complex Gaussian distributions.

\section{Signal Models for ISAC Systems with Clutter}
\label{sec:sig_model}

This section presents a unified and comprehensive signal modeling framework for ISAC systems operating in clutter-rich environments. We begin by describing the fundamental configuration and assumptions underlying the monostatic ISAC scenario, with particular emphasis on the rationale for adopting standard communication waveforms, which inherently suit communication-centric dual-functional applications. Subsequently, widely adopted multiple-input multiple-output (MIMO) orthogonal frequency-division multiplexing (OFDM) system models are introduced. These foundational models establish a basis for the clutter management strategies discussed in the subsequent sections.

\subsection{System Configuration}

We consider a monostatic ISAC system where a single BS simultaneously performs sensing and communication using separate transmit and receive antenna arrays with $N_\text{t}$ and $N_\text{r}$ elements, respectively. 
Monostatic MIMO-ISAC configurations have the advantage of simplified synchronization (with co-located transmitter and receiver) and streamlined hardware integration, making them attractive for real-world dual-functional deployments.

Multi-carrier waveforms, particularly OFDM, are the predominant choice for ISAC due to their structured design, maturity in current wireless standards, and intrinsic suitability for wideband operation. OFDM is resilient to frequency-selective fading, enables efficient frequency-domain equalization, and offers high spectral efficiency. Combining OFDM with MIMO architectures yields well-known spatial multiplexing and diversity gains for communications, and improves spatial resolution and target detection via beamforming and multi-antenna diversity for sensing. OFDM-based designs further allow direct reuse of existing baseband processors, radio frequency front ends, and protocol stacks, greatly simplifying integration of sensing and communication functionalities. In wideband MIMO-OFDM ISAC, the use of orthogonal subcarriers facilitates frequency‑selective channel management and sharpens range-Doppler resolution for sensing, while multiple antennas enable high angular resolution through spatial processing. The inclusion of a cyclic prefix (CP) further suppresses inter-symbol interference (ISI), ensuring reliable demodulation even under severe multipath and clutter. 
Although OFDM is assumed in this paper due to its likely use in future commercial ISAC systems, most clutter models and covariance-driven suppression tools reviewed later are not tied to a specific probing waveform once the received data are organized in a suitable angle-range-Doppler representation. The main OFDM-specific component in Sec. III-B is waveform de-randomization under data-embedded probing, which is unnecessary when the probing waveforms are deterministic and well designed for sensing, for example under pilot-only operation \cite{JRen-SAM2024}.

In the following description of the MIMO-OFDM transmit waveform, we will make two standard assumptions about OFDM ISAC operation: \textit{(i)} the CP duration exceeds the maximum round‑trip delay for target reflections, and \textit{(ii)} intra‑symbol Doppler frequency shifts are negligible. We formulate the downlink communication signal model and define both linear and non-linear transmit precoding strategies, and outline several metrics to evaluate the performance of joint sensing--communication designs.

\subsection{Transmit Waveform and Downlink Communication }

\subsubsection{Transmit OFDM Waveform}
We consider an OFDM waveform comprising $N$ orthogonal subcarriers indexed by $n\in\{0,1,\dots,N-1\}$, each with subcarrier spacing $\Delta_f=B/N$, where $B$ denotes the total signal bandwidth. In one coherent processing interval (CPI), the system transmits a sequence of $L$ OFDM symbols indexed by $\ell \in \{0,1,\dots,L-1\}$, each having a duration of $T_\text{sym}=1/\Delta_f+T_\text{CP}$, where $T_\text{CP}$ is the CP duration. The transmitted baseband OFDM signal vector $\mathbf{x}(t)\in\mathbb{C}^{N_\text{t}}$ can be represented as
\begin{align}\label{eq:xt-ofdm}
\mathbf{x}(t)=\frac{1}{\sqrt{N}}\!\sum_{\ell=0}^{L-1}\!\sum_{n=0}^{N-1}\!\mathbf{x}_n[\ell]e^{\jmath2\pi n\Delta_f(t-\ell T_\text{sym} -T_\text{CP})}\text{rect}\Big(\frac{t\!-\!\ell T_\text{sym}}{T_\text{sym}}\Big),
\end{align}
where $\mathbf{x}_n[\ell]\in\mathbb{C}^{N_\text{t}}$ represents the precoded frequency-domain waveform transmitted on subcarrier $n$ during the $\ell$-th OFDM symbol, and $\text{rect}\big((t - \ell T_\text{sym})/T_\text{sym}\big)$ is a rectangular pulse of length $T_\text{sym}$ that gates the $\ell$-th symbol.

\subsubsection{Downlink MU-MISO Communication Model}
We consider a multi-user multiple-input single-output (MU-MISO) scenario where each user is equipped with a single antenna. In this case, the received signal for the $k$-th user on the $n$-th subcarrier during the $\ell$-th OFDM symbol is modeled as 
\begin{align}
y_{n,k}[\ell] = \mathbf{h}^H_{n,k}\mathbf{x}_n[\ell] + n_{n,k}[\ell],
\end{align}
where $\mathbf{h}_{n,k}\in\mathbb{C}^{N_\text{t}}$ denotes the frequency-domain communication channel vector from the BS to the $k$-th user, and $n_{n,k}[\ell]\sim\mathcal{CN}(0,\sigma_\text{comm}^2)$ represents additive noise.

\subsubsection{Precoding and Communication Metrics}
Transmit beamforming, also known as precoding, shapes the transmitted signals to simultaneously meet sensing and communication performance requirements. Transmit beamforming methods can be categorized as either spatial-only linear precoding or nonlinear space-time precoding. Spatial-only linear precoding is computationally efficient and relatively straightforward to implement, whereas nonlinear precoding approaches provide enhanced degrees of freedom (DoFs) at the expense of increased complexity, improving performance in challenging scenarios with multi-user interference (MUI) by exploiting symbol-level information. Detailed formulations of linear and nonlinear precoding techniques are described next.

\textbf{Linear Precoding:}
Let $\mathbf{s}_n[\ell]\in\mathbb{C}^{N_\text{s}}$ denote the transmitted symbol vector on subcarrier $n$ at the $\ell$-th OFDM symbol. This vector consists of two parts: the $K$ modulated communication symbols $\mathbf{s}_{\text{c},n}[\ell]$ and $(N_\text{s}-K)$ additional signals $\mathbf{s}_{\text{s},n}[\ell]$ reserved for radar sensing, with $K\leq N_\text{s} \leq N_\text{t}$. The dimension $N_\text{s}$ thus represents the total number of simultaneously transmitted communication and sensing data streams. The frequency-domain transmit signal on the $n$-th subcarrier during the $\ell$-th OFDM symbol, denoted by $\mathbf{x}_n[\ell]\in\mathbb{C}^{N_\text{t}}$, is expressed as 
\begin{align}\label{eq:tx signal xnl}
\mathbf{x}_n[\ell] = \mathbf{W}_n^\text{comm}\mathbf{s}^\text{comm}_n[\ell] + \mathbf{W}_n^\text{radar}\mathbf{s}^\text{radar}_n[\ell] = \mathbf{W}_n\mathbf{s}_n[\ell],
\end{align}
where $\mathbf{W}_n^\text{comm}\in\mathbb{C}^{N_\text{t}\times K}$ and $\mathbf{W}_n^\text{radar}\in\mathbb{C}^{N_\text{t}\times (N_\text{s}-K)}$ denote the communication and sensing beamformers, respectively. The combined beamforming matrix $\mathbf{W}_n=[\mathbf{W}_n^\text{comm},~\mathbf{W}_n^\text{radar}]\in\mathbb{C}^{N_\text{t}\times N_\text{s}}$ precodes the stacked symbol vector $\mathbf{s}_n[\ell]=[\mathbf{s}^\text{comm}_n[\ell]^T,\mathbf{s}^\text{radar}_n[\ell]^T]^T\in\mathbb{C}^{N_\text{s}}$. All streams are taken to be mutually independent with unit power, i.e., $\mathbb{E}\{\mathbf{s}_n[\ell]\mathbf{s}^H_n[\ell]\}=\mathbf{I}_{N_\text{s}}$.

This structure embeds random communication data and deterministic sensing probes within the same OFDM resource, balancing the stochasticity needed for high‑rate communication with the determinism desired for accurate sensing. Allocating a subset of spatial DoFs to sensing preserves spatial/spectral diversity and yields a full‑rank illumination pattern, enhancing target observability \cite{XLiu-TSP-2020}. Accordingly, $\mathbf{W}_n^\text{comm}$ and $\mathbf{W}_n^\text{radar}$ are co‑designed to balance communication QoS and sensing performance under various system constraints.
Eq. \eqref{eq:tx signal xnl} also accommodates orthogonal resource ISAC operation, where communication and sensing are multiplexed over disjoint time-frequency resource elements. In particular, for communication resources the transmit vector specializes to $\mathbf{x}_n[\ell]=\mathbf{W}_n^{\text{comm}}\mathbf{s}_n^{\text{comm}}[\ell]$, whereas for sensing resources it becomes $\mathbf{x}_n[\ell]=\mathbf{W}_n^{\text{radar}}\mathbf{s}_n^{\text{radar}}[\ell]$, e.g., when sensing is confined to dedicated (reserved) pilot resource elements.

Typical communication performance metrics for linear precoding include the signal-to-interference-plus-noise ratio (SINR) and the achievable sum-rate. The SINR for the $k$-th user on subcarrier $n$ is  
\begin{align}\label{eq:comm SINR}
\text{SINR}_{n,k} = \frac{|\mathbf{h}_{n,k}^H\mathbf{w}_{n,k}|^2}
{\sum_{j\neq k}^{N_\text{s}}|\mathbf{h}_{n,k}^H\mathbf{w}_{n,j}|^2+\sigma_\text{comm}^2},
\end{align}
where $\mathbf{w}_{n,j}$ denotes the $j$-th column of $\mathbf{W}_n$. 
The achievable sum-rate is then given by
\begin{align}
R_\text{sum}=\sum_{k=1}^{K}\sum_{n=0}^{N-1}\log_2(1+\text{SINR}_{n,k}).
\end{align}
These metrics quantify how effectively the beamforming strategy suppresses MUI and directly affects practical communication QoS indicators such as bit error rate (BER) or symbol error rate (SER).
These communication metrics, in particular the SINR in \eqref{eq:comm SINR}, will be used in Sec. VI-A to impose communication QoS constraints in clutter-aware joint transmit and receive beamforming designs.

\textbf{Non-Linear Precoding:} 
Although linear precoders are computationally efficient, they cannot fully exploit instantaneous symbol and channel state information (CSI) to optimize performance. Symbol-level precoding (SLP) is an advanced technique that reshapes the transmitted waveform on a symbol-by-symbol basis to simultaneously enhance both communication QoS and radar sensing performance \cite{RLiu-JSAC-2022}, \cite{RLiu-JSTSP-2021}--\cite{PLi-TVT-2025}.
The SLP-designed transmit waveform at subcarrier $n$ and OFDM symbol $\ell$ can be expressed as a general nonlinear mapping $\mathcal{F}(\cdot)$ that jointly exploits instantaneous CSI $\mathbf{h}_{n,k}$ and the transmitted symbol vector~$\mathbf{s}_n[\ell]$:
\begin{align}\label{eq:xnl SLP}
\mathbf{x}_n[\ell]=\mathcal{F}\big({\mathbf{s}_n[\ell]},\{\mathbf{h}_{n,k}\}_{k=1}^{K}\big),
\end{align}
By fully harnessing the symbol-level temporal DoFs, SLP strategically aligns MUI at the receiver, effectively pushing received signals deeper into the correct decision regions and enhancing robustness to noise and interference, particularly in heavily loaded communication systems. Meanwhile, the temporal DoFs provided by SLP enable a unified precoder to simultaneously address the inherent randomness required by high-rate communications and the determinism essential for radar probing. This distinctive capability allows SLP to effectively manage the trade-off between these conflicting requirements, ensuring reliable communication while enabling precise radar waveform shaping.

To quantify the communication QoS achieved by SLP, the so-called {\em safety margin}, defined as the minimum distance between the received noise-free signal and its corresponding decision boundary, is typically employed as a performance metric. Considering $\Omega$-phase shift keying (PSK) modulation as an example, and letting $s_{n,\ell,k}$ be the intended constellation point for user $k$ on subcarrier $n$ at OFDM symbol $\ell$, the safety margin is defined as 
\begin{equation}\label{eq:SLP QoS}
\begin{aligned}
\delta_{n,k}[\ell] &= \Re\{\mathbf{h}_{n,k}^H\mathbf{x}_n[\ell]s^*_{n,\ell,k}\}\sin(\pi/\Omega)\\
&\qquad - \left|\Im\{\mathbf{h}_{n,k}^H\mathbf{x}_n[\ell]s^*_{n,\ell,k}\}\right|\cos(\pi/\Omega) .
\end{aligned}\end{equation}
Maximizing the safety margin ensures greater resilience to noise and interference, resulting in improved SER and overall communication robustness.
The SLP model will be revisited in Sec. VI-B, where it enables incorporating symbol-level communication QoS constraints into joint transmit waveform and receive STAP filter optimization for clutter-aware ISAC.

\subsection{Sensing Signal Model}
\label{sec:Sensing Model}
Given the transmit signal in \eqref{eq:xt-ofdm}, the BS performs sensing by analyzing echoes from multiple objects, including  both targets and clutter. We model the scene as a collection of discrete scatterers, each with its own radar cross‑section (RCS), propagation delay, Doppler shift, and direction.
After analog-to-digital conversion (ADC), CP removal, and a fast Fourier transform (FFT), the received signal $\mathbf{y}_n[\ell]\in\mathbb{C}^{N_\text{r}}$ at subcarrier $n$ and OFDM symbol $\ell$ is represented as 
\begin{equation}\label{eq:y_nl}
\mathbf{y}_n[\ell] =
\underbrace{\mathbf{H}^\text{t}_{n,\ell}\mathbf{x}_n[\ell]}_{\mathbf{y}_{\text{t},n}[\ell]} +
\underbrace{\mathbf{H}^\text{c}_{n,\ell}\mathbf{x}_n[\ell]}_{\mathbf{y}_{\text{cc},n}[\ell]} +
\underbrace{\mathbf{H}^\text{h}_{n,\ell}\mathbf{s}^{\text{ext}}_n[\ell]}_{\mathbf{y}_{\text{hc},n}[\ell]} +
\mathbf{z}_n[\ell],
\end{equation}
where $\mathbf{y}_{\text{t},n}$, $\mathbf{y}_{\text{cc},n}$, and $\mathbf{y}_{\text{hc},n}$ denote the target echo, cold clutter (self-echo clutter), and hot clutter (scattered external interference), respectively. The term $\mathbf{z}_n[\ell]$ accounts for other additive disturbances and is modeled as $\mathbf{z}_n[\ell]\sim\mathcal{CN}(\mathbf{0},\mathbf{R}_{\text{z},n})$.  The covariance $\mathbf{R}_{\text{z},n}$ captures receiver thermal noise as well as residual self-interference (SI), and it reduces to $\sigma^2\mathbf{I}_{N_\text{r}}$ in the ideal spatially white noise case with negligible residual SI. 
In practice, direct TX--RX leakage is mitigated through antenna/radio frequency isolation and analog/digital cancellation. Although the residual SI is typically much weaker than the transmitted signal, it is nonetheless often stronger than weak target echoes. Since the focus of this paper is environmental clutter, we absorb the residual self-interference into $\mathbf{z}_n[\ell]$ and model it through the general colored covariance $\mathbf{R}_{\text{z},n}$. This representation is also convenient in later sections where the covariance of the various disturbances explicitly enters into the clutter-aware receive filtering and transceiver optimization.

The composite clutter term can be compactly expressed as 
\begin{align}
    \mathbf{y}_{\text{c},n}[\ell] = \mathbf{y}_{\text{cc},n}[\ell] + \mathbf{y}_{\text{hc},n}[\ell].
\end{align}
It is also convenient to absorb the hot clutter and additive disturbances into an aggregate term and define 
\begin{align}\label{eq:eta def}
    \bm{\eta}_n[\ell] = \mathbf{y}_{\text{hc},n}[\ell] + \mathbf{z}_n[\ell].
\end{align}
The sensing channels associated with the target and cold clutter are given by
\begin{subequations} \label{eq:channel_tf}
\begin{align}
\mathbf{H}_{n,\ell}^{\text{t}}  & = \!\! \sum_{m=1}^M \! \alpha_{m,n} e^{\jmath 2\pi(f_{\text{D},m}\ell T_{\text{sym}}\!-\!n\Delta_f\tau_m)} \mathbf{b}_n\!(\bm{\vartheta}_m)\mathbf{a}_n^H\!(\bm{\vartheta}_m), \!\!\! \\
\mathbf{H}_{n,\ell}^{\text{c}} & = \sum_{c=1}^{C} \beta_{c,n} e^{\jmath2\pi(f_{\text{D},c}\ell T_{\text{sym}}-n\Delta_f\tau_c)}\mathbf{b}_n(\bm{\vartheta}_c)\mathbf{a}_n^H(\bm{\vartheta}_c),\label{eq:clutter channel}
\end{align}
\end{subequations}
where the frequency-dependent complex coefficients $\{\alpha_{m,n}, \beta_{c,n}\}$ capture the combined effects of RCS, path loss, and multipath-induced phase variations, and the transmit and receive steering vectors for subcarrier $n$ are denoted by $\mathbf{a}_n(\bm{\vartheta})\in\mathbb{C}^{N_\text{t}}$ and $\mathbf{b}_n(\bm{\vartheta})\in\mathbb{C}^{N_\text{r}}$. The parameter vector $\bm{\vartheta}$ may include azimuth, elevation, range, and polarization depending on the assumed array model. In the most general case, given arbitrary antenna element coordinates $\{\mathbf{r}_{\text{t},i}\}_{i=1}^{N_\text{t}}$ and $\{\mathbf{r}_{\text{r},i}\}_{i=1}^{N_\text{r}}$, the frequency-dependent array responses are defined as 
\begin{subequations}\label{eq:general_steering}
\begin{align}
\mathbf{a}_n(\boldsymbol{\vartheta}) &=
\big[e^{-\jmath\frac{2\pi}{\lambda_n}\mathbf{k}^T(\bm{\vartheta}) \mathbf{r}_{\text{t},1}},\dots,
e^{-\jmath\frac{2\pi}{\lambda_n}\mathbf{k}^T (\bm{\vartheta})\mathbf{r}_{\text{t},N_\text{t}}}\big]^T,\!\\
\mathbf{b}_n(\bm{\vartheta}) &=
\big[e^{-\jmath\frac{2\pi}{\lambda_n}\mathbf{k}^T\!(\bm{\vartheta})\mathbf{r}_{\text{r},1}},\dots,
e^{-\jmath\frac{2\pi}{\lambda_n}\mathbf{k}^T\!(\bm{\vartheta}) \mathbf{r}_{\text{r},N_\text{r}}}\big]^T,\!
\end{align}\end{subequations}
where $\lambda_n=c_0/f_n$ is the wavelength at frequency $f_n = f_0 + n\Delta_f$, $c_0$ is the speed of light, and $\mathbf{k}(\boldsymbol{\vartheta})$ denotes the unit propagation vector associated with $\boldsymbol{\vartheta}$. For simplicity, we will specialize to the far‑field, single‑polarization, azimuth-only case in which $\boldsymbol{\vartheta}$ reduces to the scalar $\theta$, representing the azimuth angle. For the special case of half-wavelength spaced uniform linear arrays (ULAs), the steering vectors simplify to
\begin{subequations}\label{eq:steering}
\begin{align}
\mathbf{a}_n(\theta)&=[1,e^{-\jmath\pi\chi_n\sin\theta},\dots,e^{-\jmath(N_\text{t}-1)\pi\chi_n\sin\theta}]^T,\\
\mathbf{b}_n(\theta)&=[1,e^{-\jmath\pi\chi_n\sin\theta},\dots,e^{-\jmath(N_\text{r}-1)\pi\chi_n\sin\theta}]^T,\!
\end{align}
\end{subequations}
where $\chi_n=1+n\Delta_f/f_0$ captures the subcarrier-dependent variation in wavelength with respect to carrier frequency $f_0$. This formulation effectively models the so-called ``beam-squint effect'', which refers to the frequency-dependence of the spatial response in wideband systems. In narrowband cases ($\Delta_f/f_0\ll 1$), the steering vectors become frequency-independent since $\chi_n\approx 1, \, \forall \, n$. While we retain the azimuth-only notation $\theta$ for brevity, the general expressions in \eqref{eq:general_steering} can be directly substituted for arbitrary array geometries or extended to include elevation, range, or polarization parameters, without altering the subsequent derivations or algorithms. 

Since the hot clutter contribution stems from external transmitters illuminating the scene, it is modeled as  
\begin{equation}\label{eq:external signal}
\begin{aligned}
    \mathbf{H}^\text{h}_{n,\ell} & = \big[\mu_{1,n}e^{\jmath 2\pi (f_{\text{D},1}\ell T_\text{sym}-n\Delta_f\tau_1)}\mathbf{b}_n(\bm{\vartheta}_1),~\dots,\\
    &\qquad  \mu_{G,n}e^{\jmath 2\pi (f_{\text{D},G}\ell T_\text{sym}-n\Delta_f\tau_G)}\mathbf{b}_n(\bm{\vartheta}_G)\big]\in\mathbb{C}^{N_\text{r}\times G} \\
\mathbf{s}^\text{ext}_n[\ell] &= [s_{n,1}[\ell],\dots,s_{n,G}[\ell]]^T\in\mathbb{C}^G, 
\end{aligned}
\end{equation}
where $\mathbf{s}^\text{ext}_n[\ell]$ collects the waveforms emitted by $G$ independent external sources, and $\mu_{g,n}$ are the corresponding gains.

Both cold and hot clutter involve environmental scattering, with the key difference being the source of illumination. Cold clutter (self-echo clutter) is generated by backscatter of the BS-controlled transmit signal $\mathbf{x}_n[\ell]$ and is therefore known at the sensing receiver. In contrast, the scattered external interference or hot clutter arises when external emissions $\mathbf{s}^{\text{ext}}_n[\ell]$ illuminate the environment and are subsequently scattered into the receiver, together with the direct path signal itself. In cooperative or in-network scenarios, the signal $\mathbf{s}^{\text{ext}}_n[\ell]$ may be available at the BS; otherwise, it must be treated as unknown, non-cooperative interference. We adopt the unified model introduced above and distinguish between these two cases in Sec. IV-E.

We reorganize the frequency-domain signals received at each subcarrier by stacking them over $L$ OFDM symbols:  
\begin{equation}\label{eq:define yn}
    \mathbf{y}_n = \big[\mathbf{y}_n^T[0], \dots, \mathbf{y}_n^T[L-1] \big]^T \in \mathbb{C}^{N_{\text{r}} L}.
\end{equation}
Defining the transmit waveform matrix associated with subcarrier $n$ as
\begin{align}\label{eq:Xn def}
\mathbf{X}_n  \triangleq \text{blkdiag}\big\{ \mathbf{I}_{N_{\text{r}}} \otimes \mathbf{x}_n^T[0], \dots, \mathbf{I}_{N_{\text{r}}} \otimes \mathbf{x}_n^T[L-1] \big\},
\end{align}
we obtain the following model for the received space-time data vector at subcarrier $n$, which integrates reflections from both targets and cold-clutter scatterers:
\begin{equation}\label{eq:yn def}
    \mathbf{y}_n = \sum_{q \in \mathcal{Q}} \gamma_{q,n} e^{-\jmath 2 \pi n \Delta_f \tau_q} \mathbf{X}_n \mathbf{v}_n(\theta_q, f_{\text{D},q}) + \boldsymbol{\eta}_n,
\end{equation}
where $\mathcal{Q}$ denotes the set of all scatterers including targets and cold clutter, $\gamma_{q,n}$ represents the reflection coefficient associated with scatterer $q$ at subcarrier $n$, and $\boldsymbol{\eta}_n$ stacks $\{\bm{\eta}_n[\ell]\}_{\ell=0}^{L-1}$ and represents the aggregate disturbance including hot clutter, receiver thermal noise and residual SI. In addition, associated with each scatterer is its Doppler steering vector
\begin{align}\label{eq:Doppler steering}
\mathbf{d}(f_{\text{D}})  \triangleq \left[1,e^{\jmath2\pi f_\text{D}T_{\text{ sym}}},\ldots,e^{\jmath2\pi f_{\text{D}}(L-1)T_{\text{sym}}}\right]^T, 
\end{align}
and spatial-temporal steering vector 
\begin{equation}\label{eq:vn def}
\mathbf{v}_n(\theta, f_{\text{D}})  \triangleq \mathbf{d}(f_{\text{D}}) \otimes \mathbf{b}_n(\theta) \otimes \mathbf{a}_n^{\ast}(\theta).
\end{equation}

We stack vectors from all $N$ subcarriers vertically into a unified frequency-space-time data vector:
$\mathbf{y}\triangleq [\mathbf{y}_0^T,\mathbf{y}_1^T,\dots,\mathbf{y}_{N-1}^T]^T$, yielding the following unified signal model spanning the full frequency 
band:
\begin{align}\label{eq:y def}
    \mathbf{y} = \sum_{q \in \mathcal{Q}} \bm{\Gamma}_q\mathbf{T}(\tau_q) \mathbf{X} \mathbf{v}(\theta_q,f_{\text{D},q}) + \bm{\eta},
\end{align}
where 
\begin{align}\label{eq:X def}
    \mathbf{X} \triangleq \text{blkdiag}(\mathbf{X}_0,\mathbf{X}_1,\dots,\mathbf{X}_{N-1}) \in \mathbb{C}^{N_\text{r}LN\times N_\text{r}LNN_\text{t}}
\end{align}
is the full-band waveform matrix, 
\begin{align}\label{eq:Ttau def}
    \mathbf{T}(\tau) = \text{diag}\{\mathbf{t}(\tau)\}\otimes\mathbf{I}_{N_\text{r}L}
\end{align}
is a diagonal delay-related matrix, 
\begin{align}\label{eq:delay steering}
    \mathbf{t}(\tau) \triangleq [1, e^{-\jmath 2\pi\Delta_f\tau},\dots,e^{-\jmath2\pi(N-1)\Delta_f\tau}]^T\in\mathbb{C}^N
\end{align}
is the frequency-delay steering vector, 
\begin{align}\label{eq:v def}
    \mathbf{v}(\theta,f_\text{D}) \triangleq [\mathbf{v}_0^T(\theta,f_\text{D}),\dots,\mathbf{v}^T_{N-1}(\theta,f_\text{D})]^T \in \mathbb{C}^{NN_\text{r}LN_\text{t}}
\end{align}
is the full-band angle-Doppler steering vector, and
\begin{align}\label{eq:Gammaq}
    \bm{\Gamma}_q \triangleq \text{diag}(\gamma_{q,0},\dots,\gamma_{q,N-1})\otimes \mathbf{I}_{N_\text{r}L}
\end{align}
accounts for the RCS of all scatterers.

In conventional radar systems, the sensing data cube is indexed by angle, range, and Doppler, with the range dimension obtained from matched filtering in the time domain. In OFDM-based ISAC, however, the same physical information is represented by angle, subcarrier, and symbol indices. Here, the subcarrier index encodes range through a linear frequency-delay phase progression, while the symbol index encodes Doppler through a linear temporal phase accumulation. The unified data model in \eqref{eq:define yn}--\eqref{eq:Gammaq} explicitly reveals this relationship via the frequency-delay steering vector $\mathbf{t}(\tau)$ defined in \eqref{eq:delay steering} and the Doppler steering vector $\mathbf{d}(f_\text{D})$ defined in \eqref{eq:Doppler steering}. 
Range focusing corresponds to projecting along $\mathbf{t}(\tau)$ over the subcarriers, while Doppler focusing corresponds to projecting along $\mathbf{d}(f_\text{D})$ over the OFDM symbols. Applying an inverse discrete Fourier transform (IDFT) across the subcarriers and a discrete Fourier transform (DFT) across the symbols jointly realize matched filtering for range and Doppler estimation directly in the modulation-symbol domain\footnote{This assumes a sufficiently long cyclic prefix, effective compensation of any residual carrier frequency offset (CFO), sampling frequency offset (SFO), and phase noise, as well as proper equalization of data-bearing subcarriers, which are typical OFDM-based ISAC requirements.}. This fundamental interpretation provides the theoretical foundation for constructing the range-Doppler map (RDM) via a two-dimensional transform, as further detailed in Sec.~III.

\begin{remark}[{\em Extensions to bistatic and multistatic scenarios}]
The model in \eqref{eq:y_nl}-\eqref{eq:external signal} is presented for a monostatic BS with co-located transmit and sensing receive arrays. The same formulation extends to bistatic and multistatic operation when the illuminator and the sensing receiver are at different locations. Consider an illuminator at $\mathbf{p}_\text{Tx}$ and a sensing receiver at $\mathbf{p}_\text{Rx}$. For a moving scatter $m$ with position $\mathbf{p}_m(t)$, the bistatic path length is $d_m(t) = \|\mathbf{p}_m(t)-\mathbf{p}_\text{Tx}\| + \|\mathbf{p}_m(t)-\mathbf{p}_\text{Rx}\|$, which yields the bistatic delay $\tau_m = d_m(0)/c_0$. The associated Doppler shift follows from the time derivative of the bistatic propagation phase and is therefore governed by the projections of the scatterer velocity onto the Tx-to-scatterer and scatterer-to-Rx look directions. With separated transmit and receive arrays, the target and clutter terms retain the same sum-of-rank-one structure as in \eqref{eq:channel_tf}, except that the array response involves distinct transmit and receive steering vectors corresponding to different departure and arrival directions. For notational clarity, we proceed with the monostatic notation in the following derivations, noting that the same stacking operations apply in bistatic and multistatic settings after substituting the corresponding steering vectors and delays. Practical waveform considerations that govern coherent processing in bistatic/multistatic sensing are discussed in Sec. III-B.
\end{remark}

\section{Sensing in MIMO-OFDM ISAC}
Building on the frequency-domain signal model established in Sec.~\ref{sec:Sensing Model}, this section develops the receiver-side processing framework for target detection and parameter estimation in MIMO-OFDM ISAC systems. The receiver processing typically follows a hierarchical two-stage structure: the first stage estimates target angles of arrival (AoAs), while the second performs range-Doppler (RD) estimation for localization and velocity retrieval.

\subsection{AoA Estimation}

Accurate AoA estimation is a critical component of multi-antenna sensing systems, including MIMO-OFDM ISAC. 
For systems employing narrowband waveforms, there are a plethora of techniques available for AoA estimation. These techniques can be classified as spectrum- or (pseudospectrum)-based methods, multidimensional parametric estimation schemes, or solutions to sparse recovery problems. The first category includes classical delay-and-sum (Bartlett) beamforming \cite{van2002optimum}, Capon or minimum variance distortionless response (MVDR) beamforming \cite{capon1969high}, and the MUSIC algorithm \cite{schmidt1986multiple}. These methods generate a function representing the received power or subspace orthogonality vs.~angle that is then searched for stationary points. Multidimensional algorithms estimate all AoAs ``simultaneously'', based for example on maximum likelihood (ML) \cite{StoicaSharman1990}, weighted subspace fitting \cite{Viberg1991}, or special array geometries such as the ESPRIT algorithm \cite{roy1989esprit}. Methods that exploit sparsity in the spatial domain recast AoA estimation as a sparse reconstruction problem \cite{malioutov2005sparse}, using algorithms such as orthogonal matching pursuit (OMP), sparse Bayesian learning (SBL) \cite{yang2018sparse}, and SPICE \cite{stoica2011spice} to reconstruct the spatial pseudo-spectrum. For wideband systems, the frequency dependence of the array response complicates AoA estimation, leading to the previously mentioned beam squint effect if such dependencies are ignored. Frequency-invariant beamforming, focusing, and averaging methods can combine data across OFDM subcarriers before determining the final target angle estimates.

The above methods have various trade-offs in terms of accuracy and complexity that are already well documented in the literature. Special array geometries such as ULAs are advantageous because statistical efficiency can often be achieved without open-ended parameter searches, using efficient procedures involving FFTs, polynomial rooting (e.g., root-MUSIC \cite{Barabell1983}, MODE \cite{Stoica1989}, or root-WSF \cite{JanssonSO98}), or eigenvalue computations as ESPRIT. However, for large receive arrays, the best performing AoA methods suffer from high computational complexity due to the need for inverting or performing an eigenvalue decomposition on a high-dimensional covariance matrix, or the requirement for large dictionaries.

\subsection{Range-Doppler Estimation} 

With AoA estimates of potential targets, the receiver can jointly estimate the range and Doppler parameters associated with each AoA. As mentioned above, in OFDM-based ISAC, range information manifests as a linear phase progression across subcarriers, while (constant) Doppler information appears as a linear phase progression across OFDM symbols. Performing an IDFT across the subcarriers and a DFT across the symbols corresponds to time-domain matched filtering executed in the modulation-symbol domain \cite{Sturm-ProcIEEE-2011}. However, OFDM transmissions carry random communication data, causing signal-dependent coupling between the transmitted waveforms and received radar echoes. This coupling complicates the RD analysis and must be carefully mitigated to ensure accurate parameter estimation. Here we present a systematic receiver processing chain comprising angle gating via spatial filtering to isolate and concentrate target returns from distinct directions, waveform de-randomization to remove the influence of random data modulation, 2D-DFT mapping to convert the received data into a structured RDM, and RD-domain detection to identify significant peaks and determine target parameters. While hot clutter is rapidly time-varying and rarely yields coherent RD peaks, target and cold clutter returns generate structured periodic signatures after 2D-DFT processing. Consequently, this stage focuses on estimating the delay and Doppler parameters of these coherent components while treating hot clutter and noise as aggregated interference.

\subsubsection{Angle Gating via Spatial Filtering}
The first step applies spatial filtering to decouple signals from different AoAs and exploit the array gain. 
Denote the estimated AoA set as $\{\hat{\theta}_{p}\}_{p = 1}^P$, where each $\hat{\theta}_p$ defines an angular sector indexed by $p$, and $P \leq M+C $ since not all scatterers are spatially resolvable.
Let $M_{p}$ and $C_p$ respectively denote the number of targets and strong clutter reflections at each estimate $\hat{\theta}_{p}$, where $M = \sum_{p=1}^{P} M_{p}$ and $C = \sum_{p=1}^{P} C_{p}$. 
The beamformed signal for the $p$-th direction is obtained as
\begin{align}
   Y_{p}(n,\ell) = \mathbf{r}_n^H(\hat{\theta}_{p}) \mathbf{y}_n[\ell],
\end{align}
where $\mathbf{r}_n(\hat{\theta}_{p})$ denotes the receive beamforming vector. Substituting $\mathbf{y}_n[\ell]$ from \eqref{eq:y_nl} yields
\begin{equation}\begin{aligned}
Y_{p}(n,\ell)
&\approx \sum_{i=1}^{M_{p}} \tilde{\alpha}_{p,i} e^{\jmath 2\pi(f_{\text{D},p,i}\ell T_{\text{sym}}-n\Delta_f\tau_{p,i})} x_{p,n,\ell}  \\
&  + \sum_{j=1}^{C_{p}} \tilde{\beta}_{p,j} e^{\jmath 2\pi(f_{\text{D},p,j}\ell T_{\text{sym}}-n \Delta_f \tau_{p,j})} x_{p,n,\ell} + \tilde{z}^{\text{h}}_{p,n,\ell}, \label{eq:y_tf_approx}
\end{aligned}\end{equation}
where the coefficients $\tilde{\alpha}_{p,i}$ and $\tilde{\beta}_{p,j}$ include both reflection magnitude and beamforming gain in angular sector $p$ for the targets and strong clutter reflections, respectively. 
The Doppler shift and propagation delay of the $i$-th target are represented by $f_{\text{D},p,i}$ and $\tau_{p,i}$, while $f_{\text{D},p,j}$ and $\tau_{p,j}$ correspond to other scatterers. The term $\tilde{z}^{\text{h}}_{p,n,\ell} = \mathbf{r}_n^H(\hat{\theta}_{p}) (\mathbf{H}_{n,\ell}^{\text{h}} \mathbf{s}^\text{ext}_n[\ell] + \mathbf{z}_n[\ell])$ denotes the residual interference after beamforming, including hot clutter, noise and residual SI. In addition, $x_{p,n,\ell}  = \mathbf{a}_n^H(\hat{\theta}_{p}) \mathbf{x}_n[\ell]$ captures the component of the transmit waveform associated with $\hat{\theta}_p$. The approximation in \eqref{eq:y_tf_approx} holds when different steering vectors are nearly orthogonal, i.e., $|\mathbf{r}_n^H(\hat{\theta}_p)\mathbf{b}_n(\theta_q)|\ll 1,~q\neq p$.

The receive beamformer $\mathbf{r}_n(\hat{\theta}_{p})$ can be realized using classical designs that balance complexity and interference suppression.
Three representative methods are maximum-ratio combining (MRC), zero-forcing (ZF), and minimum mean-square-error (MMSE) beamforming, given by \cite{Tse-book-2005, Dai-arxiv-2025}
\begin{subequations}
    \begin{align}
        \mathbf{r}_n^{\text{MRC}}(\hat{\theta}_p)
         & = \mathbf{b}_n(\hat{\theta}_p) / \| \mathbf{b}_n(\hat{\theta}_p) \|_2^2,                                                         \\
        \mathbf{r}_n^{\text{ZF}}(\hat{\theta_p})
         & = \hat{\mathbf{B}}_{p,n} \mathbf{b}_n(\hat{\theta}_p) / \| \hat{\mathbf{B}}_{p,n} \mathbf{b}_n(\hat{\theta}_p) \|_2^2,           \\
        \mathbf{r}_n^{\text{MMSE}}(\hat{\theta}_p)
         & = \hat{\mathbf{C}}_{p,n}^{-1} \mathbf{b}_n(\hat{\theta_p}) / \| \hat{\mathbf{C}}_{p,n}^{-1} \mathbf{b}_n(\hat{\theta}_p) \|_2^2,
    \end{align}
\end{subequations}
where $\hat{\mathbf{B}}_{p,n} = \mathbf{I}_{N_{\text{r}}} - \mathbf{B}_{p,n} (\mathbf{B}_{p,n}^H \mathbf{B}_{p,n})^{-1} \mathbf{B}_{p,n}^H$ projects onto the space orthogonal to the interference defined by $\mathbf{B}_{p,n} = [\mathbf{b}_n(\hat{\theta}_1), \dots, \mathbf{b}_n(\hat{\theta}_{p-1}), \mathbf{b}_n(\hat{\theta}_{p+1}), \dots, \mathbf{b}_n(\hat{\theta}_{P})]$, and $\hat{\mathbf{C}}_{p,n} = \sum_{q \neq p}^{P} \mathbf{b}_n(\hat{\theta}_q) \mathbf{b}_n^H(\hat{\theta}_q) + \mathbf{I}_{N_{\text{r}}}$ serves as an interference-plus-noise correlation template constructed from the steering vectors.

Collecting the angle gated results $Y_p(n,\ell)$ across subcarriers and OFDM symbols yields the per-angle time--frequency data matrix 
$[\mathbf{Y}_p^{\text{tf}}]_{n,\ell}\triangleq Y_p(n,\ell)$, expressed in compact form as
\begin{equation}\label{eq:Yptf}
    \mathbf{Y}_p^{\text{tf}} = \mathbf{H}_{p}^{\text{tf}} \odot \mathbf{X}_p + \mathbf{Z}_p^{\text{h}},
\end{equation}
where $[\mathbf{X}_p]_{n,\ell}\triangleq x_{p,n,\ell}$ denotes the known transmit waveform on the $p$-th angle-gated branch and $[\mathbf{Z}_p^{\text{h}}]_{n,\ell}\triangleq \tilde{z}_{p,n,\ell}^{\text{h}}$. 
The effective time-frequency sensing channel $\mathbf{H}_{p}^{\text{tf}}$ includes reflections from both targets and strong clutter components:
\begin{equation}  
        \mathbf{H}_{p}^{\text{tf}} = \!\sum_{i=1}^{M_p}\! \alpha_{p,i} \mathbf{t}(\tau_{p,i}) \mathbf{d}^T(f_{\text{D}, p, i}) + \sum_{j=1}^{C_p}\! \beta_{p,j} \mathbf{t}(\tau_{p,j}) \mathbf{d}^T(f_{\text{D}, p, j}), \!\!
\end{equation}
where $\mathbf{t}(\tau)$ and $\mathbf{d}(f_\text{D})$ represent the delay- and Doppler-domain steering vectors, respectively.

\subsubsection{Waveform De-Randomization} 
For OFDM ISAC, the effective probing signal on each time-frequency resource inherits the randomness of the communication symbols, and consequently the received echoes are data dependent for all $n$ and $\ell$ even when a deterministic sensing stream is also present. If the dependence on the random data is not compensated for, the subsequent range-Doppler focusing produces modulation-induced sidelobes and an elevated background interference floor that may mask weak targets. To mitigate this effect, we employ element-wise de-randomization.

Starting from $\mathbf{Y}^{\text{tf}}_p = \mathbf{H}^{\text{tf}}_p \odot \mathbf{X}_p + \mathbf{Z}^h_p$ in \eqref{eq:Yptf}, common de-randomization rules include reciprocal filtering (RF)~\cite{Sturm-Radarconf-2009}--\cite{Pucci-JSAC-2022}, matched filtering (MF)~\cite{Mercier-TAES-2020}, \cite{Berger-JSTSP-2010}, \cite{PLi-arxiv-2025}, and Wiener-type linear MMSE (LMMSE) filtering~\cite{Baudais-TCOM-2023}--\cite{Du-arxiv-2025}:
\begin{equation}\label{eq:derandom}
    \hat{\mathbf{H}}_p^{\text{tf}} =
    \begin{cases}
        \mathbf{Y}_p^{\text{tf}} \oslash \mathbf{X}_p                                                              & \text{RF} \\
        \mathbf{Y}_p^{\text{tf}} \odot \mathbf{X}_p^{\ast}                                                         & \text{MF} \\
        \frac{\mathbf{Y}_p^{\text{tf}} \odot \mathbf{X}_p^{\ast}}{|\mathbf{X}_p|^2 + \text{SNR}_{\text{in},p}^{-1}} & \text{LMMSE}
    \end{cases}
\end{equation}
where $[|\mathbf{X}_p|^2]_{i,j}=|[\mathbf{X}_p]_{i,j}|^2$ and $\text{SNR}_{\text{in},p} = (\sum_{i} |\alpha_{p,i}|^2 + \sum_{j} |\beta_{p,j}|^2) / \sigma^2$ denotes the effective input signal-to-noise ratio (SNR) on the $p$-th angle-gated branch. These strategies reflect different robustness/accuracy trade-offs:
\begin{itemize}
    \item \textbf{RF} completely removes the influence of the waveform and yields an ideal sinc‑like impulse response, but tends to amplify noise, especially for scenarios with low SNR or high‑order modulation.
    \item \textbf{MF} coherently integrates the signal energy, providing maximum SNR gain and robust performance at low SNR. However, it introduces random sidelobes due to symbol variations, which may mask weak targets, especially for higher-order modulation.
    \item \textbf{LMMSE} implements a linear‑MMSE compromise between sidelobe suppression and noise amplification. It approximates RF performance at high SNR and converges to MF behavior at low SNR. Its practical limitation lies in requiring accurate SNR information, which depends on the unknown scattering coefficients.
\end{itemize}

\begin{remark}[{\em Pilot-only probing}]
In pilot-only operation, sensing is confined to reserved pilot resources or repeated deterministic sequences, so that the probing symbols are known and repeatable. The receiver then reduces to standard coherent integration based on the known pilots, and de-randomization is unnecessary. The main trade-off is the sensing robustness gained from deterministic probing versus the pilot or resource overhead required to preserve communication throughput.    
\end{remark}

\begin{remark}[{\em Application to bistatic/multistatic sensing}]
The de-randomization step in \eqref{eq:derandom} presumes that the probing symbols are available at the sensing receiver. This is immediate in monostatic BS sensing, and it can also be satisfied in cooperative bistatic or multistatic deployments when the illuminator shares the probing symbols or when they can be reconstructed from the pilot pattern, applied precoders, and decoded data. In non-cooperative or passive settings, coherent processing may rely on a reference branch that captures a strong direct-path copy of the illuminator signal, as commonly done in passive radar. When neither waveform sharing nor a usable reference is available, sensing typically falls back to pilot-only operation or statistics-based processing, which incurs loss relative to waveform-aware range-Doppler focusing.    
\end{remark}

\begin{figure}[!t]
\centering
\begin{subfigure}[t]{0.49\linewidth}
\centering
\includegraphics[height=3.4cm]{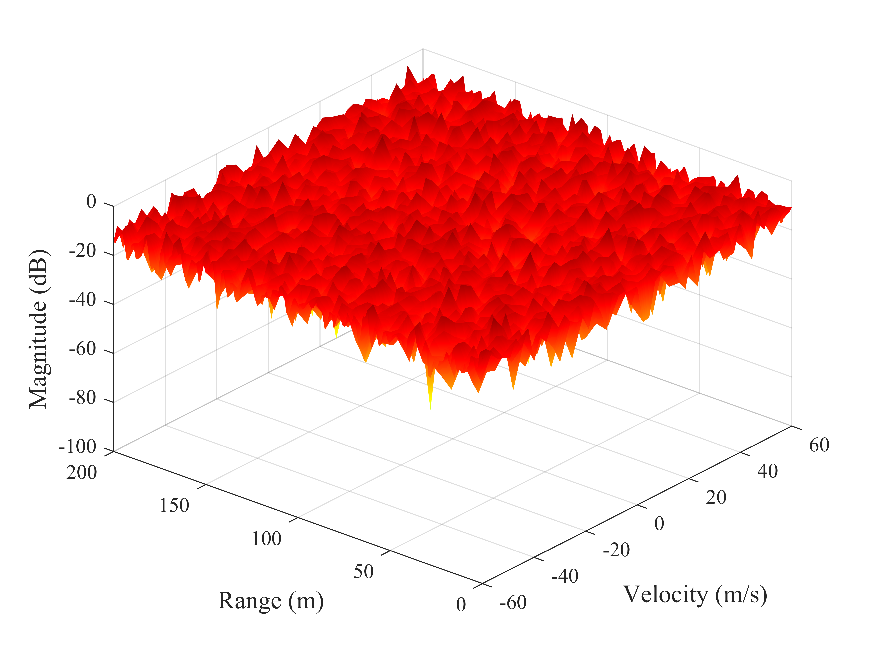}
\caption{Without de-randomization.}
\label{fig:RDM_woDerandom}
\end{subfigure} 
\begin{subfigure}[t]{0.49\linewidth}
\centering
\includegraphics[height=3.4cm]{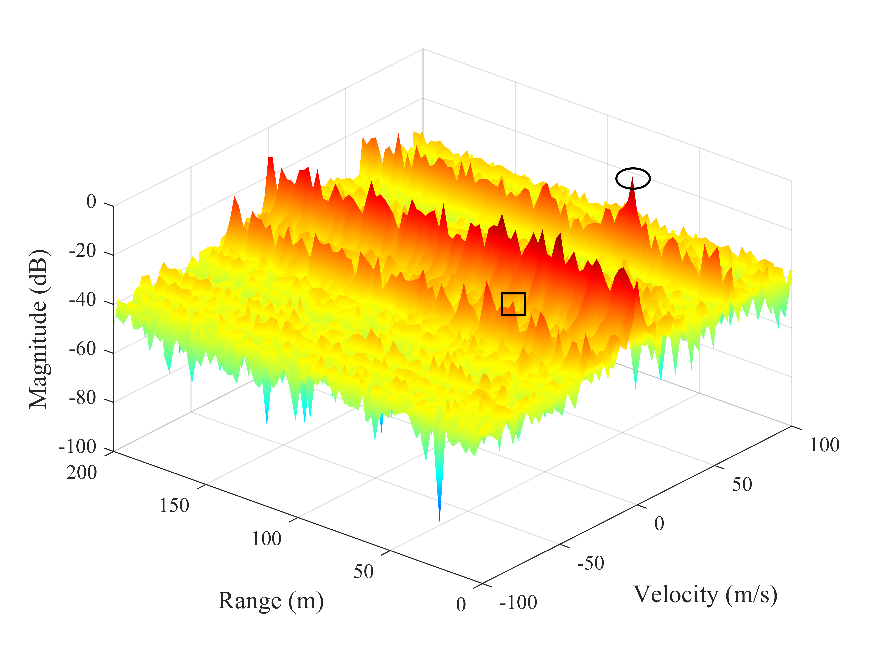}
\caption{With de-randomization.}
\label{fig:RDM_RF} \end{subfigure}
\vspace{1mm}
\caption{RDMs with and without waveform de-randomization using reciprocal filtering (rectangle: weak target of interest, ellipse: strong target, and $\text{SCNR} = -47.4$ dB).}
\label{fig:RDM_derandom}\vspace{-4mm}
\end{figure}

\subsubsection{Range-Doppler Map (RDM)}
For a de-randomized sensing channel estimate $\hat{\mathbf{H}}_p^{\text{tf}}$, the RDM is found via a 2D-DFT:
\begin{equation}\label{eq:RDM}
    \boldsymbol{\chi}_p = \mathbf{F}_N^H \hat{\mathbf{H}}_p^{\text{tf}} \mathbf{F}_L,
\end{equation}
where $\mathbf{F}_L$ and $\mathbf{F}_N^H$ are normalized DFT and inverse-DFT (IDFT) matrices of sizes $L$ and $N$, respectively. 
This operation coherently accumulates energy from identical delay-Doppler components into well-defined RD cells.

Fig.~\ref{fig:RDM_derandom} illustrates the impact of waveform de-randomization on OFDM range-Doppler focusing for a case with a weak and strong target. Using the same angle-gated measurements, we compare the RDM obtained by directly applying a 2D-DFT to $\mathbf{Y}_p^{\text{tf}}$ with the RDM obtained after waveform de-randomization in \eqref{eq:derandom} followed by the standard focusing in \eqref{eq:RDM}. Without de-randomization, the random communication symbols introduce symbol-dependent fluctuations that raise the RDM background and generate dense modulation-induced sidelobes, which obscure the targets. After de-randomization using RF, the modulation-induced artifacts are substantially reduced and the underlying delay-Doppler structure is better focused, and the target energy collapses into compact RD cells. De-randomization alone does not suppress background clutter or strong-target leakage, as indicated in Fig. \ref{fig:RDM_derandom}(b) by the pronounced Doppler clutter ridge masking the weak target (rectangle). This motivates the subsequent clutter-suppression and interference-aware processing stages.

\subsubsection{RD-Domain Detection and Estimation}
After generation of the RDM $\bm{\chi}_p$, the next step is to detect potential targets within each RD cell. The classical generalized likelihood ratio test (GLRT) provides an optimal detection framework under Gaussian and homogeneous clutter \cite{Kelly86}. Given $\mathbf{h}_p^\text{tf}=\text{vec}(\hat{\mathbf{H}}_{p}^\text{tf})\in\mathbb{C}^{NL}$, the GLRT test statistic for a hypothesized delay-Doppler cell $(\tau, f_{D})$ is 
\begin{equation}\label{eq:GLRT}
    T_{\text{GLRT}}(\tau,f_{D})=\frac{|\mathbf{v}_\text{DD}^{H}(\tau,f_{D})\mathbf{h}_p^\text{tf}|^{2}}{\|\mathbf{v}_\text{DD}(\tau,f_{D})\|_2^{2}\|\mathbf{h}_p^\text{tf}\|_2^{2}-\big|\mathbf{v}_\text{DD}^{H}(\tau,f_{D})\mathbf{h}_p^\text{tf}\big|^{2}},
\end{equation}
where $\mathbf{v}_\text{DD}(\tau,f_{D})=\mathbf{d}^{*}(f_{D})\otimes \mathbf{t}(\tau)\in\mathbb{C}^{NL\times 1}$.
Leveraging the unitarity of the 2D-DFT, the above GLRT statistic can be efficiently computed directly in the RD domain as:
\begin{equation}\label{eq:RD_GLRT}
    T_{\text{GLRT}}(n_d,n_v)=\frac{|\chi_{p}(n_d,n_v)|^{2}}{\|\chi_{p}\|_F^{2}-|\chi_{p}(n_d,n_v)|^{2}},
\end{equation}
where the corresponding cell in the RDM is  
\begin{equation}\label{eq:RD}
    \chi_{p}(n_d,n_v)=\mathbf{f}_{N}^{H}(n_d)\hat{\mathbf{H}}_{p}^\text{tf}\mathbf{f}_{L}(n_v), 
\end{equation}
and $\mathbf{f}_{N}(n_d)$, $\mathbf{f}_{L}(n_v)$ denote the $n_d$-th and $n_v$-th columns of the normalized DFT matrices of sizes $N$ and $L$, respectively, $0\leq n_d\leq N-1,\;0\leq n_v\leq L-1$.
The corresponding hypothesis test for each RD cell $(n_d,n_v)$ is
\begin{equation}\label{eq:hypothesis_test}
T_{\text{GLRT}}(n_d,n_v)\underset{H_0}{\overset{H_1}{\gtrless}}\zeta,
\end{equation}
for a given detection threshold $\zeta$.

Although the GLRT is optimal for homogeneous Gaussian interference, practical ISAC clutter is often heterogeneous and may deviate from Gaussianity. Consequently, detection with a fixed threshold can be unreliable due to local variations in the interference level. To maintain a prescribed false-alarm probability, range-Doppler constant false-alarm rate (CFAR) detectors estimate the local interference power from neighboring training  cells around the cell under test and adapt the threshold accordingly. Representative CFAR variants and comparative studies can be found in \cite{richards2005fundamentals}.

In summary, this section outlined the fundamental signal processing pipeline for MIMO-OFDM ISAC sensing, including AoA estimation, waveform de-randomization, RD-domain processing, and robust detection. However, these baseline techniques assume stationary clutter conditions with homogeneous statistics. Practical ISAC environments often deviate significantly from these idealized assumptions, resulting in performance degradation. Accurate clutter modeling and tailored clutter suppression strategies are thus crucial for reliable operation, as detailed in Sections~IV and V, respectively. 

\begin{table*}[!t]
    \centering
    \footnotesize
    \caption{Comparison of clutter amplitude distribution models}
    \renewcommand{\arraystretch}{1.4}
    \begin{tabular}{|l|c|l|l|}
        \hline
        \textbf{Distribution} & \textbf{PDF Expression} & \textbf{Characteristics} & \textbf{Pros and Cons}\\
        \hline\hline
        Rayleigh & 
        $\frac{r_a}{\sigma_\text{R}^2}\exp\left(-\frac{r_a^2}{2\sigma_\text{R}^2}\right)$ &
        Light-tailed, many small scatterers & 
        \begin{tabular}[c]{@{}l@{}}
            + Simple analytic form, good for diffuse clutter\\
            -- Poor in presence of dominant reflectors
        \end{tabular}\\
        \hline
        Log-Normal & 
        $\frac{1}{r_a\sigma\sqrt{2\pi}}\exp\left(-\frac{(\ln r_a-\mu)^2}{2\sigma^2}\right)$ &
        Moderate-tailed, multiplicative fading &
        \begin{tabular}[c]{@{}l@{}}
            + Captures shadowing/fading well\\
            -- Limited analytical convenience
        \end{tabular}\\
        \hline
        Weibull & 
        $\frac{k}{\lambda_\text{w}}\left(\frac{r_a}{\lambda_\text{w}}\right)^{k-1}\exp\left[-\left(\frac{r_a}{\lambda}\right)^k\right]$ &
        Flexible tail, empirical fit & 
        \begin{tabular}[c]{@{}l@{}}
            + Versatile tail modeling (sea, terrain)\\
            -- Requires empirical parameter fitting
        \end{tabular}\\
        \hline
        K-distribution & 
        $\frac{2b_K}{\Gamma(\nu)}\left(\frac{b_Kr_a}{2}\right)^\nu K_{\nu-1}(b_Kr_a)$ &
        Heavy-tailed, compound Gaussian & 
        \begin{tabular}[c]{@{}l@{}}
            + Accurate for spiky clutter (urban, sea)\\
            -- Higher complexity for ML estimation
        \end{tabular}\\
        \hline
    \end{tabular}
    \label{tab:clutter_models_summary}
\end{table*}

\section{Clutter Modeling and Estimation} 

Accurate clutter modeling and covariance estimation are essential for effective clutter mitigation and reliable target-parameter estimation in ISAC. Building on the wideband MIMO-OFDM model and the angle-range-Doppler data cube developed in Sections~II--III, this section surveys representative clutter statistics and practical covariance estimators under snapshot-limited operation. While the discussion is framed in the OFDM angle-range-Doppler domain, the underlying statistical and covariance modeling principles are largely waveform-agnostic. We first summarize cold-clutter statistics, then present snapshot-limited covariance learning methods for both spatial and space-time processing, including training-data selection, range focusing, and regularized/robust estimators. We further review structured covariance and sparse geometric models, extend the discussion to hot clutter statistics, and conclude with scenario-specific modeling guidelines. These interference statistics support the adaptive receive processing in Sec. V and provide the environmental signatures needed by the  waveform-adaptive transceiver designs in Sec. VI.

\subsection{Statistical Characterization of Cold Clutter}
Statistical models characterize clutter through amplitude distributions and multi-dimensional correlation structures, forming the foundation for detector design and performance prediction in ISAC systems. These models represent clutter as random processes with specific probability density functions (PDFs) and correlation properties across the spatial, temporal, and frequency domains. 
To keep the discussion focused, we summarize representative clutter amplitude distributions and their typical operating regimes in Table I, and then emphasize the space-frequency correlation structures that are particularly relevant to MIMO-OFDM ISAC. Throughout this paper, complex Gaussian modeling is used as a tractable baseline for covariance-based analysis and design. When clutter measurements exhibit heavy tails and heterogeneity, compound-Gaussian/SIRV models provide a more general description at the cost of increased estimation and processing complexity.

\subsubsection{Amplitude Distributions}

Clutter amplitude statistics depend on the underlying scattering mechanisms and are often characterized by their tail behavior, which directly impacts CFAR thresholding and false-alarm performance. Representative models ranging from light-tailed Rayleigh to heavy-tailed compound-Gaussian distributions are summarized in Table I. Rayleigh models provide a convenient and analytically tractable baseline for diffuse clutter with many comparable scatterers. Log-normal captures multiplicative effects such as shadowing and partial blockages and yields moderately heavy tails. Weibull offers flexible tail behavior and is widely used as an empirical fit for sea/terrain clutter and grazing-angle returns \cite{Schleher76}. The K-distribution and related compound-Gaussian models are well suited for spiky clutter with intermittent large excursions, commonly observed in urban and maritime environments \cite{Jakeman76}. Other amplitude models and parameter fitting procedures are well documented in the radar literature (e.g., \cite{greco2014clutter,RichardsPMR2010}). While amplitude statistics govern CFAR behavior and outlier prevalence, most suppression and beamforming designs depend on the structure of the second-order statistics in the spatial and frequency domains, as discussed next.

\subsubsection{Second-Order Space-Frequency Statistics}
Clutter returns are typically modeled as the coherent superposition of numerous independent scattering returns from the environment. According to the central limit theorem (CLT), when the number of i.i.d. scatterers is sufficiently large, the aggregate clutter echo in a given location can be accurately approximated by a complex Gaussian distribution. Under this assumption, the clutter is fully characterized by its covariance structure. 
Consequently, many of the receiver designs discussed later, e.g., sample covariance matrix (SCM)-based MVDR/STAP and MMSE-type processing, are formulated under this Gaussian baseline, since they depend primarily on second-order statistics and admit tractable covariance-based formulations.
For non-Gaussian heavy-tailed observations, the normalized shape matrix together with texture statistics should be used instead, as discussed in Sec. IV-A.3. Since in general array responses and clutter reflectivity are frequency dependent, the clutter space-frequency covariance matrix varies across the full OFDM signal spectrum, but there often exists significant correlation between adjacent subcarriers. Accurately modeling these space-frequency dependencies is essential for effective clutter suppression and robust sensing.

For the wideband MIMO-OFDM ISAC model in Sec.~\ref{sec:sig_model}, the space-frequency covariance matrix between subcarriers $n$ and $n'$ captures the joint spatial-spectral correlation of the clutter:

{\small
\begin{equation} 
\begin{aligned}\label{eq:Rcnn}
\mathbf{R}_\text{cc}[n,n'] = \!\!\sum_{c=1}^{C}\!\rho_c[n, n']\mathbf{b}_n(\theta_c)\mathbf{a}^H_n(\theta_c)\mathbf{R}_X[n,n']\mathbf{a}_{n'}(\theta_c)\mathbf{b}_{n'}^H(\theta_c),
\end{aligned}\end{equation}} 
{\flushleft{where}} $\rho_c[n,n'] = \mathbb{E}\{\beta_{c,n}\beta_{c,n'}^*\}$ and $\mathbf{R}_X[n,n'] = \mathbb{E}\{\mathbf{x}_n[\ell]\mathbf{x}^H_{n'}[\ell]\}$. This covariance captures three key mechanisms: (\textit{i}) frequency-selective scatterer reflectivity through  $\rho_c[n,n']$, (\textit{ii}) beam squint via frequency-dependent steering vectors $\mathbf{b}_n(\theta_c)$ and $\mathbf{a}_n(\theta_c)$, and (\textit{iii}) waveform-induced correlation through $\mathbf{R}_X[n,n']$. Each component presents unique challenges for ISAC systems.

\textbf{Frequency-Selective Gaussian Clutter Reflectivity}: The complex scattering coefficient $\beta_{c,n}$ for clutter patch $c$ at subcarrier frequency $f_n=f_0+n\Delta_f$ is modeled as a zero-mean complex Gaussian random variable:
\begin{equation}
\beta_{c,n} \sim \mathcal{CN}\left(0, \sigma_c^2(f_n)\right),
\end{equation}
where the variance $\sigma_c^2(f)$ represents the strength of the frequency-dependent RCS. The Gaussian assumption is valid for resolution cells encompassing numerous independent scatterers \cite{greco2014clutter,SSen-TSP-2011}. To characterize the frequency dependence, the empirical land-clutter model in \cite{RichardsPMR2010} can be adopted, which provides the clutter reflectivity $\sigma_0$ as a function of wavelength $\lambda$, depression angle $\phi_\text{dep}$, and surface roughness $\sigma_h$:
\begin{align} \sigma_0[\text{dB}] = & 10\log_{10} \Big( A_\text{GIT}(\phi_\text{dep} + C_\text{GIT})^{B_\text{GIT}}   \nonumber \\ & \hspace{1.5 cm}   \times \exp\left[-D_\text{GIT}\left(1+0.1\frac{\sigma_h}{\lambda}\right)\right]\Big), \end{align}
\noindent where $A_\text{GIT},B_\text{GIT},C_\text{GIT},D_\text{GIT}$ are terrain-specific constants \cite{greco2014clutter}. Subsequently, the frequency-dependent clutter variance at frequency $f$ is obtained by mapping the reflectivity to receiver power through a system-specific scaling factor $\xi_c$:
\begin{equation}
\sigma_c^2(f) = \xi_c \sigma_0(f;\phi_\text{dep},\sigma_h).
\end{equation}

The cross-frequency correlation of the clutter scattering coefficients is characterized by:
\begin{equation}
\mathbb{E}\{\beta_{c,n}\beta_{c,n'}^*\} = \sqrt{\sigma_c^2(f_n)\sigma_c^2(f_{n'})}\rho_f(|f_n - f_{n'}|),
\end{equation}
where $\rho_f(\Delta f)$ denotes the normalized coherence function, satisfying $|\rho_f(\Delta f)|\leq 1$ and $\rho_f(0)=1$. In typical propagation environments, empirical evidence from frequency-agile radar measurements shows that coherence decays rapidly with frequency separation \cite{greco2014clutter}. This can be modeled as an exponential decay $\rho_f(\Delta f) = \exp\left(-{|\Delta f|}/{B_\text{c}}\right)$, where $B_\text{c}$ represents the coherence bandwidth governed by the multipath delay spread. The value of $B_\text{c}$ can be determined from measured data or inferred from the average power delay profile (PDP) specific to the given environment \cite{greco2014clutter,SSen-TSP-2011}.

Classical scattering theory indicates that the frequency dependence of the clutter is closely tied to the scatterer's electrical size across the signal bandwidth. A practical engineering measure for this purpose is the bandwidth-induced electrical size change \cite{knott1993rcs}:
\begin{equation}
\Delta(ka)= 2\pi D_\text{dom}B/c_0,
\end{equation}
where $D_{\text{dom}}$ is the size of the scattering object. When $\Delta(ka)\ll 1$, the clutter RCS and array responses can be approximated as frequency-flat, simplifying full-band covariance estimation. Conversely, when $\Delta(ka)\gtrsim \mathcal{O}(1)$, the frequency-dependent scattering characteristics necessitate subband-based modeling and estimation strategies. As a practical illustration for ISAC, a scatterer dimension $D_{\text{dom}}=1$m and bandwidth $B=400$MHz leads to $\Delta(ka)\approx 8.4$, indicating that frequency selectivity is a relevant issue.

\textbf{Per-Subcarrier Spatial Covariance}: 
The diagonal blocks $\mathbf{R}_{\text{cc},n}$ represent the spatial covariance at each subcarrier: 
\begin{equation}\begin{aligned}
\mathbf{R}_{\text{cc},n} &= \mathbf{R}_\text{cc}[n,n]\\
& = \sum_{c=1}^{C}\sigma_{c,n}^2\mathbf{b}_n(\theta_c)\mathbf{a}_n^H(\theta_c)\mathbf{R}_{X,n}\mathbf{a}_n(\theta_c)\mathbf{b}_n^H(\theta_c),
\end{aligned} \end{equation} 
where $\sigma_{c,n}^2 = \mathbb{E}\{| \beta_{c,n}|^2\}$ is the power of scatterer $c$ on subcarrier $n$ and $\mathbf{R}_{X,n}\triangleq \mathbf{R}_X[n,n]$. The frequency-dependent steering vectors cause systematic variations in spatial structure across the OFDM bands. This frequency dependence impacts ISAC processing in three ways. \textit{(i)} The eigenvectors of $\mathbf{R}_{\text{cc},n}$ ``rotate'' from one subcarrier to the next. \textit{(ii)} The effective rank of $\mathbf{R}_{\text{cc},n}$ also varies with frequency as beam squint redistributes clutter energy across the spatial dimension, potentially concentrating interference towards communication users while leaving sensing directions relatively clear, or vice versa. \textit{(iii)} Furthermore, at the millimeter-wave frequencies proposed for 5G/6G ISAC, the resulting narrower beamwidths reduce clutter angular spread but increase sensitivity to beam misalignment, requiring more frequent covariance updates as communication users move. These frequency-dependent effects violate the traditional assumption of frequency-invariant spatial processing. For ISAC systems, this necessitates per-resource-block covariance tracking, increasing computational and memory requirements compared to narrowband approaches. 

\textbf{Cross-Frequency Correlation}: 
The off-diagonal blocks $\mathbf{R}_\text{cc}[n,n']$ for $n\neq n'$ capture cross-frequency correlation in the clutter returns. Standard OFDM communication systems employ independent data symbols across subcarriers, yielding
\begin{align}
    \mathbf{R}_X[n,n']=\mathbf{0},\quad n\neq n'.
\end{align}
Under Gaussian clutter assumptions, this independence produces a block-diagonal covariance structure that simplifies per-subcarrier processing.
ISAC systems, however, require coherent processing across subcarriers to achieve adequate range resolution. Sensing waveforms must maintain phase relationships through structured pilots or repeated sequences, inducing transmit correlation:
\begin{equation}
\mathbf{R}_X[n,n'] = \mathbb{E}\{\mathbf{x}_n[\ell]\mathbf{x}_{n'}^H[\ell]\} \neq \mathbf{0}, \quad n \neq n'.
\end{equation} 
These coherent waveforms introduce notable off-diagonal terms in the clutter covariance structure, forming a banded matrix concentrated around the main diagonal. The correlation bandwidth of these off-diagonal elements depends on waveform design parameters and sensing resolution requirements.  
For ISAC processing, this cross-frequency correlation presents both opportunities and challenges, enabling improved clutter estimation through subcarrier averaging, but producing a banded space-frequency covariance that breaks strict per-subcarrier independence. These space-frequency correlation properties also motivate the subband/overlapping-window covariance learning strategies discussed in Sec. IV-B.

\subsubsection{Non-Gaussian Clutter Modeling} 

In high-resolution scenarios, particularly at low grazing angles or in heterogeneous environments such as urban canyons or sea surfaces, clutter frequently deviates from classical Gaussian statistics, exhibiting prominent heavy-tailed behavior and intermittent high-intensity spikes. Such situations motivate the adoption of compound-Gaussian models, which separate slowly varying local power fluctuations (texture) from rapidly changing coherent scattering (speckle). 
Compared with the Gaussian baseline, SIRV/compound-Gaussian models \cite{greco2014clutter} can better match empirical clutter statistics observed in fine resolution cells, as the limited number of scatterers per cell weakens the Gaussian assumption. However, these models typically make inference more challenging and motivate robust shape-matrix (covariance) estimation methods, as discussed in Sec. IV-B.4.

In SIRV models, the non-Gaussian clutter is represented as 
\begin{equation}
\mathbf{y}[\ell] = \sqrt{\kappa[\ell]}\mathbf{g}[\ell], 
\end{equation}
where the scalar texture parameter $\kappa[\ell]>0$ captures the slow temporal variations in clutter power, and the vector $\mathbf{g}[\ell]\sim\mathcal{CN}(\mathbf{0},\bm{\Sigma})$ represents fast coherent fluctuations (speckle). The spatial correlation structure is embedded in the shape (covariance) matrix $\bm{\Sigma}$, identifiable only up to a scale factor and conventionally normalized as $\text{tr}(\bm{\Sigma})=N_\text{r}$. In this formulation, the clutter covariance naturally factorizes as $\mathbf{R}=\mathbb{E}\{\kappa\}\bm{\Sigma}$. This model can be interpreted as a simplified, short time-on-target instance of the broader class of spherically invariant random processes (SIRP) model. Commonly used texture distributions such as the K-distribution flexibly capture clutter heterogeneity and tail heaviness, while the shape matrix $\bm{\Sigma}$ characterizes instantaneous spatial correlation.

\subsection{Practical Covariance Estimation in MIMO-OFDM}

Covariance estimation is the bridge between the statistical clutter models in Sec. IV-A and the adaptive processing/design tasks considered later. On each subcarrier $n$, the disturbance seen by the receiver can be decomposed into waveform-dependent cold clutter, whose covariance is colored by the realized probing matrix $\mathbf{X}_n$, and a waveform-independent component $\bm{\eta}_n$ that aggregates hot clutter, receiver thermal noise, and residual interference. Two estimation products are useful in clutter-aware ISAC: (\textit{i}) an instantaneous interference-plus-noise covariance for receiver-side adaptation within a CPI, and (\textit{ii}) a waveform-independent scene signature (e.g., an inner clutter kernel) that enables predicting the interference covariance for arbitrary candidate probing waveforms. Wideband OFDM further complicates both because range-frequency coupling spreads returns from a given delay across all subcarriers, making training-snapshot selection critical. We therefore describe a practical snapshot-formation pipeline for target-free and target-present conditions (including range focusing when needed), review regularized/robust covariance estimators for snapshot-limited operation, and finally discuss how separate estimates of $\mathbf{R}_{\bm{\eta},n}$ can be leveraged to learn waveform-independent clutter kernels.

\begin{remark}[{\em Spatial vs. space-time snapshots}] 
Unless otherwise stated, Sections~IV-B.1--IV-B.4 focus on \emph{spatial} snapshots
$\mathbf{y}_n[\ell]\in\mathbb{C}^{N_\text{r}}$ and the corresponding $N_\text{r}\times N_\text{r}$ disturbance covariance
$\mathbf{R}_{I,n}=\mathbb{E}\{\mathbf{y}_n[\ell]\mathbf{y}_n^H[\ell]\}$.
The extension to STAP is obtained by stacking $L$ slow-time snapshots into
$\mathbf{y}^{\text{st}}_n[m]\in\mathbb{C}^{N_\text{r}L}$ and estimating the $N_\text{r}L\times N_\text{r}L$
space--time covariance $\mathbf{R}^{\text{st}}_{I,n}=\mathbb{E}\{\mathbf{y}^{\text{st}}_n[m](\mathbf{y}^{\text{st}}_n[m])^H\}$,
where the training index $m$ enumerates \emph{secondary} snapshots (e.g., training range gates and/or multiple CPIs),
since $\ell$ is already absorbed into the stacked vector.
\end{remark}

\subsubsection{Target-Free Case}
When no targets are present within the training snapshots, i.e., $\mathbf{y}_n[\ell] = \mathbf{y}_{\text{c},n}[\ell] + \bm{\eta}_n[\ell]$, the process of estimating the interference-plus-noise covariance is considerably simplified and can be performed directly from frequency-domain snapshots without range focusing, since no target excision is required.
Under the assumption of local-homogeneity over the selected training snapshots, one can employ classical SCM-based estimation \cite{Kay1993} on each subcarrier:
\begin{equation}\label{eq:SCM no target}
\widehat{\mathbf{R}}_{I,n}
=\frac{1}{N_\text{tr}}\sum_{\ell\in\mathcal{T}}\mathbf{y}_{n}[\ell]\mathbf{y}^H_{n}[\ell],
\end{equation}
where $\mathcal{T}$ is the training set and $|\mathcal{T}| = N_\text{tr}$. To ensure reliable estimation, the Reed-Mallett-Brennan (RMB) guideline recommends a minimum of $N_\text{tr}\ge2D$ independent samples, where $D$ denotes the DoFs processed. Specifically, $D=N_\text{r}$ applies for spatial-only processing, while $D=N_\text{r}L$ applies for space-time processing involving $L$ symbols. 
In addition to sample support, covariance estimation quality also depends on the slow-time structure of the probing waveform used to collect the secondary data, especially for space-time covariance learning. Deterministic sensing waveforms are generally preferable, since temporally white communication symbols tend to whiten the slow-time statistics after averaging and weaken clutter Doppler selectivity.
In practice, only a small number of slow-time symbols ($L\approx 12-14$) may be available per CPI. Thus, if one relies only on per-symbol snapshots for spatial SCM learning, $N_\text{tr}\leq L$ can be insufficient when $N_\text{r}$ is large. For STAP, the adaptive dimension grows to $D=N_\text{r}L$ and reliable estimation of $\mathbf{R}_{I,n}^{\text{st}}$ typically requires additional independent secondary snapshots, e.g., across training range gates or multiple CPIs. Note that the clutter statistics often vary relatively slowly with frequency, and in such cases it is not necessary to estimate the clutter covariance at every subcarrier. Instead, one can divide the total bandwidth into multiple subbands and estimate a covariance that can be applied at the subband level:
\begin{equation}\label{eq:SCMb no target}
\widehat{\mathbf{R}}_{I,b}
=\frac{1}{N_\text{tr}N_\text{sb}}\sum_{n\in\mathcal{N}_b}\sum_{\ell\in\mathcal{T}}\mathbf{y}_{n}[\ell]\mathbf{y}_{n}^H[\ell],
\end{equation}
where $\mathcal{N}_b$ is the set of subcarriers in the $b$-th subband and $N_\text{sb} = |\mathcal{N}_b|$. 
This reduces computation without significantly sacrificing accuracy when the subband width is chosen within the clutter coherence bandwidth. To obtain an individual covariance estimate at each subcarrier, one can implement the subband averaging using a sliding window with overlapping frequency windows, trading off estimation variance versus frequency resolution in a controlled manner.

\subsubsection{Target-Present Case}
When a target echo may be present in the data, care must be taken to eliminate the target contribution before estimating the clutter covariance, to avoid target nulling during adaptive beamforming. In classical pulsed radars this is achieved by range gating after time-domain matched filtering; range cells under interrogation that may contain a target are excluded and the covariance is formed from neighboring cells. In OFDM sensing, by contrast, range is encoded as a linear phase progression across subcarriers, so a target at delay $\tau$ leaves a coherent signature on all subcarriers. As a result, simple per-subcarrier excision is ineffective. The remedy is to convert the subcarrier-domain measurements to the delay domain, and remove the contributions of potential target range cells before forming the clutter covariance.
Alternatively, if sufficiently accurate target parameters are available to synthesize a reliable target replica, one may first perform target cancellation in the frequency domain and then apply the per-subcarrier SCM in \eqref{eq:SCM no target}, although in the discussion below we emphasize the range-gated training approach.

To realize this target-aware training mechanism, the receiver follows a slightly modified version of the sensing pipeline described in Sec. III. After obtaining an initial AoA estimate that defines the angular sector $p$, the received samples on each subcarrier are retained without applying spatial beamforming, so that all spatial DoFs remain available for subsequent adaptive clutter suppression. The preserved samples at subcarrier $n$ and OFDM symbol $\ell$ can be written as 
\begin{equation}\label{eq:array_preserving}
\begin{aligned}
\widetilde{\mathbf{y}}_{p,n}[\ell] &= \sum_{i=1}^{M_{p}} \alpha_{p,i} e^{\jmath 2\pi(f_{\text{D},p,i}\ell T_{\text{sym}}-n\Delta_f\tau_{p,i})} x_{p,n,\ell}\mathbf{b}_n(\hat{\theta}_p)  \\
&  + \sum_{j=1}^{C_{p}} \beta_{p,j} e^{\jmath 2\pi(f_{\text{D},p,j}\ell T_{\text{sym}}-n \Delta_f \tau_{p,j})} x_{p,n,\ell}\mathbf{b}_n(\hat{\theta}_p) \!+\! \mathbf{z}^{\text{h}}_{p,n,\ell}.
\end{aligned}\end{equation}
Subsequently, the de-randomization in \eqref{eq:derandom} is applied to remove the known modulation-induced component $x_{p,n,\ell}$. We denote the resulting de-randomized spatial snapshot as $\mathbf{y}_{p,n}[\ell]$. For deterministic/pilot-only probing, $\mathbf{x}_{p,n}[\ell]$ is known and fixed, and this de-randomization step is unnecessary. Range focusing is performed next to coherently aggregate energy dispersed across subcarriers into discrete range bins. For each angular sector $p$, discrete range gate $\tau$, and OFDM symbol $\ell$, the corresponding range-gated spatial snapshot is obtained through the following frequency-to-delay transformation:
\begin{equation}\label{eq:range_focusing} 
\mathbf{y}_{p,\tau}[\ell] = \frac{1}{\sqrt{N}}\sum_{n=0}^{N-1} e^{\jmath2\pi n\Delta_f\tau}\mathbf{y}_{p,n}[\ell]\in\mathbb{C}^{N_\text{r}}.
\end{equation} 
This focusing operation results in a structured three-dimensional radar data cube, indexed by angular sector $p$, range gate $\tau$, and slow-time $\ell$. This localizes any target to a small number of range gates whose data can be excised as needed.

For each range cell $\tau_r$ of interest, a number of guard cells are introduced to mitigate target leakage into the training set. Training snapshots are then collected from neighboring range cells, assuming local homogeneity of the clutter. The interference-plus-noise covariance associated with $\tau_r$ is then estimated using the classical SCM:
\begin{equation}\label{eq:SCMp}
\widehat{\mathbf{R}}_{I,p}(\tau_r) = \frac{1}{|\mathcal{S}(p,\tau_r)|}\sum_{(\ell,\tau)\in\mathcal{S}(p,\tau_r)}
\mathbf{y}_{p,\tau}[\ell]\mathbf{y}^H_{p,\tau}[\ell],
\end{equation}
where $\mathcal{S}(p,\tau_r)$ is the set of selected training indices $(\ell,\tau)$ drawn from neighboring cells, with cardinality $|\mathcal{S}(p,\tau_r)|$.

After performing range focusing over the entire OFDM bandwidth, coherent accumulation across subcarriers significantly improves the signal-to-clutter-and-noise ratio (SCNR). However, this full-band processing inherently averages out the frequency-selective characteristics intrinsic to wideband clutter, such as those introduced by beam squint and frequency-dependent RCS variations. When the clutter coherence bandwidth $B_\text{c}$ is substantially smaller than the signal bandwidth $B$, a single full-band covariance estimate becomes mismatched, and covariance estimation should be carried out over a narrower frequency subband, as described above. 
To implement the range gating method described above, range focusing is performed for each subband $b$ using only the subcarriers in $\mathcal{N}_b$, yielding subband-specific range-gated snapshots: 
\begin{equation} 
\mathbf{y}_{p,b,\tau}[\ell] = \frac{1}{\sqrt{N_\text{sb}}}\sum_{n \in \mathcal{N}_b} e^{\jmath2\pi n\Delta_f \tau}\mathbf{y}_{p,n}[\ell] \in \mathbb{C}^{N_\text{r}}. 
\end{equation} 
The clutter covariance at each range gate is then estimated within each subband by applying SCM to the corresponding target-free training set:
\begin{equation} \label{eq:SCM pb}
\widehat{\mathbf{R}}_{I,p,b}(\tau_r) = \frac{1}{|\mathcal{S}(p,\tau_r)|}\sum_{(\tau,\ell) \in \mathcal{S}(p,\tau_r)} \!\!\mathbf{y}_{p,b,\tau}[\ell]\mathbf{y}_{p,b,\tau}^H[\ell].
\end{equation}  

\subsubsection{Regularized Estimators}

In practical ISAC deployments, particularly those aligned with standardized protocols such as 5G NR, only a limited number of slow-time snapshots may be available per CPI. This can lead to poorly conditioned SCM estimates especially for high-dimensional space-time covariance matrices. In this section we discuss methods to improve the conditioning of SCM estimates obtained using any of the methods in \eqref{eq:SCM no target}, \eqref{eq:SCMb no target}, \eqref{eq:SCMp}, or \eqref{eq:SCM pb}. We will focus on spatial covariance estimates of dimension $N_\text{r}\times N_\text{r}$, but identical methods can be applied for space-time covariance estimation where the dimension is $N_\text{r}L$. When subcarrier-wise processing is adopted, each subcarrier $n$ can be associated with either a shared full-band estimate under frequency-stationarity, or a local/subband estimate under frequency selectivity.

Shrinkage estimators offer a statistically sound approach to regularization in small-sample scenarios by combining the empirical SCM with a structured shrinkage component \cite{Ledoit04}:
\begin{equation}
\widehat{\mathbf{R}}_{\text{shr}}
=(1-\alpha)\widehat{\mathbf{R}}_{\text{SCM}}+\alpha\,\mu\,\mathbf{I}_{N_\text{r}}, \quad
\mu= \text{tr}(\widehat{\mathbf{R}}_{\text{SCM}})/N_\text{r},
\end{equation}
where the shrinkage factor $\alpha$ balances estimation bias and variance. A widely used method for determining $\alpha$ is the oracle-approximating shrinkage (OAS) estimator \cite{YChen-TSP-2010}:
\begin{equation}
\alpha_{\text{OAS}}=
\frac{\bigl(1-\tfrac{2}{N_\text{r}}\bigr)\text{tr}(\widehat{\mathbf{R}}_{\text{SCM}}^{2})
+\text{tr}^{2}(\widehat{\mathbf{R}}_{\text{SCM}})}
{(N_\text{ts}+1-\tfrac{2}{N_\text{r}})\bigl[\text{tr}(\widehat{\mathbf{R}}_{\text{SCM}}^{2})
-\tfrac{1}{N_\text{r}}\text{tr}^{2}(\widehat{\mathbf{R}}_{\text{SCM}})\bigr]},
\end{equation}
which is subsequently projected onto the interval $[0,1]$. The scalar $N_\text{ts}$ denotes the number of training samples, i.e., $N_\text{ts}=N_\text{tr}$ or $N_\text{ts} = |\mathcal{S}(p,\tau_r)|$. This data-driven approach minimizes the expected Frobenius risk and typically outperforms fixed diagonal loading methods, especially when $L < N_\text{r}$.

When the array geometry exhibits structural symmetries, additional spatial processing can be employed to enhance covariance estimation without requiring more data. For instance, ULAs that possess centro-Hermitian symmetry can benefit from forward-backward averaging 
(FBA) \cite{van2002optimum}: 
\begin{equation}
\widehat{\mathbf{R}}_{\text{FB}} 
=\tfrac{1}{2}\bigl(\widehat{\mathbf{R}}_{\text{SCM}} 
+\mathbf{J}\widehat{\mathbf{R}}_{\text{SCM}}^{*}\mathbf{J}\bigr),
\end{equation}
where $\mathbf{J}$ is the anti-identity (exchange) matrix. This averaging operation enforces persymmetry and improves the variance and condition number of the covariance estimate.

Environments with strong specular clutter reflections frequently result in (near) rank deficient covariance matrices. For ULAs, spatial smoothing (SS) partitions the array into subarrays and averages the subarray covariances to provide additional statistical diversity \cite{van2002optimum}. For a chosen subarray length $F$, the number of forward subarrays is $\tilde{F}=N_\text{r}-F+1$, and the SS covariance is given by  
\begin{equation}
\widehat{\mathbf{R}}_{\text{SS}}
=\frac{1}{\tilde{F}}\sum_{i=1}^{\tilde{F}}\mathbf{S}_{i}
\widehat{\mathbf{R}}_{\text{SCM}} 
\mathbf{S}_{i}^{H},
\end{equation}
where $\mathbf{S}_{i}\!\in\!\{0,1\}^{F\times N_\text{r}}$ is a selection matrix. In ISAC systems, it is critical to choose $F$ large enough to decorrelate coherent clutter components while maintaining sufficient spatial DoFs to accommodate multi-user communication beamforming. If sensing and communication occupy separate time-frequency resources, this constraint can be relaxed.

\subsubsection{Robust Estimation under Heavy-tailed Samples}
When the amplitude statistics are heavy-tailed or training snapshots contain outliers and heterogeneity, Gaussian SCM-based learning can be unreliable. Under the SIRV model in Sec. IV-A.3, robust estimators of the normalized shape matrix provide improved resilience.
Clutter covariance estimation within the SIRV framework typically proceeds in two steps, first estimating the normalized shape matrix $\bm{\Sigma}$ and then determining the mean texture $\mathbb{E}\{\kappa\}$ or related parameters. Robust covariance estimators such as Tyler's M-estimator and its structured or regularized variants are especially effective in handling heavy-tailed or heterogeneous training samples and remain robust even in small-sample scenarios, as detailed in \cite{Zoubir2018}. In practice, employing the SIRV model is justified when empirical data exhibit pronounced non-Gaussian features, such as excess kurtosis or notable heterogeneity across neighboring range-angle cells. Under such conditions, SIRV-based estimators consistently outperform Gaussian-based SCM methods.

\subsubsection{Waveform-Independent Inner Clutter Covariance Estimation in ISAC Systems}
The SCM-type estimators in Sections~IV-B.1--IV-B.4 learn the \emph{spatial} disturbance covariance $\mathbf{R}^{\text{sp}}_{I,n}\in\mathbb{C}^{N_\text{r}\times N_\text{r}}$ from per-symbol snapshots $\mathbf{y}_n[\ell]$. The extension to STAP replaces $\mathbf{y}_n[\ell]$ by stacked space-time snapshots and yields $\mathbf{R}^{\text{st}}_{I,n}\in\mathbb{C}^{N_\text{r}L\times N_\text{r}L}$, whose estimation typically relies on independent secondary data indexed by training range gates and/or multiple CPIs. Unlike conventional radar systems that employ an identical periodically pulsed waveform, ISAC systems that perform joint sensing and communication on the same resource blocks (RBs) employ waveforms that change with each CPI, due to the presence of random communication data. Consequently, even if the physical scene is unchanged, the cold-clutter covariance observed at the receiver changes with the illuminating waveform. 
Kernel learning is therefore most reliable when the secondary data are collected using probing waveforms with explicit slow-time structure, since temporally independent data symbols tend to whiten the slow-time statistics after averaging and weaken the clutter Doppler signature in the second-order statistics.
For joint transmit--receive optimization, it is therefore desirable to factor the cold-clutter covariance into a waveform-dependent term and an underlying waveform-independent clutter signature. Since this signature depends on whether the receiver performs STAP or spatial-only processing, we next discuss two corresponding kernel formulations: a space-time inner clutter kernel associated with the space-time waveform $\mathbf{X}_n$, and a spatial kernel associated with spatial-only  beamforming via $\mathbf{W}_n$.

Based on the signal model and definitions in \eqref{eq:y_nl}--\eqref{eq:yn def}, the stacked cold-clutter on subcarrier $n$ can be written as 
\begin{align}
    \mathbf{y}_{\text{cc},n} = \sum_{c=1}^C\beta_{c,n}e^{-\jmath 2\pi n\Delta_f\tau_c}\mathbf{X}_n\mathbf{v}_n(\theta_c,f_{\text{D},c}),
\end{align}
where $\beta_{c,n}$ is the reflection coefficient of clutter patch $c$. Under the standard uncorrelated patch model, $\mathbb{E}\{\beta_{c,n}\beta_{c',n}^*\}=0,~\forall c\neq c'$, and $\mathbb{E}\{|\beta_{c,n}|^2\} = \sigma_{c,n}^2$. 
The space-time cold-clutter covariance on subcarrier $n$ is defined as $\mathbf{R}^{\text{st}}_{\text{cc},n}\triangleq \mathbb{E}\{\mathbf{y}_{\text{cc},n}\mathbf{y}_{\text{cc},n}^H\}$, which in practice can be estimated from target-free secondary snapshots. For a given CPI with the probing matrix $\mathbf{X}_n$, the SCM estimate of $\mathbf{R}^{\text{st}}_{\text{cc},n}$ is given by:
\begin{equation}\label{eq:RccXVX}
\widehat{\mathbf{R}}^{\text{st}}_{\text{cc},n}(\mathbf{X}_n) \approx \mathbf{X}_n \mathbf{V}_{\text{cc},n}^{\text{st}}\mathbf{X}_n^H,
\end{equation}
where the waveform-independent inner clutter kernel is
\begin{align}\label{eq:Vccn def}
\mathbf{V}_{\text{cc},n}^{\text{st}} \triangleq \sum_{c=1}^{C}\sigma_{c,n}^{2}
\mathbf{v}_n(\theta_c, f_{\text{D},c}) \mathbf{v}_n^{H}(\theta_c, f_{\text{D},c}).
\end{align}
The ``inner'' clutter kernel $\mathbf{V}_{\text{cc},n}^{\text{st}}$ depends only on the scene geometry and the angle-Doppler scattering power, but not on $\mathbf{X}_n$. Once $\mathbf{V}_{\text{cc},n}^{\text{st}}$ is learned, the cold-clutter covariance induced by any candidate probing waveform follows immediately via~\eqref{eq:RccXVX}, which is particularly useful for the joint waveform-receiver optimization approaches discussed in Sec. VI-B. In practice, $\mathbf{V}^{\text{st}}_{\text{cc},n}$ should be learned from training range bins outside the cell under test, with guard cells to prevent target leakage. Otherwise, if target returns contaminate the training set, the subsequent adaptive processor may inadvertently suppress the target along with the clutter. Moreover, kernel learning is most meaningful when hot clutter is absent or can be accounted for separately, since hot clutter is independent of $\mathbf{X}_n$.

A practical approach is covariance fitting across multiple probing realizations. Let $\widehat{\mathbf{R}}_{I,n}^{\text{st},i}$ denote a sample estimate obtained from target-free training cells in the $i$-th CPI, and let $\mathbf{X}_n^{(i)}$ be the corresponding known probing waveform. When an estimate of the waveform-independent disturbance covariance $\mathbf{R}_{\bm{\eta},n}^{\text{st}}$ is available, we first isolate the cold-clutter contribution $\widehat{\mathbf{R}}_{\text{cc},n}^{\text{st},i}\approx \widehat{\mathbf{R}}_{I,n}^{\text{st},i}-\widehat{\mathbf{R}}_{\bm{\eta},n}^{\text{st}}$, and then recover the kernel by solving
\begin{align}
    \widehat{\mathbf{V}}_{\text{cc},n}^{\text{st}} = \arg\min_{\mathbf{V} \succeq 0}
\sum_{i=1}^{I}\|\widehat{\mathbf{R}}_{\text{cc},n}^{\text{st},i}
- \mathbf{X}_n^{(i)} \mathbf{V} (\mathbf{X}_n^{(i)})^H\|_F^2
+ \lambda \mathcal{P}(\mathbf{V}),
\end{align}
where $\mathcal{P}(\mathbf{V})$ can encode priors such as low-rank and Kronecker structures reviewed in the next section to improve sample efficiency. The resulting $\widehat{\mathbf{V}}_{\text{cc},n}^{\text{st}}$ can then be used to predict  $\mathbf{R}^{\text{st}}_{\text{cc},n}(\mathbf{X}_n)$ for any candidate waveform during transceiver optimization. In Sec. IV-E we discuss how $\mathbf{R}_{\bm{\eta},n}^{\text{st}}$  can be obtained in cooperative and non-cooperative hot-clutter settings, enabling the isolation of $\mathbf{R}^{\text{st}}_{\text{cc},n}$ for kernel learning.

For spatial-only processing, the relevant waveform-dependent object is the $N_\text{r}\times N_\text{r}$ receive-side spatial clutter covariance on each subcarrier. Under the downlink ISAC transmit model $\mathbf{x}_n[\ell] = \mathbf{W}_n\mathbf{s}_n[\ell]$ with $\mathbb{E}\{\mathbf{s}_n[\ell]\mathbf{s}^H_n[\ell]\} = \mathbf{I}$, the per-symbol transmit covariance is $\mathbf{R}_{X,n}\triangleq \mathbb{E}\{\mathbf{x}_n[\ell]\mathbf{x}^H_n[\ell]\} = \mathbf{W}_n\mathbf{W}_n^H$. The resulting spatial cold-clutter covariance on subcarrier $n$ can be expressed as
\begin{align}\label{eq:RcWn}
    \mathbf{R}^{\text{sp}}_{\text{cc},n}(\mathbf{W}_n) = \sum_{c=1}^C\sigma^2_{c,n}\mathbf{a}_n^H(\theta_c)\mathbf{W}_n\mathbf{W}_n^H\mathbf{a}_n(\theta_c)\mathbf{b}_n(\theta_c)\mathbf{b}_n^H(\theta_c).
\end{align}
The dependence of the clutter on the transmit precoder is clear from~\eqref{eq:RcWn}, which also separates the influence of the precoder from the scene-dependent parameters. The mapping from $\mathbf{W}_n\mathbf{W}_n^H$ to $\mathbf{R}^{\text{sp}}_{\text{cc},n}$ can be written explicitly as
\begin{align}
    \text{vec}\{\mathbf{R}^{\text{sp}}_{\text{cc},n}(\mathbf{W}_n)\} = \mathbf{V}^{\text{sp}}_{\text{cc},n}\text{vec}\{\mathbf{W}_n\mathbf{W}_n^H\},
\end{align}
where the spatial kernel depends only on the environment:
\begin{align}
\mathbf{V}^{\text{sp}}_{\text{cc},n} \triangleq \sum_{c=1}^C\sigma_{c,n}^2\text{vec}\{\mathbf{b}_n(\theta_c)\mathbf{b}_n^H(\theta_c)\}\text{vec}^H\{\mathbf{a}_n(\theta_c)\mathbf{a}_n^H(\theta_c)\}.    
\end{align}
Compared with the space-time kernel in \eqref{eq:Vccn def}, the spatial-only formulation captures the receive-side angular covariance after temporal averaging, and it depends on the probing strategy only through the transmit covariance $\mathbf{W}_n\mathbf{W}_n^H$.

In summary, Sec. IV-B has reviewed covariance learning for ISAC, including training-data selection under target-free/target-present conditions, range focusing, and regularized/robust estimators, and highlighted the need for waveform-independent clutter signatures in adaptive transmission designs. The methods above treat the covariance as largely unstructured. We next introduce structured covariance and sparse geometric models that incorporate physical constraints to further improve sample efficiency and interpretability.

\subsection{Structured Covariance Models}

Imposing physically motivated structure on the clutter covariance estimate, such as separable space-frequency correlation, spatial stationarity, or low-rank-plus-noise representations, greatly reduces the number of free parameters compared with an unstructured covariance. This reduction lowers the required training support and enables numerically stable adaptive suppression. The models in this subsection are introduced for frequency-domain snapshots prior to full-band range focusing. After full-band range focusing, frequency selectivity is no longer explicitly tracked in the range-gated covariance, and spatial-only structures often suffice. Below we review five representative structures: Kronecker separability, Toeplitz spatial stationarity, low-rank-plus-noise, space-time autoregressive (STAR) models, and selection-induced (thinned) covariance models.

\subsubsection{Kronecker Model}
When spatial and frequency correlations arise from approximately independent physical mechanisms, such as mild beam squint combined with near wide-sense stationary uncorrelated scattering, the space-frequency covariance is approximately separable:
\begin{equation}\label{eq:Kronecker model}
\mathbf{R}_{\text{sf}}
~\triangleq~
\mathbb{E}\big\{\text{vec}(\mathbf{Y})\,\text{vec}(\mathbf{Y})^{H}\big\}
~\approx~
\mathbf{R}_{\text{sp}} \otimes \mathbf{R}_{\text{fr}},
\end{equation}
where $\mathbf{Y}\!\in\!\mathbb{C}^{N_\text{r}\times N_{\text{sb}}}$ stacks antenna observations across subcarriers and $N_\text{sb}$ is the number of subcarriers in the considered subband. 
Such structures have been extensively studied in high‑dimensional covariance estimation and shown to yield improved statistical convergence and robustness compared with unstructured estimators \cite{Tsiligkaridis13}.
By leveraging this separability, the Kronecker model reduces the number of free parameters from $\mathcal{O}(N_\text{r}^2 N_{\text{sb}}^2)$ in the unstructured case to $\mathcal{O}(N_\text{r}^2 + N_{\text{sb}}^2)$, while preserving the dominant correlation geometry that governs clutter statistics.

\subsubsection{Toeplitz Spatial Model}  

For a ULA receiver with locally wide-sense stationary clutter, the spatial covariance can be approximated by a Hermitian Toeplitz matrix with entries $[\mathbf{R}_{\text{sp}}]_{i,j}=r_{\text{sp}}(|i-j|)$ \cite{van2002optimum}. This form reflects the spatial stationarity along the array aperture and admits a straightforward and statistically efficient estimator that averages the diagonals of the sample covariance matrix:
\begin{align}
r_{\text{sp}}(k)=\frac{1}{N_\text{r}-k}\sum_{i=1}^{N_\text{r}-k}[\widehat{\mathbf{R}}_{\text{SCM}}]_{i,i+k},
\end{align}
where $k = 0,1,\dots,N_\text{r}-1$.
When the spatial coherence is limited, a banded Toeplitz model that retains only the first few diagonals reduces estimation variance while preserving the physically meaningful correlations. The Toeplitz approximation is less accurate in sectorized or spatially nonstationary scenarios, e.g., near-field regimes typical of extra-large MIMO or terahertz arrays, where the spatial phase progression is no longer shift-invariant. In such cases, subarray-based modeling or explicit near-field parameterizations are more appropriate.

\subsubsection{Low-Rank plus Noise Representation} In clutter environments dominated by a few strong scatterers, the spatial covariance can be accurately modeled by a low‑rank signal term superimposed on an isotropic noise floor \cite{wax1985detection}:
\begin{equation}
\mathbf{R}_{\text{sp}} \approx \mathbf{U}_r \bm{\Lambda}_r \mathbf{U}_r^{H} + \sigma^2 \mathbf{I}, \quad r \ll N_\text{r},
\end{equation}
where $\mathbf{U}_r$ contains the eigenvectors associated with the $r$ largest eigenvalues $\bm{\Lambda}_r$, and $\sigma^2$ denotes the spatially white noise power. The effective rank $r$ can be determined by analyzing the eigenvalue spectrum through tools such as the Akaike Information Criterion (AIC) or the Minimum Description Length (MDL) approach \cite{wax1985detection}. Diagonal loading or shrinkage methods can be used to ensure numerical stability. 

\subsubsection{Parametric AutoRegressive Model} 
The STAR model provides a compact parametric clutter description that assumes a low-order vector autoregressive (VAR) relationship in slow time \cite{PParker-TAES-2003}. Let $\mathbf{y}_n[\ell]\in\mathbb{C}^{N_\text{r}}$ denote the $N_\text{r}$-antenna array snapshot on subcarrier $n$ at slow-time index $\ell$ in an interval where the clutter can be treated as approximately stationary. STAR assumes that there exist matrices $\{\mathbf{H}_i\}_{i=0}^{L_\text{AR}-1}$, $\mathbf{H}_i\in\mathbb{C}^{M'\times N_\text{r}}$ with $M'\leq N_\text{r}$, such that 
\begin{equation}\label{eq:STAR model}
\sum_{i=0}^{L_\text{AR}-1}\mathbf{H}_i\mathbf{y}_n[\ell+i]\approx \mathbf{0},\quad \ell=1,\dots,L_{\text{win}},
\end{equation}
where $L_\text{AR}$ is the model order and $L_{\text{win}}$ is the number of available secondary slow-time samples. To express \eqref{eq:STAR model} compactly, define $\mathbf{H}\triangleq[\mathbf{H}_0\ \mathbf{H}_1\ \cdots\ \mathbf{H}_{L_\text{AR}-1}]\in\mathbb{C}^{M'\times N_\text{r}L_\text{AR}}$, and form $\mathbf{Y}_\ell\triangleq[\mathbf{y}_n[\ell],\dots,\mathbf{y}_n[\ell+L_\text{AR}-1]]\in\mathbb{C}^{N_\text{r}\times L_\text{AR}}$, and $\mathbf{E}\triangleq[\text{vec}(\mathbf{Y}_1),\dots,\text{vec}(\mathbf{Y}_{L_\text{win}})]\in\mathbb{C}^{N_\text{r}L_\text{AR}\times L_\text{win}}$. Then \eqref{eq:STAR model} becomes $\mathbf{H}\mathbf{E}\approx\mathbf{0}$, and $\mathbf{H}$ can be estimated via a constrained least-squares fit that selects the left singular vectors associated with the smallest singular values of $\mathbf{E}$, yielding a basis that is nearly orthogonal to the clutter subspace. Compared with unstructured space-time covariance estimation, STAR reduces the effective number of parameters by exploiting recursive space-time structure, thereby lowering the required amount of secondary data. In practice, $L_\text{AR}$ and $M'$ can be chosen using information criteria such as AIC or MDL, to balance model complexity and residual prediction error.

\subsubsection{Selection-Induced (Thinned) Covariance Model}

Beyond imposing algebraic models on the clutter covariance, a complementary ``structure'' arises when the receiver deliberately reduces the observation dimension by selecting only a subset of space-time channels prior to covariance estimation. This is referred to as {\em thinned} STAP with joint antenna-pulse selection, where thinning refers to removing redundancy across space and slow time \cite{XWang-SPL-2015,XWang-IET-2016}. Let $\mathbf{y}\in\mathbb{C}^{N_rL}= \text{vec}\{\mathbf{Y}\}$ denote the full stacked snapshot, and define a binary selection matrix $\mathbf{S}_\text{sel}\in\{0,1\}^{K\times N_rL}$, where $K\ll D$. Then,  
\begin{align}
\widetilde{\mathbf{y}} &= \mathbf{S}_\text{sel}\mathbf{y}, \\
\widetilde{\mathbf{R}} &\triangleq \mathbb{E}\{\widetilde{\mathbf{y}}\widetilde{\mathbf{y}}^H\} = \mathbf{S}_\text{sel}\mathbf{R}\mathbf{S}_\text{sel}^H,
\end{align}
where $\mathbf{R} = \mathbb{E}\{\mathbf{yy}^H\}$ is the full-dimensional clutter covariance. The resulting $\widetilde{\mathbf{R}}$ is a principal submatrix determined by the known sampling pattern, which can thus be estimated with reduced training support. It also reduces subsequent MVDR/STAP processing since the dimension of the matrix inverse is smaller. Thinning is beneficial when the interference is effectively low rank, e.g., clutter concentrated on an angle-Doppler ridge, so $K$ can be chosen on the order of the effective clutter rank while preserving the needed adaptive DoFs. In wideband ISAC, the same selection approach extends naturally to antennas, subcarriers, and/or OFDM symbols, and it can be combined with the structured priors in \eqref{eq:Kronecker model}--\eqref{eq:STAR model} to further improve sample efficiency and numerical stability.

\subsection{Parametric Sparse (Geometric) Models}
In scenarios characterized by sparse clutter originating from a limited number of dominant reflectors, parametric models offer substantial advantages over the purely statistical approaches described earlier. Sparse parametric models estimate geometric parameters such as angles, delays, and Doppler frequencies associated with individual clutter sources such as buildings, terrain features, and prominent infrastructure. Using the MIMO-OFDM ISAC framework introduced in Sec. II-C, clutter echoes can be represented by discrete scatterers described by parameters $(\theta_c, \tau_c, f_{\text{D},c})$ and frequency-dependent reflection coefficients $\beta_{c,n}$. These parameters can be estimated using established methods, potentially enabling deterministic nulling when dominant scatterers are well resolved. Sparse recovery methods may suffer from dictionary coherence and off-grid mismatch, motivating gridless or refined parameterizations when high resolution is required.

\subsection{Hot Clutter in ISAC} \label{sec:hot_model}

We now switch from studying cold clutter that is coherent with the probing waveform to considering hot clutter generated by external interference that is not under the control of the ISAC system. Hot clutter arises when external signals illuminate the environment and reach the sensing receiver both directly and after multipath scattering. When the source is non-cooperative and the interference is unknown, the hot clutter cannot be de-randomized or decomposed as in~\eqref{eq:RccXVX}, and thus it exhibits stronger nonstationarity than cold clutter. Hot clutter possesses three distinctive characteristics: (\textit{i}) spatially distributed reflections, (\textit{ii}) pronounced cross-frequency (fast-time) correlation shaped jointly by the emitter spectrum and the multipath delay spread, and (\textit{iii}) temporal nonstationarity across slow-time CPIs \cite{Kogon1996}--\cite{Abramovich1998}. These properties motivate the development of models that capture coupled dependencies.

In cellular ISAC deployments, a common source of hot-clutter is leakage from nearby cellular transmissions. This includes adjacent-channel interference and out-of-band (OOB) emissions from co-sited transmitters at the same BS site, e.g., other sectors/carriers and potentially other operators, as well as from neighboring BSs. Since such leakage is typically spectrally shaped and non-flat due to transmitter filtering/spectral masks and spectral regrowth, it naturally yields a frequency-colored covariance across OFDM subcarriers. 
 
Following the OFDM ISAC model in \eqref{eq:y_nl}, the received hot clutter on subcarrier $n$ and OFDM symbol $\ell$ is expressed as 
\begin{equation} 
\mathbf{y}_{\text{hc},n}[\ell] = \sum_{g=1}^{G}
\mu_{g,n} e^{\jmath 2\pi f_{\text{D},g} \ell T_{\text{sym}}} \mathbf{b}_n(\theta_{g}) s_{n,g}[\ell] e^{-\jmath 2\pi n \Delta_f \tau_{g}}, \label{eq:hot_clutter_model} 
\end{equation} 
where $G$ is the number of dominant hot-clutter components induced by one or more external emitters, and $(\mu_{g,n},\tau_{g},f_{\text{D},g},\theta_{g})$ denote the frequency-dependent amplitude, delay, Doppler, and AoA associated with the $g$-th component.
The term $s_{n,g}[\ell]$ is the corresponding external waveform component captured within the sensing band on subcarrier $n$ and symbol $\ell$, which may include adjacent-channel/OOB leakage after receiver filtering. Depending on the deployment, $s_{n,g}[\ell]$ may be available at the BS in cooperative/in-network settings; otherwise it is treated as unknown. 
The second-order statistics of hot clutter exhibit an inherently three-dimensional structure across space, frequency, and slow time: 
\begin{equation} \begin{aligned} 
\mathbf{R}_{\text{hc}}[n,n';\Delta_\ell] = & \sum_{g} \sigma_{g}^{2} e^{-\jmath 2\pi (n-n')\Delta_f\tau_{g}} e^{-\jmath 2\pi f_{\text{D},g}\Delta_\ell T_{\text{sym}}} \\
&  \mathbf{b}_n(\theta_{g})\mathbf{b}_{n'}^{H}(\theta_{g}) \mathbb{E}\{s_{n,g}[\ell] s_{n',g}^*[\ell+\Delta_\ell]\},
\label{eq:Rhc_model} 
\end{aligned} \end{equation} 
where $\sigma^2_g$ denotes the power of $\mu_{g,n}$ within the considered band. 
This representation highlights its structure in both the frequency and slow-time dimensions: delays induce cross-frequency correlation, while emitter Dopplers introduce correlation across OFDM symbols. 
In particular, under adjacent-channel/OOB leakage, the interference power spectral density within the sensing band is generally non-flat, so $\mathbf{R}_\text{hc}$ is frequency-colored, e.g., the per-subcarrier interference power $\mathbf{R}_\text{hc}[n,n;0]$ varies with $n$, and the cross-frequency term $\mathbb{E}\{s_{n,g}[\ell] s_{n',g}^*[\ell+\Delta_\ell]\}$ is typically not frequency-white.

Learning the hot clutter statistics largely depends on what information is available. For hot clutter due to cooperative in-network sources (e.g., another coordinated BS), the interference waveform on subcarrier $n$ may be available at the sensing receiver, in which case one could exploit a kernelized representation similar to cold clutter:
\begin{align}
\label{eq:hot-decompose}
    \widehat{\mathbf{R}}_{\text{hc},n} \approx \mathbf{X}_{\text{h},n}\mathbf{V}_{\text{hc},n}\mathbf{X}_{\text{h},n}^H,
\end{align}
where $\mathbf{X}_{\text{h},n}$ is the known interference waveform over the CPI (constructed analogously to $\mathbf{X}_n$), and $\mathbf{V}_{\text{hc},n}$ captures the angle-Doppler scattering of the clutter patches illuminated by the hot-clutter source. In practice, $\mathbf{V}_{\text{hc},n}$ or $\mathbf{R}_{\text{hc},n}$ can be estimated offline during ``quiet'' periods in which the interferer is active while the ISAC BS is muted and no uplink users are present, so that cold clutter is absent and the received snapshots are dominated by hot clutter and background noise. As in standard STAP training, the estimation should exclude range bins for which target detection is performed to avoid self-nulling.

For non-cooperative emitters, the interference waveform is unknown at the BS and the factorization above cannot be used. In this case, the full hot-clutter covariance or the disturbance covariance $\mathbf{R}_{\bm{\eta},n}$ can be estimated directly from secondary data collected during quiet periods where only the interference is present. In this case, one cannot exploit the factorization in~\eqref{eq:hot-decompose} to handle the clutter nonstationarity. Thus,  covariance learning should be performed over short quasi-stationary windows, and can significantly benefit from the structured estimators reviewed in Sections~IV-B--IV-C to reduce training requirements.

\begin{table*}[htbp]\small
\centering
\renewcommand{\arraystretch}{1.3}
\setlength{\tabcolsep}{5pt}
\caption{Recommended Clutter Models for Various ISAC Environments}
\label{tab:clutter_summary}
\begin{tabular}{|l|l|l|}
\hline
\textbf{Environment} & \textbf{Dominant Clutter Features} & \textbf{Recommended Models} \\
\hline \hline
\textbf{V2X/UMi-Street} & Heavy-tailed; dual-mode Doppler; hot clutter & Geometric + K/SIRV hybrid \\
\hline
\textbf{Indoor} & LoS-dominant; sparse multipath; micro-Doppler & Geometric/Sparse parametric \\
\hline
\textbf{UAV/Aerial} & Range-Doppler gradients; heterogeneous terrain & Hybrid geometric-statistical \\
\hline
\textbf{Industrial IoT} & Metallic strong scatterers; structured geometry & Sparse geometric + structured covariance \\
\hline
\end{tabular}
\end{table*}

\begin{figure}\centering
\includegraphics[width=\linewidth]{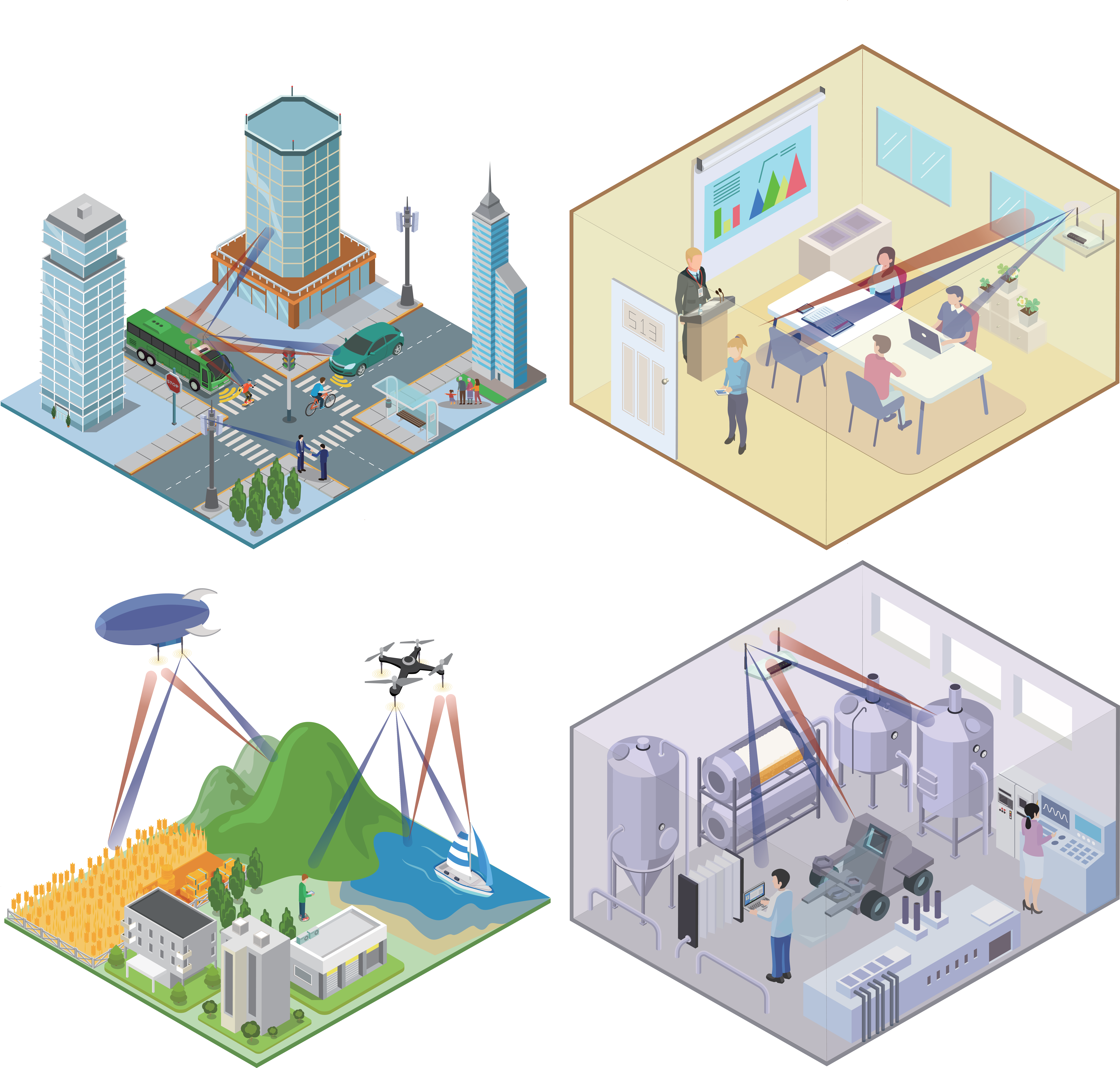}
\vspace{-0.2 cm}
\caption{Representative ISAC environments.}
\label{fig:application scenarios}\vspace{-0.4 cm}
\end{figure}

\subsection{Scenario-Specific Clutter Modeling Guidelines} \label{sec:III-E}

Practical ISAC deployments span diverse environments, each characterized by unique scattering dynamics and clutter characteristics. Consequently, a single universal clutter model is insufficient, necessitating customized approaches. In this subsection, we provide modeling guidelines for several representative ISAC scenarios studied in the literature. 

\subsubsection{Vehicle-to-Everything (V2X) Environments}
V2X scenarios typically feature clutter environments dominated by static urban infrastructure (e.g., buildings, road signs, barriers) and dynamic scatterers such as moving vehicles. Static infrastructure produces strong specular reflections exhibiting heavy-tailed, non-Gaussian amplitude distributions. Depending on the directionality of the transmit waveform (e.g., side- vs.~forward- or rear-looking), an ISAC receiver on a moving vehicle will see clutter from static objects spread across the Doppler spectrum. Mobile targets will exhibit relative Doppler shifts distinct from the predictable patterns due to static objects. Effective clutter modeling in V2X scenarios thus benefit from hybrid geometric-statistical approaches: deterministic geometric models represent dominant specular paths, while the residual clutter background is captured using statistical models, e.g., K-distributions or general compound-Gaussian frameworks. Moreover, frequency-selective behaviors become significant when $B/B_\text{c} \ge 1$ or when $\Delta(ka)=2\pi D_{\text{dom}} B_\text{c} \gtrsim \mathcal{O}(1)$, prompting subband or resource-block-specific (per-RB) covariance estimation and beam-squint compensation strategies.

\subsubsection{Indoor and Residential Environments} 
Indoor scenarios, including residential and office settings (e.g., corresponding to 3GPP Indoor Hotspot (InH)-Office/Home scenarios), exhibit considerable multipath scattering dominated by stable line-of-sight (LoS) and first-order reflective paths. These multipath components generally experience only micro-Doppler effects, primarily driven by limited human motion. The relatively slowly changing environment facilitates clutter characterization, and it is often possible to exploit detailed environmental maps of indoor environments to predict clutter effects. Deterministic geometric modeling can also characterize the dominant propagation paths in such environments. Statistical modeling, typically employing log-normal or Weibull distributions, can complement geometric models for residual diffuse scattering, particularly in environments with diverse materials.

\subsubsection{UAV and Aerial Platforms}
ISAC systems using UAV or other aerial platforms often encounter heterogeneous clutter over urban, suburban, rural, or maritime regions within their coverage areas. The spatial heterogeneity can lead to significant range-dependent variations in clutter statistics. Like V2X applications, mobile airborne platforms spread static ground clutter across the Doppler spectrum. Effective clutter modeling thus requires a hybrid geometric-statistical approach, combining deterministic Doppler adjustments derived from the platform's known kinematics with segmented statistical clutter models adapted to different terrain types. 

\subsubsection{Smart Factories and Industrial IoT}
Industrial IoT environments such as factory floors generally involve large indoor spaces with structured geometries and strong dominant scatterers, typically metallic machinery and equipment. Clutter in these settings predominantly originates from strong specular reflections from metallic surfaces, supplemented by weaker diffuse scattering from construction materials like concrete. Sparse geometric modeling, enhanced by environmental priors derived from CAD-based factory layouts and ray-tracing simulations, can effectively capture these dominant scattering paths. Residual diffuse components can be statistically modeled using, e.g., Gaussian or log-normal distributions.

In summary, effective clutter modeling in ISAC systems should integrate physical environmental insights with adaptive statistical estimation methods. Employing hybrid models that fuse statistical, sparse geometric, and structured covariance frameworks according to scene-specific dynamics and system bandwidth considerations provides a robust approach. Table~\ref{tab:clutter_summary} summarizes recommended clutter modeling strategies for the representative scenarios shown in Fig. \ref{fig:application scenarios}.

\section{Clutter Suppression in ISAC Systems}\label{sec:clutter-suppression} 
 
Building on the clutter models and estimation methods in Sec. IV, this section focuses on receiver-side clutter suppression that operates on clutter-contaminated ISAC measurements. The core ingredients mirror classical radar processing and include slow-time filtering that exploits Doppler separation, spatial beamforming and nulling that exploit angular separation, and joint space-time adaptive filtering enabled by interference covariance estimation. These operations can be interpreted as linear differencing, projection, and adaptive weighting in the Doppler, angle, and joint domains. The methods below assume the MIMO-OFDM waveforms introduced earlier, but they also apply to other sensing waveforms once suitable space-time snapshots are available. For wideband processing, frequency-domain snapshots are also required.

ISAC introduces additional issues that shape both algorithm design and performance. The probing waveform is often data dependent, and may only be partially known at the sensing receiver. Cold clutter and hot clutter may coexist and evolve with the environment and network traffic, and the secondary data for covariance estimation are frequently limited and heterogeneous. Wideband operation further yields subcarrier-dependent array responses, which motivates per-subcarrier or subband adaptation and careful recombination across frequency. We therefore organize the discussion by the available processing dimensions. We start from slow-time suppression, move to spatial-domain methods, and then review space-time and space-frequency-time processing for coupled angle-Doppler scenarios. We close with knowledge-aided and learning-based approaches. For each case, we summarize applicable scenarios, sample-support requirements, and implementation issues, and conclude with application guidelines.

\subsection{Slow-Time-Domain Processing}

Slow-time processing methods exploit the Doppler contrast between (quasi-)stationary clutter and moving targets by applying one-dimensional temporal filtering along the slow-time index.  Building on the angle-gating and waveform de-randomization framework in Sec. III, let $\mathbf{y}_{p,n}[\ell]$ denote the clutter-contaminated return at angular sector $p$, subcarrier $n$, and OFDM symbol $\ell$ within a CPI of length $L$. Most slow-time suppression methods reviewed below can be written in the unified form $\widetilde{\mathbf{y}}_{p,n}[\ell] = \mathbf{y}_{p,n}[\ell]-\widehat{\mathbf{y}}_{\text{c},p,n}[\ell]$, where $\widehat{\mathbf{y}}_{\text{c},p,n}[\ell]$ is a clutter/background estimate obtained via (\textit{i}) moving target indication (MTI)-based high-pass filtering, (\textit{ii}) background estimation and subtraction, or (\textit{iii}) model-based adaptive tracking such as Kalman filtering.

\subsubsection{MTI-based Methods (High-pass Filtering)}

MTI methods suppress low-Doppler clutter by applying discrete-time high-pass filtering along slow time. A classical approach is the single-delay canceller (SDC) \cite{Wang-JSAC-2024,SShen-ICCT-2024}:
\begin{equation}
\widetilde{\mathbf{y}}_{p,n}[\ell] = \mathbf{y}_{p,n}[\ell] - \mathbf{y}_{p,n}[\ell - G_d],\quad G_d\in\mathbb{Z}_+,
\end{equation}
whose Doppler response is given by:
\begin{equation}
H_{\text{SDC}}(f_{\text{D}}) = 1 - e^{-\jmath2\pi f_{\text{D}} G_d T_{\text{sym}}},
\end{equation}
with $f_\text{D}$ interpreted over the unambiguous interval $[-1/(2T_\text{sym}),1/(2T_\text{sym}))$. While SDC provides strong rejection around $f_\text{D}\approx 0$, it exhibits spectral nulls at 
\begin{equation}
f_\text{D}^{\text{blind}} = \frac{m}{G_d T_{\text{sym}}}, \quad m \in \mathbb{Z}.
\end{equation}
These nulls create blind Doppler regions that can attenuate targets whose Dopplers fall near the null locations.
Moreover, the noise variance inherently increases; two-pulse SDC doubles it, and higher-order differences further amplify it, necessitating careful threshold calibration. To alleviate suppression of low-Doppler targets (e.g., pedestrians), selecting larger delay intervals $G_d > 1$ or employing infinite impulse response (IIR) high-pass filters can provide flexible trade-offs between clutter suppression depth and target distortion.

MTI-based clutter suppression has been effectively demonstrated in recent ISAC work. For instance, \cite{DLuo-WCNC-2024} implemented an IIR high-pass filter over slow-time in a beam-scanning framework, enhancing subsequent root-MUSIC and GLRT performance in dynamic vehicular scenarios. To address synchronization challenges posed by timing and carrier frequency offsets in asynchronous vehicular networks, \cite{Wang-JSAC-2024} introduced a cyclically-shifted clutter-map correlation (CMCC) synchronization scheme prior to MTI processing. This approach leverages environmental clutter maps as reference fingerprints, and yields improvements over traditional recursive mean averaging (RMA) methods in asynchronous conditions.

\subsubsection{Background Estimation and Subtraction Methods}
Background subtraction improves target visibility by predicting the slowly varying clutter component and subtracting it from the observation \cite{YGeng-ICC-2024}. The main design choice is how to construct the background estimate from the slow-time history. In the following, we review a batch estimate based on symbol-wise averaging, a recursive estimate based on exponential smoothing, and a one-step predictor based on differencing. 

\textbf{Symbol-wise Averaging and Subtraction}: This method assumes that clutter remains stationary or exhibits minimal variation within a CPI. Clutter estimates are computed by averaging the received signals across multiple symbols:
\begin{align}
\widehat{\mathbf{y}}_{\text{c},p,n}^{\text{stat}} &= \frac{1}{L}\sum_{\ell=0}^{L-1}\mathbf{y}_{p,n}[\ell], \\
\widetilde{\mathbf{y}}_{p,n}[\ell]&=\mathbf{y}_{p,n}[\ell]-\widehat{\mathbf{y}}_{\text{c},p,n}^{\text{stat}}.
\end{align}
Symbol-wise averaging has been effectively integrated into multi-subcarrier joint detection approaches, leading to improved angle-Doppler estimation by exploiting frequency diversity \cite{Luo-TWC-2024,YWang-WCNC-2024}. The YOLO scheme in \cite{HLuo-TWC-2024} applies symbol-wise averaging in a beam-squint-enabled architecture, demonstrating robust performance for wide-angle clutter suppression.

\textbf{Recursive Mean Averaging (RMA)}: When clutter statistics drift slowly over time, RMA updates a recursive clutter-map estimate via exponential smoothing:
\begin{align}  
\widehat{\mathbf{y}}_{\text{c},p,n}[\ell] & = \rho\widehat{\mathbf{y}}_{\text{c},p,n}[\ell-1]+(1-\rho)\mathbf{y}_{p,n}[\ell], \\ \quad \widetilde{\mathbf{y}}_{p,n}[\ell] & =\mathbf{y}_{p,n}[\ell]-\widehat{\mathbf{y}}_{\text{c},p,n}[{\ell}],
\end{align}
where the parameter $0<\rho<1$ trades off convergence speed against stability. 
In the frequency domain, the resulting residual sequence behaves as an IIR high-pass filter with a tunable notch around $f_\text{D}\approx 0$, enabling improved robustness compared with batch averaging for slowly drifting clutter. A detailed investigation of RMA in perceptive mobile networks suggests an optimal selection of $\rho\approx0.99\sim0.995$ for balancing convergence speed and stability \cite{MLRahman-TAES-2020}.

\textbf{Consecutive-Symbol Differencing (CSD)}: CSD is a special case of SDC with $G_d=1$ and can be interpreted as a one-step clutter predictor $\widehat{\mathbf{y}}_{\text{c},p,n}[\ell] = \mathbf{y}_{p,n}[\ell-1]$. It attempts to minimize the prediction error
\begin{equation}
\widetilde{\mathbf{y}}_{p,n}[\ell]=\mathbf{y}_{p,n}[\ell]-\mathbf{y}_{p,n}[\ell-1],
\end{equation}
which effectively suppresses static or slowly varying clutter but increases the residual noise and may attenuate very low-Doppler targets, as in other differencing-based MTI filters. Its robust ISAC clutter suppression performance has been experimentally confirmed in \cite{SShen-ICCT-2024}, particularly when integrated with subsequent 2D-IFFT and MUSIC-based parameter estimation.

\subsubsection{Adaptive Filtering Techniques}

Adaptive filtering techniques, such as the Kalman filter (KF), address temporally correlated clutter by modeling its slow-time evolution as a first-order autoregressive process. After angle gating and waveform de-randomization as previously described, the clutter component $\mathbf{y}_{\text{c},p,n}[\ell]$ is represented by 
\begin{align}
\mathbf{y}_{\text{c},p,n}[\ell]= a_c \mathbf{y}_{\text{c},p,n}[\ell-1] + \bm{\epsilon}_{p,n}[\ell],  
\end{align}
where $a_c$ characterizes the temporal correlation and $\bm{\epsilon}_{p,n}[\ell]$ denotes the process noise representing clutter evolution uncertainty. The corresponding observation model, comprising clutter, target signals, and noise, is expressed as
\begin{equation}
\mathbf{y}_{p,n}[\ell] = \mathbf{y}_{\text{t},p,n}[\ell] + \mathbf{y}_{\text{c},p,n}[\ell]+ \mathbf{z}_{p,n}[\ell].
\end{equation}
The Kalman filter recursively estimates the clutter state through standard prediction and update steps. Subsequently, clutter suppression is achieved by subtracting the estimated clutter state $\widehat{\mathbf{y}}_{\text{c},p,n}[\ell]$ from the observed signal:
\begin{equation}
\widetilde{\mathbf{y}}_{\text{t},p,n}[\ell]= \mathbf{y}_{p,n}[\ell] - \widehat{\mathbf{y}}_{\text{c},p,n}[\ell].
\end{equation}
Practical implementations may further enhance KF robustness by integrating communication-aided estimation methods or neural-network-based predictors, adjusting model parameters to manage dynamic clutter conditions.

Each of the slow-time clutter suppression methods discussed above offers distinct advantages and limitations. MTI-based methods provide straightforward yet effective clutter suppression in stationary environments but risk attenuating low-Doppler targets and amplifying noise. Background estimation techniques (symbol-wise averaging and RMA) robustly handle slowly drifting clutter but require careful tuning and stable system synchronization. Adaptive Kalman filtering provides enhanced flexibility for nonstationary clutter, but with increased computational complexity and synchronization requirements. Selecting suitable slow-time processing strategies necessitates thorough analysis of clutter characteristics, target dynamics, and practical system constraints.

\subsubsection{Case Studies}
\begin{figure*}[!t]
\centering
\begin{subfigure}[t]{0.3\textwidth}
  \centering
  \includegraphics[height=4.4cm]{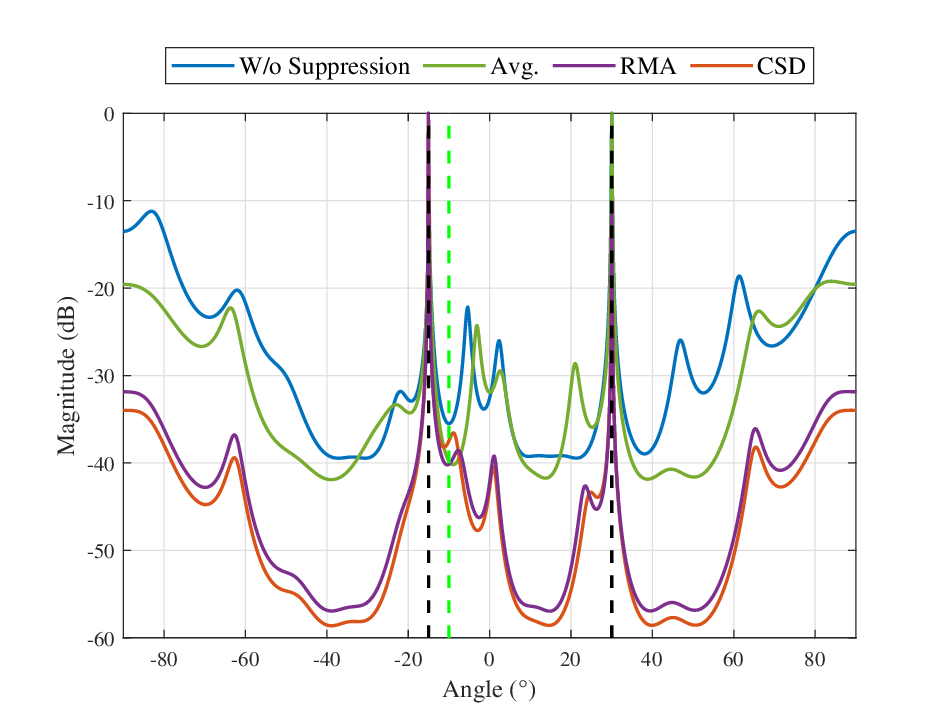}
  \caption{Spatial pseudo-spectrum.}
  \label{fig:AoA_cold_backsub}
\end{subfigure}\hfill
\begin{subfigure}[t]{0.33\textwidth}
  \centering
  \includegraphics[height=4.8cm]{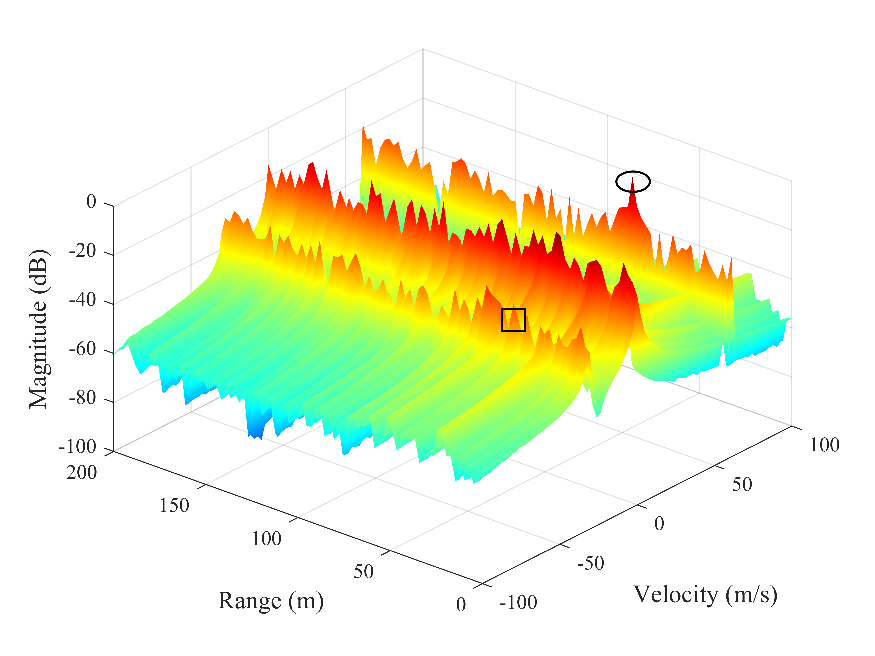}
  \caption{RDM before clutter suppression.}
  \label{fig:RDM_cold}
\end{subfigure}\hfill
\begin{subfigure}[t]{0.33\textwidth}
  \centering
  \includegraphics[height=4.8cm]{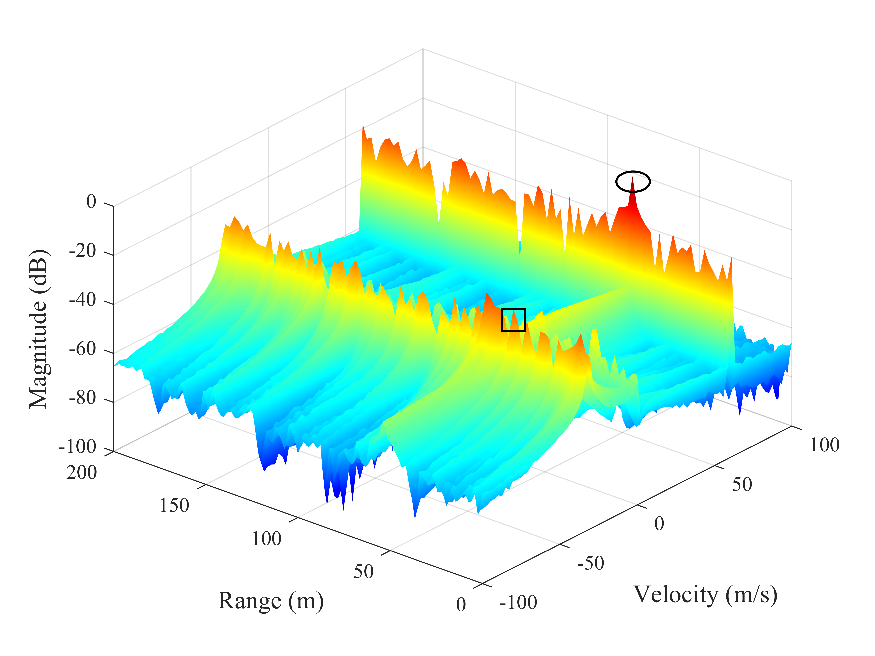}
  \caption{RDM after using RMA.}
  \label{fig:RDM_cold_backsub}
\end{subfigure}

\vspace{1mm}
\caption{Spatial pseudo-spectra and RDMs before and after slow-time filtering in the cold-clutter-only case (no external emitter). The target of interest is indicated by the green dashed line and rectangle, and the strong UAV targets are indicated by the black dashed lines and ellipses. $\text{SCNR} = -45.9$ dB.}
\label{fig:AOA_RDM_cold}
\vspace{-5mm}
\end{figure*}

\begin{figure*}[!t]\centering
\captionsetup[subfigure]{skip=2pt}
\begin{subfigure}[t]{0.30\textwidth}
    \centering
    \includegraphics[height=4.4cm]{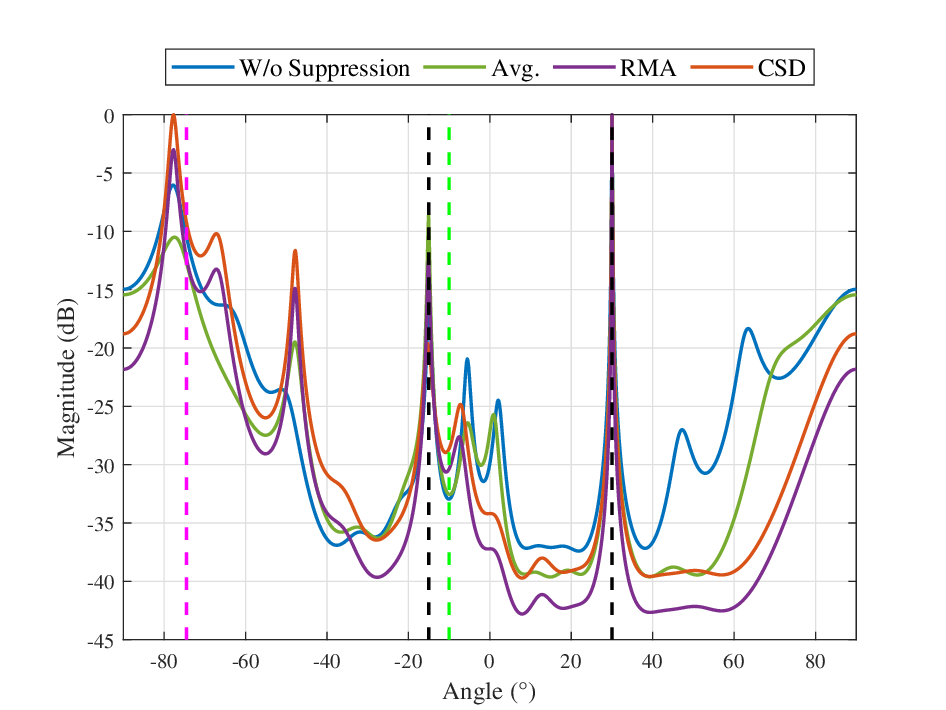}
    \caption{Spatial pseudo-spectrum.}
    \label{fig:AoA_mixed_backsub}
    \end{subfigure}\hfill
    \begin{subfigure}[t]{0.33\textwidth}
        \centering
        \includegraphics[height=4.8cm]{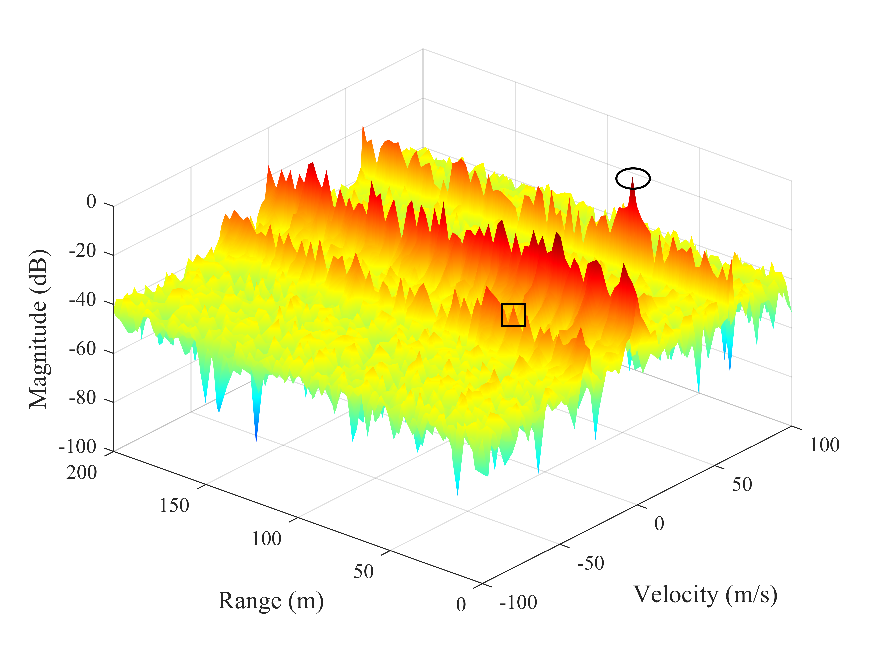}
        \caption{RDM before clutter suppression.}
        \label{fig:RDM_mixed}
    \end{subfigure}\hfill
    \begin{subfigure}[t]{0.33\textwidth}
        \centering
        \includegraphics[height=4.8cm]{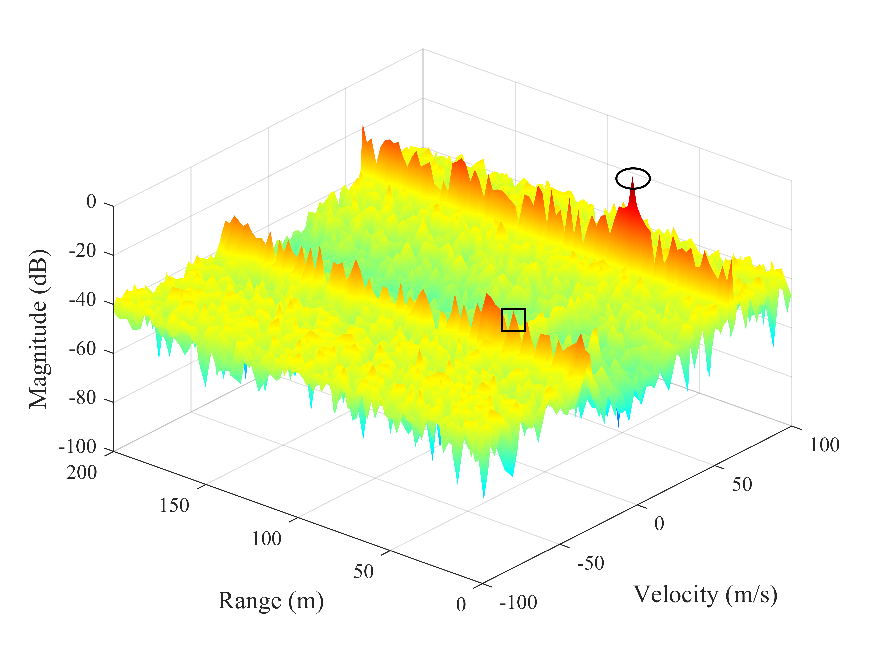}
        \caption{RDM after using RMA.}
        \label{fig:RDM_mixed_backsub}
    \end{subfigure}

    \vspace{1mm}
    \caption{Spatial pseudo-spectra and RDMs before and after slow-time filtering with both cold and hot clutter. The target of interest is marked by the green dashed line and rectangle, the strong UAV targets are marked by the black dashed lines and ellipses, and the external emitter is marked by the purple dashed line. $\text{SCNR} = -47.4$ dB.}
    \label{fig:AOA_RDM_mixed}
    \vspace{-4mm}
\end{figure*}

\begin{figure}[!t]
    \centering
    \includegraphics[width=0.95\linewidth]{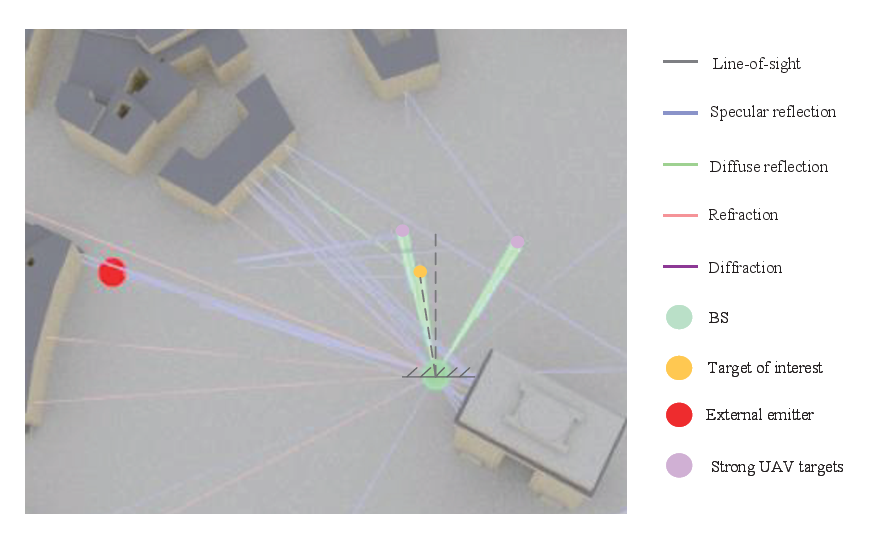}
    \vspace{-0.3 cm}
    \caption{Illustration of an ISAC propagation environment with realistic clutter generated by the Sionna ray-tracing simulator.}
    \label{fig:sionna_map}\vspace{-0.4 cm}
\end{figure}

\begin{figure*}[!t]\centering
\captionsetup[subfigure]{skip=2pt}
\begin{subfigure}[t]{0.30\textwidth}
    \centering
    \includegraphics[height=4.4cm]{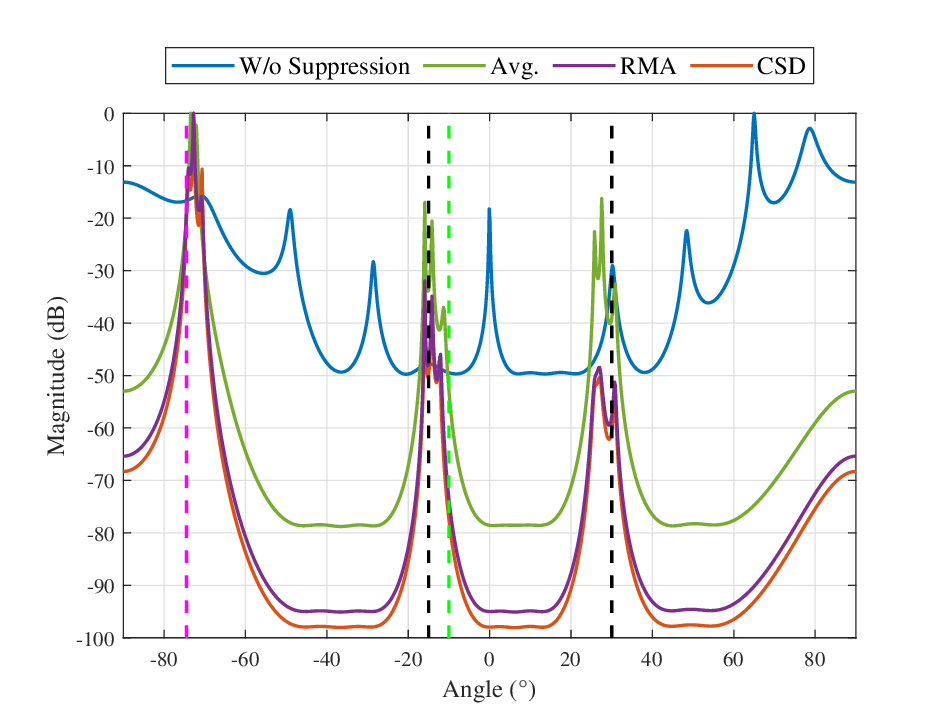}
    \caption{Spatial pseudo-spectrum.}
    \label{fig:AoA_mixed_sionna}
    \end{subfigure}\hfill
    \begin{subfigure}[t]{0.33\textwidth}
        \centering
        \includegraphics[height=4.8cm]{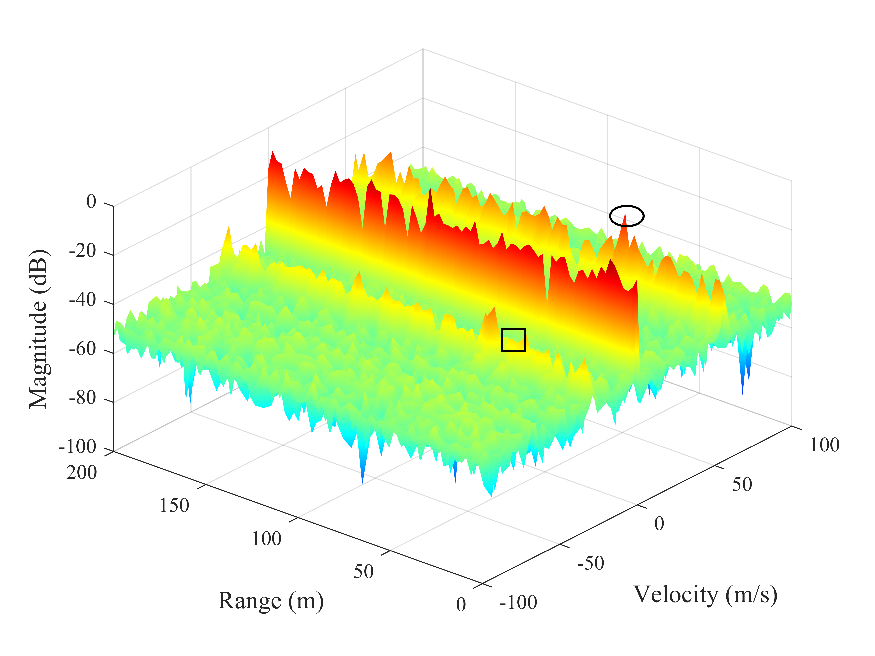}
        \caption{RDM before clutter suppression.}
        \label{fig:RDM_mixed_sionna}
    \end{subfigure}\hfill
    \begin{subfigure}[t]{0.33\textwidth}
        \centering
        \includegraphics[height=4.8cm]{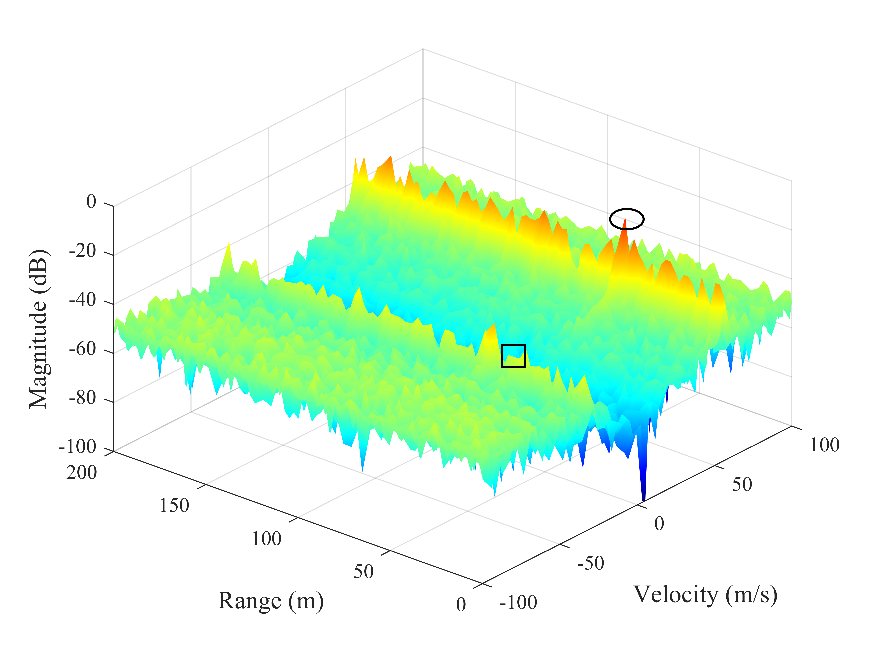}
        \caption{RDM after using RMA.}
        \label{fig:RDM_mixed_backsub_sionna}
    \end{subfigure}

    \vspace{1mm}
    \caption{Spatial pseudo-spectra and RDMs before and after slow-time filtering in the Sionna RT-generated mixed-clutter case. The target of interest is marked by the green dashed line and rectangle, the strong UAV targets are marked by the black dashed lines and ellipses, and the external emitter is marked by the purple dashed line. $\text{SCNR} = -63.5$ dB.}
    \label{fig:AOA_RDM_mixed_sionna}
    \vspace{-4mm}
\end{figure*}

To illustrate the behavior of slow-time clutter suppression in different interference regimes, we consider two representative settings and evaluate performance using the MUSIC spatial pseudo-spectrum and the corresponding RDM. Unless otherwise specified, all simulations in Figs.~\ref{fig:AOA_RDM_cold}--\ref{fig:post_STAP_sionna} use the common parameters summarized in Table~\ref{tab:sim_params}. For the stochastic clutter examples in Figs.~\ref{fig:AOA_RDM_cold}, \ref{fig:AOA_RDM_mixed}, \ref{fig:beampattern_rx}, and \ref{fig:post_STAP_ADM}, the cold clutter is modeled as a collection of $C=100$ scatterers uniformly distributed over four iso-range rings whose radii bracket the target range, with two rings on the near-range side and two on the far-range side. The scatterer azimuth angles are independently drawn from a uniform distribution over $[-90^\circ,90^\circ]$, and their radial velocities are uniformly distributed over $[-1,1]$~m/s to represent slow-moving environmental clutter. 
We consider a multi-target scene consisting of a weak target of interest (ToI) and two mobile UAVs with stronger reflections. While the UAVs are readily detectable in isolation, their dominant echoes and sidelobe leakage can mask the ToI in its range cell and in the angle--Doppler plane. UAV-1 is close to the ToI in azimuth but well separated in Doppler, which stresses angular resolvability under strong-target leakage. UAV-2 is well separated in azimuth but shares the Doppler of the ToI, which creates Doppler overlap and elevates the sidelobe floor along the Doppler slice of interest. This scenario provides a challenging test case for weak-target extraction. When present, hot clutter is generated by an interference source located at an azimuth angle of $-74.5^\circ$ and a range of $122.4$~m. For MVDR and STAP processing, the interference-plus-noise covariance is estimated using the SCM formed from $N_\text{tr}$ independent training snapshots drawn from periodically structured communication blocks rather than temporally independent data symbols, so that the SCM retains the clutter angle-Doppler structure.

The wideband signal model in Sec. II inherently captures the beam-squint effect via the subcarrier-dependent steering factor $\chi_n$. The simulation scenarios consider typical cellular configurations with modest fractional bandwidths and conventional far-field array geometries. Consequently, beam-squint-induced angular spread is not visually prominent in the displayed pseudo-spectra, and the observed angular distributions mainly reflect the propagation geometry and the adopted sectorized illumination and beamforming strategy.

\begin{table}[t]
\centering
\begin{small}
\caption{Common simulation parameters.}
\label{tab:sim_params}
\begin{tabular}{l l}
\hline
Parameter & Value \\
\hline
Carrier frequency                  & $f_\text{c} = 28$~GHz \\
Tx/Rx arrays                       & $N_\text{t}=16$, $N_\text{r}=16$\\
OFDM subcarrier spacing            & $\Delta_f = 120$~kHz \\
\# subcarriers                     & $N = 512$ \\
\# OFDM symbols per CPI            & $L = 256$ \\
Modulation                         & 64-QAM \\
Training snapshots                 & $N_\text{tr} = 256$  \\
Weak target (ToI)  $(\theta, r, v)$          & $(-10^\circ,~~~41.8~\text{m},~ -31.2~\text{m/s})$ \\
Strong target (UAV-1) $(\theta, r, v)$       & $(-15^\circ,~~~53.2~\text{m},~~ 61~\text{m/s})$ \\
Strong target (UAV-2) $(\theta, r, v)$       & $(~~30^\circ,~~~55.4~\text{m},~~ -31.2~\text{m/s})$ \\
External emitter  $(\theta, r, v)$   & $(-74.5^\circ,~ 122.4~\text{m},~ 0~\text{m/s})$ \\
\hline\vspace{-8mm}
\end{tabular}\end{small}
\end{table}

Fig.~\ref{fig:AOA_RDM_cold} shows the cold-clutter-only case corresponding to $\text{SCNR} = -45.9$~dB. In the MUSIC pseudo-spectrum of Fig.~\ref{fig:AOA_RDM_cold}(a), the unsuppressed response exhibits a pronounced clutter pedestal and dense sidelobes, and it is dominated by the UAV reflections near $-15^\circ$ and $30^\circ$. The weak ToI near $-10^\circ$ is clearly obscured when only angular information is used. Symbol-wise averaging provides limited improvement because it mainly targets quasi-stationary background components and does not mitigate slow-moving clutter and stronger moving-target returns. RMA and CSD reduce the pedestal and sidelobe levels by placing a notch around $f_\text{D}\approx 0$, but the strong UAV peak remains since it is not associated with near-zero-Doppler background scatterers. The RDMs in Fig.~\ref{fig:AOA_RDM_cold}(b) and Fig.~\ref{fig:AOA_RDM_cold}(c) further confirm this limitation. Before suppression, a near-zero-Doppler clutter ridge is visible, together with a strong peak around $61$~m/s from UAV-1, and elevated sidelobes along the Doppler slice $v=-31.2$~m/s caused by UAV-2, which shares the ToI Doppler and leaks through range sidelobes. After applying RMA, the near-zero-Doppler ridge is strongly attenuated, but the strong-target components and their leakage remain evident. This shows that slow-time-only processing is insufficient when dominant moving objects occupy non-zero Doppler bins or overlap the Doppler of the target of interest.

Fig.~\ref{fig:AOA_RDM_mixed} illustrates performance for the mixed cold and hot clutter scenario, where $\text{SCNR} = -47.4$~dB. Compared with Fig.~\ref{fig:AOA_RDM_cold}(a), the ToI is harder to identify in Fig.~\ref{fig:AOA_RDM_mixed}(a) because the external emitter produces a strong LoS response and raises the pseudo-spectrum floor through scattering, which raises sidelobe levels and reduces ToI contrast. The same behavior appears in the RDM of Fig.~\ref{fig:AOA_RDM_mixed}(b), where the emitter-induced interference increases the range-Doppler background and masks the weak ToI, even though the UAV targets remain visible. After applying RMA in Fig.~\ref{fig:AOA_RDM_mixed}(c), the near-zero-Doppler ridge is attenuated, but pronounced residual interference and leakage persist because the hot-clutter contribution is not confined to a narrow band around $f_\text{D}=0$ and cannot be removed by slow-time background subtraction alone. These results motivate the joint space-time suppression strategies discussed later.

In addition to the stochastic examples, we also consider a site-specific propagation environment generated by the NVIDIA Sionna ray-tracing simulator \cite{SionnaRT}, as illustrated in Fig.~\ref{fig:sionna_map}. The scenario includes a monostatic ISAC BS, a weak ToI, two stronger mobile UAV targets, and an external emitter. The BS is deployed at a height of $13.5$~m, the targets are placed at heights between $10$~m and $15$~m, and the emitter is at $5$~m. Unless otherwise specified, the array and waveform settings follow Table~\ref{tab:sim_params}. The key difference is that the background cold-clutter returns from surrounding buildings are generated by ray tracing over a detailed 3D scene. The resulting observations include static background clutter from buildings, echoes from the ToI and the strong UAVs, and emitter-induced interference and hot-clutter contributions associated with the external source. The maximum interaction depth for Sionna is set to three, so each ray undergoes at most three interactions with scene objects. In Fig.~\ref{fig:sionna_map}, colored line segments show rays arriving at the BS, with the color indicating the underlying propagation mechanism. Sionna defines the LoS as a direct Tx--Rx path, so for monostatic sensing the target echoes appear as reflected and scattered paths that interact with the target object, even when a geometric LoS exists. Although propagation is simulated in full 3D, the sensing processing in this example assumes azimuth estimation only since the BS employs a ULA. The ToI and UAVs are modeled as simplified 3D mesh objects imported into Sionna. These meshes are obtained by simplifying publicly available 3D models in Blender and exporting them in Sionna's XML format, which keeps the ray-tracing scene lightweight while enabling flexible material assignment.

Fig.~\ref{fig:AOA_RDM_mixed_sionna} presents the MUSIC spatial pseudo-spectrum and the corresponding RDMs for the Sionna RT scene in Fig.~\ref{fig:sionna_map}. The overall behavior is consistent with the mixed-clutter example in Fig.~\ref{fig:AOA_RDM_mixed}, but the detection task is more challenging due to the much lower SCNR. Before suppression, the sensing signal is dominated by strong building backscatter, strong UAV echoes, and emitter-induced hot clutter, so the ToI near $-10^\circ$ is not discernible in either the spatial pseudo-spectrum or the RDM. Applying RMA largely removes the quasi-static building contribution and improves the visibility of the ToI signature. However, pronounced residual peaks and sidelobe leakage associated with the moving UAVs and the external emitter remain, and the weak ToI return is still strongly masked in the RDM. This RT case therefore reinforces that slow-time filtering alone is insufficient in realistic urban scenes with mobile cold clutter and emitter-induced hot clutter.

\subsection{Spatial-Domain Processing}

Spatial-domain clutter suppression leverages the receive array to attenuate interference from undesired directions while maintaining gain toward the target. One can apply a spatial combiner to the $N_\text{r}$-dimensional snapshots in~\eqref{eq:y_nl} or to range-gated snapshots obtained after waveform de-randomization and range focusing, depending on how the cell under test and the training data are defined. In wideband MIMO-OFDM, the steering vector $\mathbf{b}_n(\theta)$ varies with the subcarrier index due to beam squint, so the combiner is typically designed per subcarrier or within narrow subbands where the array response and interference statistics are approximately stationary. Building on the angle-gating role of spatial filtering in Sec. III-B, we next summarize clutter-aware spatial designs that either impose explicit null constraints or adapt to an estimated interference covariance.
In this subsection, $\mathbf{R}_{I,n}\triangleq\mathbb{E}\{\mathbf{y}_n[\ell]\mathbf{y}^H_n[\ell]\}\in\mathbb{C}^{N_\text{r}\times N_\text{r}}$ denotes the per-subcarrier spatial disturbance covariance or its range-gated counterpart when $\mathbf{y}_n[\ell]$ is replaced by $\mathbf{y}_{p,\tau}[\ell]$.

\subsubsection{Deterministic Beam Nulling}
Deterministic beam nulling is a straightforward and computationally efficient solution when a small number of dominant clutter directions are known a priori or can be reliably estimated, for example from a clutter map \cite{Xu-ICCC-2024,ZXu-arxiv-2025} or from a high-SCNR spatial pseudo-spectrum. The fundamental approach involves placing deep nulls towards clutter directions while preserving distortionless reception in the target direction \cite{XZhang-Radarconf-2020}. Let $\{\theta_{\text{c},1},\dots,\theta_{\text{c},C_0}\}$ denote the set of clutter AoAs selected for nulling, and define clutter steering matrix $\mathbf{B}_{\text{c},n} = [\mathbf{b}_n(\theta_{\text{c},1}), \dots, \mathbf{b}_n(\theta_{\text{c},C_0})]\in\mathbb{C}^{N_\text{r}\times C_0}$. The projector orthogonal to the subspace spanned by  $\mathbf{B}_{\text{c},n}$ is
\begin{align}
\mathbf{P}_{\perp,n} = \mathbf{I}_{N_\text{r}}-\mathbf{B}_{\text{c},n}\mathbf{B}_{\text{c},n}^\dagger,
\end{align}
where $(\cdot)^\dagger$ denotes the Moore-Penrose pseudoinverse. When $\mathbf{B}_{\text{c},n}$ has full column rank, $\mathbf{B}_{\text{c},n}^\dagger =(\mathbf{B}_{\text{c},n}^H\mathbf{B}_{\text{c},n})^{-1}\mathbf{B}_{\text{c},n}^H$. When $\mathbf{B}_{\text{c},n}^H\mathbf{B}_{\text{c},n}$ is ill-conditioned, a regularizer can improve numerical stability. A distortionless spatial combiner with unit gain toward the target AoA $\theta_t$ is then given by
\begin{align}\label{eq:u det}
    \mathbf{u}_{\text{det},n} = \frac{\mathbf{P}_{\perp,n}\mathbf{b}_n(\theta_t)}{\mathbf{b}_n^H(\theta_t)\mathbf{P}_{\perp,n}\mathbf{b}_n(\theta_t)}.
\end{align}
This design enforces $\mathbf{u}_{\text{det},n}\mathbf{b}_n(\theta_t)=1$ while placing nulls towards the dominant clutter sources. When the target AoA is near the clutter subspace, the denominator in \eqref{eq:u det} becomes small and the combiner becomes sensitive to steering vector mismatch and calibration errors. In such situations, spatial-only suppression may not provide adequate target discrimination and additional processing dimensions are needed.

\subsubsection{Subspace Projection}
When clutter AoAs are not explicitly known, or when clutter occupies an extended angular region, it is more effective to estimate the clutter subspace \cite{SRivetti-arxiv-2025}. Given the dominant eigenvectors $\mathbf{U}_{\text{c},n}\in\mathbb{C}^{N_\text{r}\times r_{\text{c},n}}$ of the  covariance estimate $\widehat{\mathbf{R}}_{I,n}\in\mathbb{C}^{N_\text{r}\times N_\text{r}}$ for subcarrier $n$, the corresponding orthogonal projector can be written as
\begin{align}
    \mathbf{P}_{\perp,n} = \mathbf{I}_{N_\text{r}}-\mathbf{U}_{\text{c},n}\mathbf{U}_{\text{c},n}^H.
\end{align}
The resulting subspace-projection combiner takes the same form as in \eqref{eq:u det}. 
Compared with deterministic nulling, the performance of subspace projection depends on selecting an appropriate clutter subspace dimension $r_{\text{c},n}$. Overestimating $r_{\text{c},n}$ increases the risk of target self-nulling, while underestimating $r_{\text{c},n}$ increases the residual clutter. In practice, $r_{\text{c},n}$ can be determined using eigenvalue thresholds, energy capture ratios, or information-theoretic criteria such as MDL or AIC, as discussed in Sec. IV-C.

\subsubsection{Covariance-Adaptive Beamforming}
Covariance-adaptive spatial beamforming directly incorporates $\widehat{\mathbf{R}}_{I,n}$ and can achieve near-optimal output SCNR when the covariance estimate is accurate and the training data are locally homogeneous. The MVDR combiner is obtained by minimizing the output interference power subject to a distortionless constraint toward $\theta_t$:
\begin{align}\label{eq:u MVDR}
   \mathbf{u}_{\text{MVDR},n} = \frac{\widehat{\mathbf{R}}_{I,n}^{-1}\mathbf{b}_n(\theta_t)}{\mathbf{b}_n^H(\theta_t)\widehat{\mathbf{R}}_{I,n}^{-1}\mathbf{b}_n(\theta_t)}.
\end{align} 
The linearly constrained minimum variance (LCMV) beamformer generalizes MVDR by allowing multiple linear constraints to protect target or communication signals \cite{YNiu-TCCN-2025}:
\begin{align}\label{eq:u LCMV}
 \mathbf{u}_{\text{LCMV},n} = \widehat{\mathbf{R}}_{I,n}^{-1}\mathbf{C}_n(\mathbf{C}_n^H\widehat{\mathbf{R}}_{I,n}^{-1}\mathbf{C}_n)^{-1}\mathbf{f},  
\end{align}
where $\mathbf{C}_n$ collects the steering vectors for which the response should be constrained, and $\mathbf{f}$ specifies the desired gains, e.g., equal to 1 or 0 for desired directions and interference, respectively. 
The MVDR combiner in \eqref{eq:u MVDR} is recovered as a special case of \eqref{eq:u LCMV} with $\mathbf{C}_n = \mathbf{b}_n(\theta_t)$ for target direction $\theta_t$ and $\mathbf{f}=1$. In snapshot-limited settings, the robustness of MVDR and LCMV can be significantly improved by the regularized and structured covariance estimators in Sec. IV.

\begin{figure}[!t]
    \centering
    \includegraphics[width=0.85\linewidth]{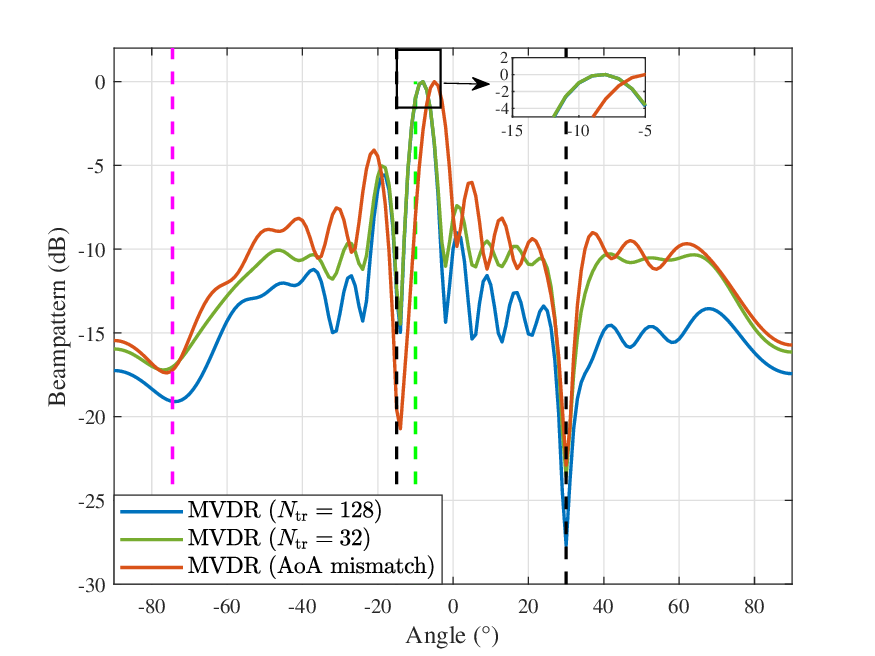}
    \vspace{-0.1 cm}
    \caption{Receive beampattern comparison for the stochastic model (green dashed line: target of interest; black dashed lines: strong UAV targets; purple dashed line: external emitter).}
    \label{fig:beampattern_rx}\vspace{-0.5 cm}
\end{figure}

\begin{figure}[!t]
    \centering
    \includegraphics[width=0.85\linewidth]{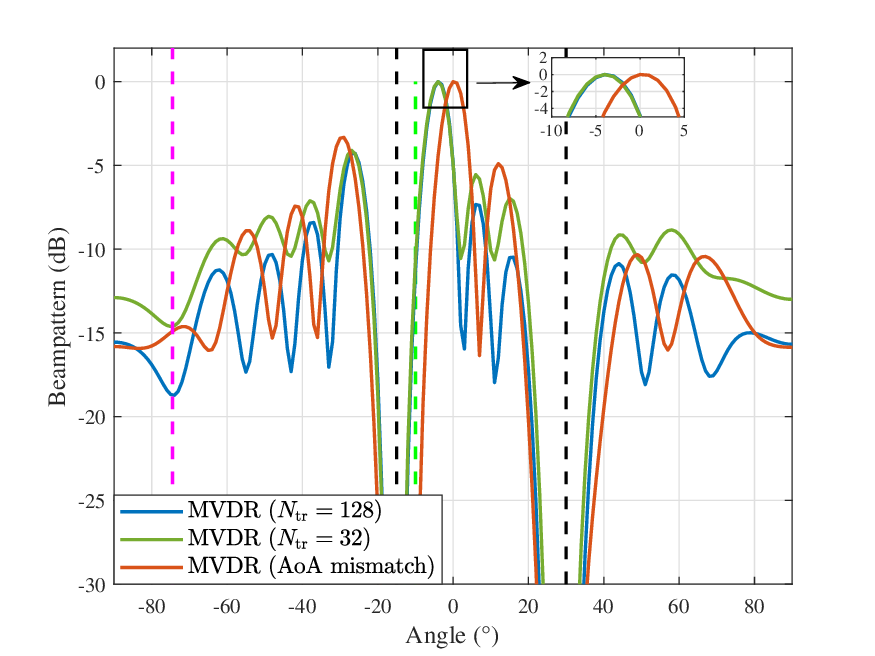}
    \vspace{-0.1 cm}
    \caption{Receive beampattern comparison using Sionna-RT simulations (green dashed line: target of interest; black dashed lines: strong UAV targets; purple dashed line: external emitter).}
    \label{fig:beampattern_rx_sionna}\vspace{-0.2 cm}
\end{figure}

Figs.~\ref{fig:beampattern_rx} and \ref{fig:beampattern_rx_sionna} compare MVDR receive beampatterns obtained with SCM-based covariance estimation for the stochastic clutter and Sionna RT models, respectively. With sufficient training ($N_{\mathrm{tr}}=128$), MVDR preserves the desired response at the ToI and forms nulls toward the UAVs. In Fig.~\ref{fig:beampattern_rx}, the suppression of UAV-1 is limited compared to UAV-2 because it is near the ToI. In Fig.~\ref{fig:beampattern_rx_sionna}, the sidelobe floor is higher and the mainlobe peak exhibits a more noticeable bias relative to the nominal ToI direction. This is expected in the RT setting because the scattered UAV echoes occupy an angular sector rather than a single AoA. As a result, spatial-only MVDR cannot carve a deep notch around UAV-1 without incurring stronger mainlobe distortion, which leads to higher residual sidelobes and a larger apparent mainlobe shift. When the amount of training is reduced to $N_{\mathrm{tr}}=32$, SCM estimation errors further elevate sidelobes in both figures, with the most severe degradation occurring near UAV-1 where angular separability is smallest. When the target AoA is imprecisely known (assumed to be $-11^\circ$ instead of $-10^\circ$, shown by the red curves), the response for the target AoA is reduced by $-10$dB due to the null for the nearby UAV. If the ToI return had been present in the training data, the actual target AoA would have been even more deeply attenuated. These results show that spatial-only filtering can be fragile, further highlighting the need for joint space-time processing to exploit Doppler separation in addition to spatial discrimination.

\subsection{Space-Time Adaptive Processing (STAP)}

STAP suppresses clutter by jointly exploiting the available spatial and slow-time DoFs \cite{Guerci2003}. In MIMO-OFDM ISAC, STAP operates on the stacked snapshot $\mathbf{y}_n\in\mathbb{C}^{N_\text{r}L}$ in \eqref{eq:yn def} for each subcarrier $n$ or on the full-band stacked snapshot $\mathbf{y}\in\mathbb{C}^{N_\text{r}NL}$ in \eqref{eq:y def}. It coherently processes measurements collected from the receive array over multiple OFDM symbols. This joint processing is effective for clutter with coupled angle-Doppler structure that is difficult to mitigate using angle- or Doppler-only suppression. STAP is particularly useful when platform motion or moving scatterers induce Doppler-spread clutter, the clutter spans a wide angular sector and leaves noticeable residual sidelobes after one-dimensional processing, or when hot clutter occupies a broad region in the angle-Doppler plane. These gains come with increased computational and sample complexity because STAP relies on estimating and inverting a large $N_\text{r}L\times N_\text{r}L$ or $N_\text{r}NL\times N_\text{r}NL$ interference covariance matrix, which often necessitates reduced-complexity and structured formulations. We first review classical MVDR-STAP, discuss the wideband SFTAP extension, present reduced-complexity and structured STAP variants, and finally move to bistatic and multistatic generalizations.
In this section, $\mathbf{R}_{I,n}\triangleq\mathbb{E}\{\mathbf{y}_n\mathbf{y}_n^H\}\in\mathbb{C}^{N_\text{r}L\times N_\text{r}L}$ denotes the space-time covariance associated with the stacked snapshot $\mathbf{y}_n$ in \eqref{eq:yn def}, and $\mathbf{R}_I\in\mathbb{C}^{N_\text{r}NL\times N_\text{r}NL}$ for the full-band stacked snapshot $\mathbf{y}$ in \eqref{eq:y def}.

\subsubsection{Classical STAP}
The classical narrowband STAP method designs a spatial-temporal adaptive filter for each hypothesized angle-Doppler bin $(\theta,f_{\text D})$ using the per-subcarrier stacked snapshot $\mathbf{y}_n$. Recall that the stacked space-time snapshot $\mathbf{y}_n\in\mathbb{C}^{N_\text{r}L}$ admits the form
\begin{align}\label{eq:yn for STAP}
    \mathbf{y}_n = \sum_{q\in\mathcal{Q}}\gamma_{q,n}e^{-\jmath 2\pi n \Delta_f\tau_q}\mathbf{X}_n\mathbf{v}_n(\theta_q, f_{\text{D},q}) + \bm{\eta}_n,
\end{align}
where $\mathbf{X}_n$ is constructed from the transmitted signals $\{\mathbf{x}_n[\ell]\}_{\ell=0}^{L-1}$ as in \eqref{eq:Xn def}, $\mathbf{v}_n(\theta,f_\text{D})$ is the space-time steering vector defined in \eqref{eq:vn def}, and $\bm{\eta}_n$ is the aggregated disturbance. 
For ISAC-STAP design, it is convenient to define the effective waveform-dependent space-time steering vector
\begin{align}
    \widetilde{\mathbf{v}}_n(\theta,f_\text{D}) \triangleq \mathbf{X}_n\mathbf{v}_n(\theta,f_\text{D})\in\mathbb{C}^{N_\text{r}L}.
\end{align}
The STAP weight for bin $(\theta,f_\text{D})$ is obtained from the standard MVDR problem
\begin{equation}\label{eq:STAP prob}
\begin{aligned}
\min_{\mathbf{w}_n} \quad & \mathbf{w}_n^{H}\widehat{\mathbf{R}}_{I,n}\mathbf{w}_n \\
\text{s.t.} \quad & \mathbf{w}_n^{H}\widetilde{\mathbf{v}}_n(\theta,f_\text{D}) = 1,
\end{aligned}
\end{equation}
where $\widehat{\mathbf{R}}_{I,n}\in\mathbb{C}^{N_\text{r}L\times N_\text{r}L}$ is the space-time interference-plus-noise covariance estimate. 
The solution to \eqref{eq:STAP prob} is the normalized MVDR form
\begin{equation} 
\mathbf{w}_{\text{MVDR},n} =\frac{\widehat{\mathbf{R}}_{I,n}^{-1} \widetilde{\mathbf{v}}_n(\theta,f_\text{D})}{\widetilde{\mathbf{v}}_n^H(\theta,f_\text{D})\widehat{\mathbf{R}}_{I,n}^{-1}\widetilde{\mathbf{v}}_n(\theta,f_\text{D})},
\end{equation}
Substituting $\widetilde{\mathbf{v}}_n(\theta,f_\text{D}) = \mathbf{X}_n\mathbf{v}_n(\theta,f_\text{D})$ yields an explicit waveform-dependent expression:
\begin{equation}\label{eq:w MVDR}
\mathbf{w}_{\text{MVDR},n}=\frac{\widehat{\mathbf{R}}_{I,n}^{-1} \mathbf{X}_n\mathbf{v}_n(\theta,f_\text{D})}{\mathbf{v}^H_n(\theta,f_\text{D})\mathbf{X}_n^H \widehat{\mathbf{R}}_{I,n}^{-1}\mathbf{X}_n\mathbf{v}_n(\theta,f_\text{D})}.
\end{equation}
The corresponding output SCNR for a unit-power target is 
\begin{equation}\label{eq:SCNR STAP} 
\text{SCNR}_n=\mathbf{v}^H_n(\theta,f_\text{D}) \mathbf{X}^H_n\widehat{\mathbf{R}}_{I,n}^{-1}\mathbf{X}_n\mathbf{v}_n(\theta,f_\text{D}).
\end{equation}
Note that both $\widehat{\mathbf{R}}_{I,n}$ and $\mathbf{w}_{\text{MVDR},n}(\theta,f_\text{D})$ depend on the transmitted waveform through $\mathbf{X}_n$. Effective clutter mitigation is thus not only a receiver-side problem. Joint transmit--receive optimization can shape the illumination pattern and reduce waveform-induced clutter sidelobes. We further examine this proactive optimization perspective in Sec. VI-B.

Classical STAP achieves near-optimal suppression when the interference statistics are stationary within the CPI and when sufficient homogeneous training data are available for reliable covariance estimation. Its practical deployment is constrained by the cubic complexity of inverting an $N_\text{r}L$ dimensional matrix and by the sample support required for stable estimation of $\mathbf{R}_{I,n}$. Classical narrowband processing also becomes mismatched in wideband scenarios since beam squint and frequency-selective clutter induce frequency-dependent steering vectors and covariance matrices. In OFDM systems, this frequency dependence necessitates subcarrier-specific STAP filters. However, when individually processed subcarrier results are combined for delay/range focusing, residual mismatches arising from per-subcarrier adaptation can degrade coherent integration performance. These limitations motivate the development of space-frequency-time adaptive processing.

\subsubsection{Space-Frequency-Time Adaptive Processing (SFTAP)}
SFTAP extends classical STAP by adapting jointly across the spatial, slow-time, and frequency dimensions. This extension addresses wideband effects by accounting for frequency-dependent array responses and by explicitly incorporating delay structure across the subcarriers. 
Based on the signal model $\mathbf{y} =\sum_{q \in \mathcal{Q}}\bm{\Gamma}_q \mathbf{T}(\tau_q)\mathbf{X}\mathbf{v}(\theta_q, f_{\text{D},q}) + \bm{\eta}$, and the common simplification that the frequency-dependent reflectivity term $\bm{\Gamma}_q$ can be treated as approximately flat over the processed band/subband, the response of a hypothesized scatterer at $(\theta,f_\text{D},\tau)$ is proportional to the effective space-frequency-time steering vector
\begin{align}
    \widetilde{\mathbf{v}}(\theta,f_\text{D},\tau) \triangleq \mathbf{T}(\tau)\mathbf{Xv}(\theta,f_\text{D}),
\end{align}
where $\mathbf{T}(\tau)$ defined in \eqref{eq:Ttau def} captures the frequency-delay phase progression, $\mathbf{X}$ in \eqref{eq:X def} is the full-band waveform matrix, and $\mathbf{v}(\theta,f_{\text D})$ in \eqref{eq:v def} is the full-band angle-Doppler steering vector. If $\widehat{\mathbf{R}}_I\in\mathbb{C}^{N_\text{r}NL\times N_\text{r}NL}$ is an estimate of the full-dimensional interference-plus-noise covariance, the MVDR-SFTAP filter is  
\begin{equation}\begin{aligned}
\mathbf{w}_{\text{SFTAP}}&=\frac{\widehat{\mathbf{R}}_I^{-1}\widetilde{\mathbf{v}}(\theta,f_\text{D},\tau)}{\widetilde{\mathbf{v}}^H(\theta,f_\text{D},\tau)\widehat{\mathbf{R}}_I^{-1}\widetilde{\mathbf{v}}(\theta,f_\text{D},\tau)}\\
& =\frac{\widehat{\mathbf{R}}_I^{-1}\mathbf{T}(\tau)\mathbf{Xv}(\theta,f_\text{D})}{\mathbf{v}^H(\theta,f_\text{D})\mathbf{X}^H\mathbf{T}^H(\tau)\widehat{\mathbf{R}}_I^{-1}\mathbf{T}(\tau)\mathbf{Xv}(\theta,f_\text{D})}.
\end{aligned}\end{equation}
SFTAP aligns naturally with OFDM-based ISAC architectures because it preserves frequency-dependent clutter structure across the signal bandwidth while maintaining coherent processing over slow time. Its main limitation is computational and statistical scalability. The matrix inversion now has dimension $N_\text{r}NL$, and direct matrix inversion leads to complexity of order $\mathcal{O}\{(N_\text{r}NL)^3\}$. In addition, the enlarged covariance matrix is more difficult to estimate robustly from limited training data. These challenges motivate reduced-complexity and structured STAP variants, which we discuss next.

\subsubsection{Reduced-Complexity STAP Methods}
Reduced-complexity STAP variants aim to retain most of the interference rejection capability of full-dimensional MVDR filtering while reducing the cost of high-dimensional covariance inversion and the amount of training data needed for reliable covariance estimation. In wideband OFDM settings, these ideas can also be combined with subband processing so that the adaptive dimension scales with the local coherence bandwidth. We briefly summarize three representative approaches.

\textbf{Reduced-Dimension (RD)-STAP}:
RD-STAP decreases the computational load by projecting the $N_\text{r}L$-dimensional space-time snapshot onto a lower-dimensional subspace. The subspace is typically defined in beamspace, in the Doppler domain, or using prior knowledge of the angle-Doppler region where clutter and targets are expected to lie. Let $\mathbf{T}_\text{RD}\in \mathbb{C}^{N_\text{r}L\times d}$ denote a basis matrix with $d\ll N_\text{r}L$. Projecting the covariance, steering vector, and the filter weights at subcarrier $n$ on this basis yields

\begin{small}
\begin{equation}
\mathbf{w}_{\text{RD},n} = 
\frac{\mathbf{T}_\text{RD}(\mathbf{T}_\text{RD}^H\widehat{\mathbf{R}}_{I,n}\mathbf{T}_\text{RD})^{-1}\mathbf{T}_\text{RD}^H\mathbf{X}_n\mathbf{v}_n(\theta,f_\text{D})}
{\mathbf{v}_n^H(\theta,f_\text{D})\mathbf{X}_n^H\mathbf{T}_\text{RD}(\mathbf{T}_\text{RD}^H\widehat{\mathbf{R}}_{ I,n}\mathbf{T}_\text{RD})^{-1}\mathbf{T}_\text{RD}^H\mathbf{X}_n\mathbf{v}_n(\theta,f_\text{D})}, 
\end{equation}
\end{small}

\noindent which reduces the problem dimension from $N_\text{r}L$ to $d$. The basis $\mathbf{T}_\text{RD}$ can be constructed from parametric models such as STAR \cite{PParker-TAES-2003} or from prior knowledge of the clutter ridge and the target search region.

\textbf{Reduced-Rank (RR)-STAP Methods}:
RR-STAP is similar to RD-STAP, but exploits the eigenstructure of the interference covariance and adapts only within a dominant low-rank subspace of $\widehat{\mathbf{R}}_{I,n}$. A common approach is principal-components STAP based on the approximation
\begin{equation}
\widehat{\mathbf{R}}_{I,n} \approx \mathbf{U}_r\mathbf{\Lambda}_r\mathbf{U}_r^H,
\end{equation}
where $\mathbf{U}_r\in\mathbb{C}^{N_\text{r}L\times r}$ contains the $r$ dominant eigenvectors and $\bm{\Lambda}_r\in\mathbb{C}^{r\times r}$ the corresponding eigenvalues. The associated weight vector is then 
\begin{equation}
\mathbf{w}_{\text{RR},n} = \frac{\mathbf{U}_r\bm{\Lambda}_r^{-1}\mathbf{U}_r^{H}\mathbf{X}_n\mathbf{v}_n(\theta,f_\text{D})}{\mathbf{v}_n^H(\theta,f_\text{D})\mathbf{X}_n^H\mathbf{U}_r\bm{\Lambda}_r^{-1}\mathbf{U}_r^H\mathbf{X}_n\mathbf{v}_n(\theta,f_\text{D})}.
\end{equation}
RD- and RR-STAP can approach full-rank STAP performance when the selected subspace captures the effective clutter while leaving predominantly white noise behind. In practice, choosing $r$ and efficiently computing the dominant eigenvectors are the main implementation issues when $N_\text{r}L$ is large.

\textbf{Structured STAP}:
In this approach a structured model for the covariance is imposed to reduce estimation variance and computational cost. One example assumes approximate separability between the spatial and slow time dimensions, which yields the Kronecker model
\begin{equation}
\widehat{\mathbf{R}}_{I,n} \approx \widehat{\mathbf{R}}_{\text{t},n} \otimes \widehat{\mathbf{R}}_{\text{s},n},
\end{equation}
where $\widehat{\mathbf{R}}_{\text{s},n}\in \mathbb{C}^{N_\text{r}\times N_\text{r}}$ and $\widehat{\mathbf{R}}_{\text{t},n}\in\mathbb{C}^{L\times L}$ denote the spatial and slow-time covariances. When the waveform-aware steering vector is also separable, the MVDR solution factorizes into a Kronecker product of two lower-dimensional MVDR filters:
\begin{equation}
    \mathbf{w}_{\text{struct},n} = \frac{\widehat{\mathbf{R}}_{\text{t},n}^{-1}\mathbf{t}_n(\theta,f_\text{D})}{\mathbf{t}_n^H(\theta,f_\text{D})\widehat{\mathbf{R}}_{\text{t},n}^{-1}\mathbf{t}_n(\theta,f_\text{D})} \otimes \frac{\widehat{\mathbf{R}}_{\text{s},n}^{-1}\mathbf{b}_n(\theta)}{\mathbf{b}_n^H(\theta)\widehat{\mathbf{R}}_{\text{s},n}^{-1}\mathbf{b}_n(\theta)},
\end{equation}
where $\mathbf{t}_n(\theta,f_\text{D}) = [\mathbf{a}^H_n(\theta)\mathbf{x}_n[0],\dots \mathbf{a}^H_n(\theta)\mathbf{x}_n[L-1]]^T\odot\mathbf{d}(f_{\text{D}})$, and $\mathbf{X}_n\mathbf{v}_n(\theta,f_\text{D}) = \mathbf{t}_n(\theta,f_\text{D})\otimes \mathbf{b}_n(\theta)$. This Kronecker structure reduces the computation of the matrix inverse from $\mathcal{O}\{(N_\text{r}L)^3\}$ to $\mathcal{O}\{N_\text{r}^3+L^3\}$ and improves numerical stability with limited training data. However, strong space-time coupling, heterogeneous clutter, or rapid dynamics can violate the model and reduce suppression performance.

\subsubsection{Extension to Bistatic/Multistatic STAP}

The STAP formulation in \eqref{eq:w MVDR} is waveform-aware because the effective space-time steering vector takes the form $\widetilde{\mathbf{v}}_n=\mathbf{X}_n\mathbf{v}_n(\cdot)$ and depends on the instantaneous probing symbols $\{\mathbf{x}_n[\ell]\}_{\ell=0}^{L-1}$ through $\mathbf{X}_n$ in \eqref{eq:Xn def}. In bistatic and multistatic sensing, the feasibility of STAP is largely determined by whether the sensing receiver has access to such symbol-level waveform information. Following Remark 3, we consider two cases according to the availability of such information.

When the sensing waveform is available at the receiver, the extension of \eqref{eq:w MVDR} and \eqref{eq:SCNR STAP} follows by replacing the monostatic space-time steering vector with its bistatic counterpart. Knowledge of the waveform may come from a cooperative transmitter that shares $\{\mathbf{x}_n[\ell]\}$, or from a reference link that enables waveform reconstruction. With non-colocated transmit and receive arrays, the space-time steering vector must account for distinct transmit angles-of-departure (AoDs) and receive AoAs. For subcarrier $n$, define the bistatic steering vector as
\begin{equation}
\mathbf{v}^\text{bi}_n(\theta_\text{T},\theta_\text{R},f_\text{D})\triangleq \mathbf{d}(f_\text{D})\otimes \mathbf{b}_n(\theta_\text{R})\otimes \mathbf{a}_n^{*}(\theta_\text{T}),
\end{equation} 
where $\theta_\text{T}$ and $\theta_\text{R}$ denote the transmit AoD and receive AoA, respectively. 
Replacing the steering vector in \eqref{eq:w MVDR}--\eqref{eq:SCNR STAP} by $\mathbf{v}^\text{bi}_n(\theta_\text{T},\theta_\text{R},f_\text{D})$ yields the corresponding waveform‑aware bistatic/multistatic STAP weights and SCNR.

When the instantaneous transmit waveform is not available, coherent waveform-aware STAP in the form of \eqref{eq:w MVDR} is generally infeasible. This situation arises with non-cooperative sources or under stringent inter-node signaling constraints. An alternative is to formulate STAP using second-order transmit statistics, which are substantially cheaper to exchange than per-symbol waveforms and are often the only information available under limited backhaul or fronthaul. Define the stacked transmit vector on subcarrier $n$ over one CPI as
\begin{align}
    \widetilde{\mathbf{x}}_n \triangleq [\mathbf{x}_n^T[0],\dots,\mathbf{x}_n^T[L-1]]^T\in\mathbb{C}^{N_\text{t}L},
\end{align}
and its space-time covariance $\mathbf{R}_{\widetilde{\mathbf{x}},n}\triangleq \mathbb{E}\{\widetilde{\mathbf{x}}_n\widetilde{\mathbf{x}}_n^H\}$.
For transmit direction $\theta_\text{T}$, the corresponding effective slow‑time waveform seen by a scatterer can be expressed as
\begin{equation}\begin{aligned}
\mathbf{x}_{\text{prb},n}(\theta_\text{T}) & \triangleq [\mathbf{a}_n^H(\theta_\text{T})\mathbf{x}_n[0],\dots,\mathbf{a}_n^H(\theta_\text{T})\mathbf{x}_n[L-1]]^T \\
& = [\mathbf{I}_L \otimes \mathbf{a}_n^H(\theta_\text{T})]\widetilde{\mathbf{x}}_n,
\end{aligned}\end{equation}
whose covariance is
\begin{align}
\mathbf{R}_{\text{prb},n}(\theta_\text{T})&\triangleq 
\mathbb{E}\{\mathbf{x}_{\text{prb},n}(\theta_\text{T})\mathbf{x}^H_{\text{prb},n}(\theta_\text{T})\} \\
& = (\mathbf{I}_L \otimes \mathbf{a}_n^{H}(\theta_\text{T}))\,
\mathbf{R}_{\widetilde{\mathbf{x}},n}(\mathbf{I}_L \otimes \mathbf{a}_n(\theta_\text{T})).
\end{align}
The target return on subcarrier $n$ can be written as 
\begin{align}
    \mathbf{y}_{t,n} = \alpha_{t,n} [\mathbf{D}(f_{\text{D},t})\mathbf{x}_{\text{prb},n}(\theta_{\text{T},t})]\otimes\mathbf{b}_n(\theta_{\text{R},t}),
\end{align}
where $\mathbf{D}(f_\text{D})\triangleq \text{diag}(\mathbf{d}(f_\text{D}))$, and $\alpha_{t,n}\sim \mathcal{CN}(0,\sigma_{t,n}^2)$ represents the target RCS together with the associated path-loss. For a target hypothesized at
$(\theta_{\text{T},t},\theta_{\text{R},t},f_{\text{D},t})$, the corresponding space-time target covariance is
\begin{align}\label{eq:target covariance}
\mathbf{R}_{t,n} & = \sigma^2_{t,n}
[\mathbf{D}(f_{\text{D},t})\mathbf{R}_{\text{prb},n}(\theta_{\text{T},t})\mathbf{D}^{H}(f_{\text{D},t})] \nonumber \\
& \qquad \otimes [\mathbf{b}_n(\theta_{\text{R},t})\mathbf{b}_n^{H}(\theta_{\text{R},t})].
\end{align}
Given $\mathbf{R}_{t,n}$ and the interference-plus-noise covariance $\mathbf{R}_{I,n}$ (or its SCM estimate), STAP can be posed as a maximum-SCNR design through the generalized Rayleigh quotient
\begin{equation}
\max_{\mathbf{w}_n\neq \mathbf{0}} \ \text{SCNR}_n(\mathbf{w}_n)
= \frac{\mathbf{w}_n^{H}\mathbf{R}_{t,n}\mathbf{w}_n}
{\mathbf{w}_n^{H}\mathbf{R}_{I,n}\mathbf{w}_n},
\end{equation}
whose maximizer is the dominant generalized eigenvector of $(\mathbf{R}_{t,n},\mathbf{R}_{I,n})$, or equivalently the dominant eigenvector of
$\mathbf{R}_{I,n}^{-1}\mathbf{R}_{t,n}$, and the optimal SCNR is
\begin{equation}
\text{SCNR}_n^{\star}=\lambda_{\max}(\mathbf{R}_{I,n}^{-1}\mathbf{R}_{t,n}).
\end{equation}
When the waveform is available, $\mathbf{R}_{t,n}$ is rank-one and the resulting solution satisfies $\mathbf{w}_n\propto\mathbf{R}_{I,n}^{-1}\widetilde{\mathbf{v}}_n$, which matches the waveform-aware solution in \eqref{eq:w MVDR} up to an irrelevant scaling.

This covariance-driven formulation highlights a key trade-off. For many ISAC waveforms, including OFDM, the slow-time samples in one CPI are often modeled as having temporally white second-order statistics. In this case, $\mathbf{R}_{\text{prb},n}(\theta_{\text{T}})\propto \mathbf{I}_L$ and different slow-time samples are uncorrelated. This situation arises when the exchanged transmit information is limited to a per-symbol spatial covariance $\mathbf{R}_{X,n}\triangleq\mathbb{E}\{\mathbf{x}_n[\ell]\mathbf{x}^H_n[\ell]\}\in\mathbb{C}^{N_\text{t}\times N_\text{t}}$, which is commonly used in spatial beamforming design. If the data symbols are temporally independent across OFDM symbols and the per-symbol transmit covariance is approximately constant within a CPI, then 
\begin{align}
    \mathbf{R}_{\widetilde{\mathbf{x}},n} \approx \mathbf{I}_L\otimes \mathbf{R}_{X,n}.
\end{align}
This implies $\mathbf{R}_{\text{prb},n}(\theta_\text{T})\approx \left(\mathbf{a}^H_n(\theta_\text{T})\mathbf{R}_{X,n}\mathbf{a}_n(\theta_\text{T})\right)\mathbf{I}_L$. Doppler does not change the second-order statistics in this case, so $\mathbf{D}(f_{\text{D},t})\mathbf{R}_{\text{prb},n}(\theta_{\text{T},t})\mathbf{D}^{H}(f_{\text{D},t})=\mathbf{R}_{\text{prb},n}$. Thus, $f_{\text{D}}$ disappears from the target covariance in \eqref{eq:target covariance}, and Doppler discrimination cannot be achieved using second-order statistics alone.

Doppler selectivity can be recovered within the covariance-driven framework when the available transmit statistics retain nontrivial slow-time correlation. One low-overhead option is to exchange or infer a separable space-time covariance model
\begin{align}
\mathbf{R}_{\widetilde{\mathbf{x}},n}\approx\mathbf{R}_{\text{st},n}^{\text{tx}}\otimes\mathbf{R}_{X,n},
\end{align}
where $\mathbf{R}_{\text{st},n}^{\text{tx}}\in\mathbb{C}^{L\times L}$ summarizes the slow‑time correlation induced by structured transmissions such as repeated reference sequences and deterministic training. In this model, $\mathbf{R}_{\text{prb},n}(\theta_\text{T})\approx \left(\mathbf{a}^H_n(\theta_\text{T})\mathbf{R}_{X,n}\mathbf{a}_n(\theta_\text{T})\right)\mathbf{R}^\text{tx}_{\text{st},n}$, and the Doppler dependence in \eqref{eq:target covariance} appears through $\mathbf{D}(f_\text{D})\mathbf{R}_{\text{st},n}^{\text{tx}}\mathbf{D}^H(f_\text{D})$.

For multistatic deployments with multiple illuminators indexed by $i$, the same construction applies by using the corresponding $\mathbf{R}^{(i)}_{\widetilde{\mathbf{x}},n}$ or a low‑overhead factorization thereof to form $\mathbf{R}^{(i)}_{\text{t},n}$. The overall target covariance can then be obtained by aggregating the per-illuminator covariances. If the illuminators are mutually coherent, the cross-covariance terms can be included. Otherwise, an additive superposition of the per-illuminator covariances provides an accurate approximation.

\begin{figure*}[!t]\centering
\captionsetup[subfigure]{skip=2pt}
\begin{subfigure}[t]{0.33\textwidth}
    \centering
    \includegraphics[height=4.8cm]{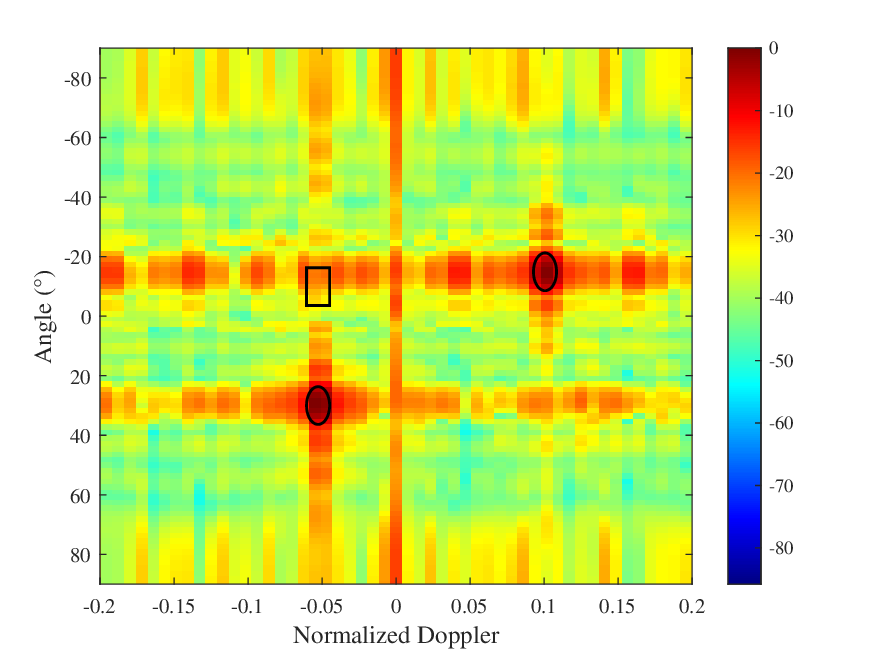}
    \caption{Before STAP.}
    \label{fig:post_MF}
    \end{subfigure}\hfill
    \begin{subfigure}[t]{0.33\textwidth}
        \centering
        \includegraphics[height=4.8cm]{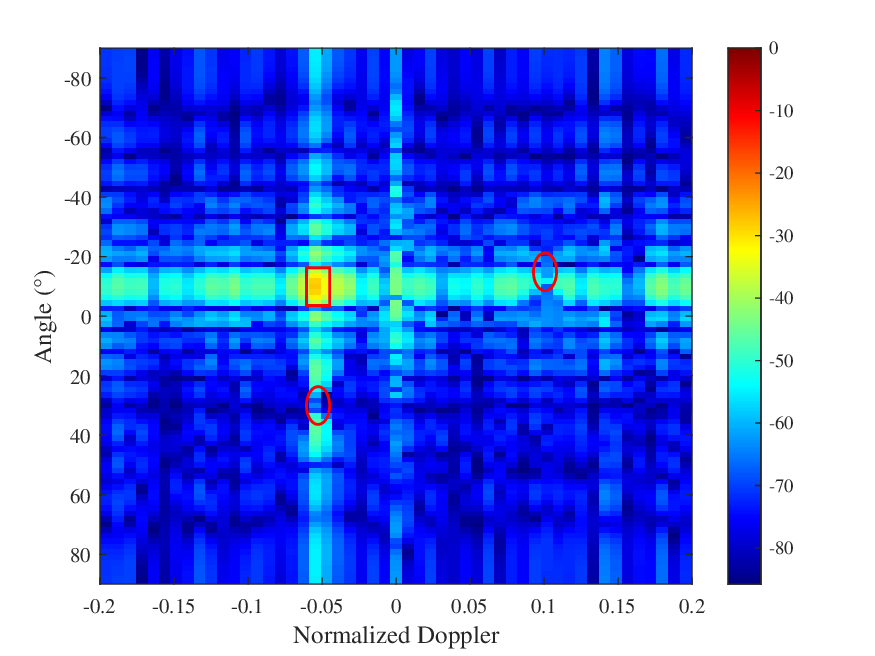}
        \caption{Classical STAP.}
        \label{fig:poSTAP_classic}
    \end{subfigure}\hfill
    \begin{subfigure}[t]{0.33\textwidth}
        \centering
        \includegraphics[height=4.8cm]{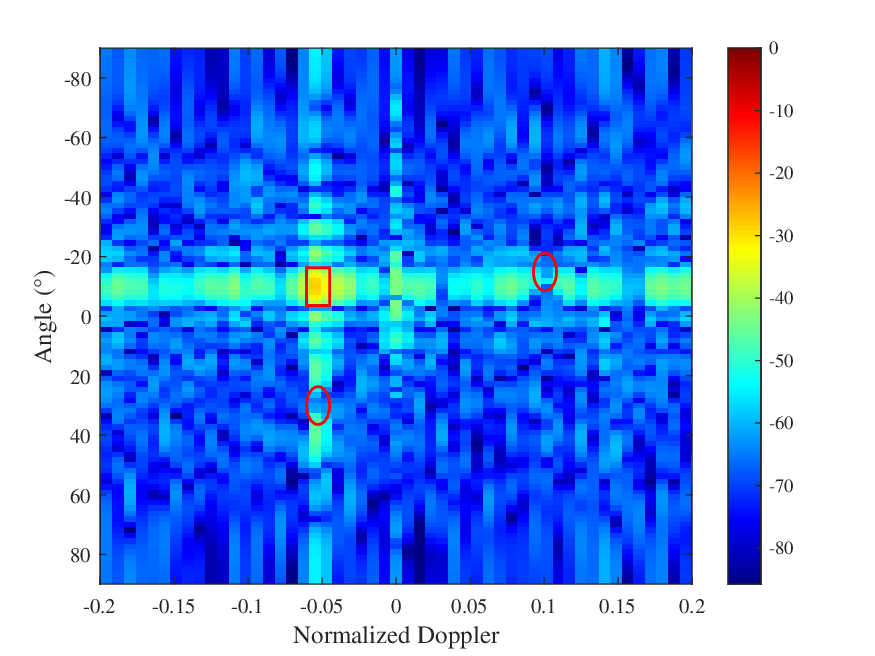}
        \caption{RR-STAP.}
        \label{fig:postSTAP_RR}
    \end{subfigure}

    \vspace{1mm}
    \caption{Post-STAP angle-Doppler maps under cold and hot clutter (rectangle: target of interest, ellipse: strong UAV targets).}
    \label{fig:post_STAP_ADM}
    \vspace{-2mm}
\end{figure*}

\begin{figure*}[!t]\centering
\captionsetup[subfigure]{skip=2pt}
\begin{subfigure}[t]{0.33\textwidth}
    \centering
    \includegraphics[height=4.8cm]{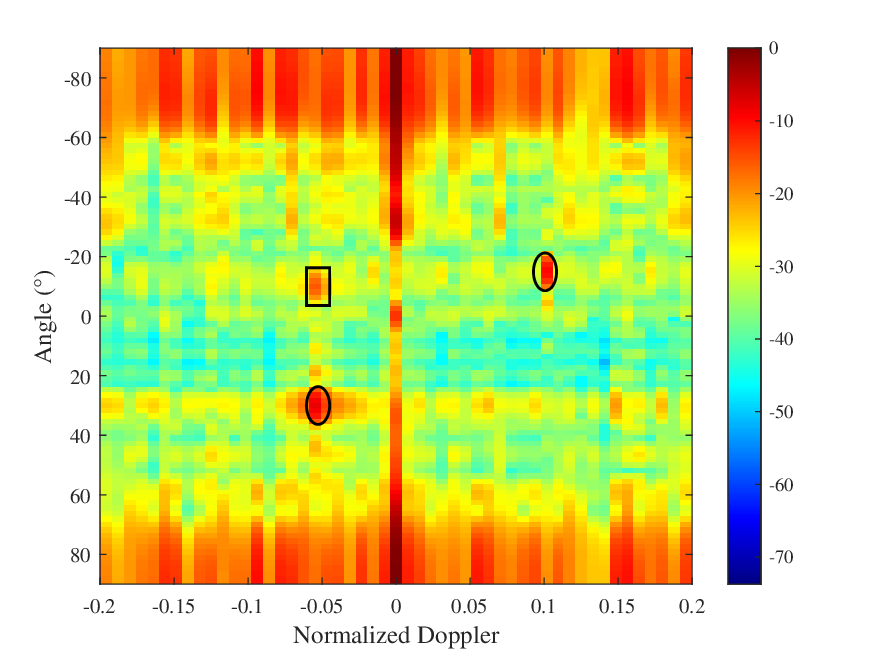}
    \caption{Before STAP.}
    \label{fig:post_MF_sionna}
    \end{subfigure}\hfill
    \begin{subfigure}[t]{0.33\textwidth}
        \centering
        \includegraphics[height=4.8cm]{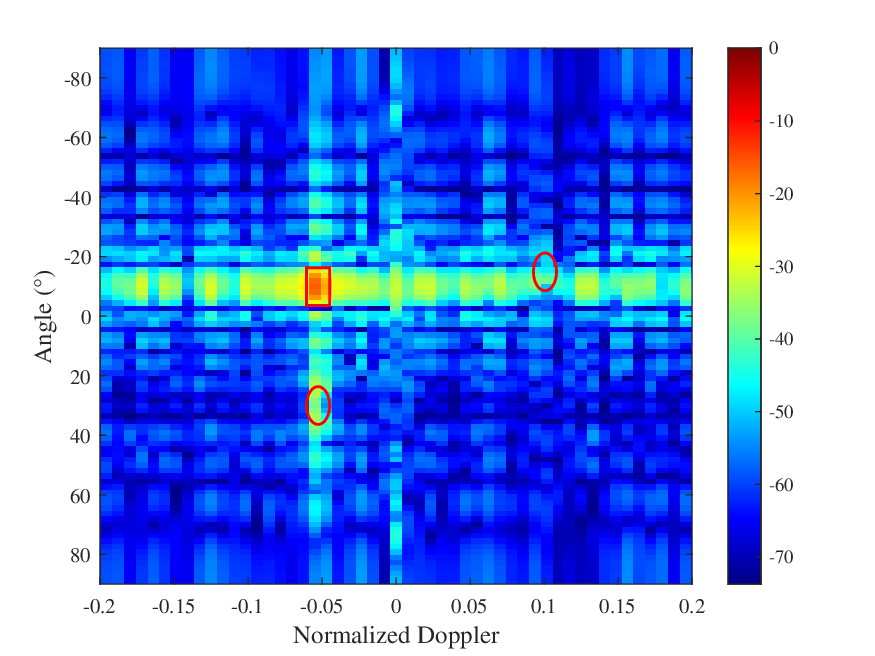}
        \caption{Classical STAP.}
        \label{fig:poSTAP_classic_sionna}
    \end{subfigure}\hfill
    \begin{subfigure}[t]{0.33\textwidth}
        \centering
        \includegraphics[height=4.8cm]{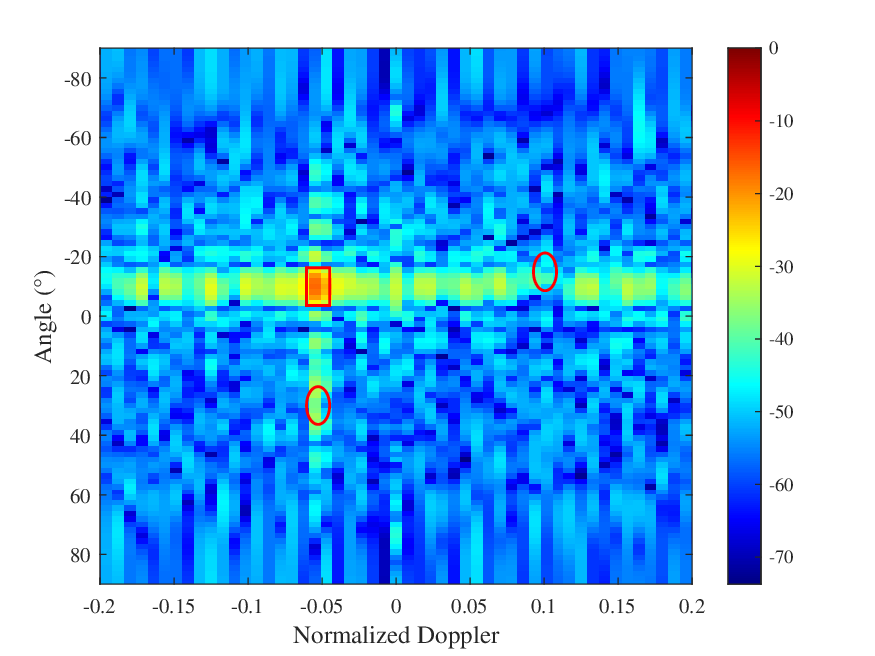}
        \caption{RR-STAP.}
        \label{fig:postSTAP_RR_sionna}
    \end{subfigure}

    \vspace{1mm}
    \caption{Post-STAP angle-Doppler maps under cold and hot clutter generated by Sionna RT simulation (rectangle: target of interest, ellipse: strong UAV targets).}
    \label{fig:post_STAP_sionna}
    \vspace{-2mm}
\end{figure*}

\subsubsection{Case Studies}

Figs.~\ref{fig:post_STAP_ADM} and \ref{fig:post_STAP_sionna} show the angle-Doppler maps at the ToI range gate for the mixed cold- and hot-clutter scenario, under the stochastic scatterer model and the site-specific Sionna RT scene, respectively. In both cases, we adopt a wideband per-subcarrier processing procedure. For each scanned $(\theta,f_{\mathrm{D}})$, the matched-filter baselines in Figs.~\ref{fig:post_STAP_ADM}(a) and \ref{fig:post_STAP_sionna}(a) apply a space-time matched filter on each subcarrier steered to $(\theta,f_{\mathrm{D}})$, i.e., the weight is chosen proportional to the effective steering vector $\widetilde{\mathbf{v}}_n(\theta,f_{\mathrm{D}})$ and is applied to the stacked snapshot $\mathbf{y}_n$ without SCM-based interference whitening. For the classical STAP and RR-STAP results in Figs.~\ref{fig:post_STAP_ADM}(b)--(c) and \ref{fig:post_STAP_sionna}(b)--(c), we instead form the STAP weights using the SCM estimate $\widehat{\mathbf{R}}_{I,n}$ and $\widetilde{\mathbf{v}}_n(\theta,f_{\mathrm{D}})$. The resulting per-subcarrier outputs are then coherently fused across frequency using the delay steering vector associated with the ToI range gate to obtain each map value.

In the stochastic case, the cold clutter is generated by $C=100$ scatterers whose AoAs are independently drawn from a uniform distribution over $[-90^\circ, 90^\circ]$. This produces a spatially rich near-zero-Doppler clutter ridge that is relatively uniform over angle. In Fig.~\ref{fig:post_STAP_ADM}(a), the matched-filter map is interference dominated. A pronounced ridge near $f_{\mathrm{D}}\approx 0$ arises from quasi-static cold clutter, while hot clutter and the strong UAV echoes create additional peaks and leakage that raise the background and mask the weak ToI. This outcome also reflects the limitations of one-dimensional suppression. Slow-time filtering mainly attenuates the near-zero-Doppler region and cannot eliminate strong mobile scatterers at non-zero Doppler, whereas spatial-only filtering has difficulty when a dominant scatterer lies close to the ToI in angle and the interference exhibits coupled angle-Doppler structure. Fig.~\ref{fig:post_STAP_ADM}(b) shows that STAP resolves these issues by jointly exploiting spatial and slow-time DoFs. It suppresses the zero-Doppler clutter and places deep notches at the UAV bins marked by ellipses, which lowers the local background and makes the ToI clearly visible in the angle-Doppler plane. Fig.~\ref{fig:post_STAP_ADM}(c) shows the RR-STAP result, which shows similar performance but with a higher residual background consistent with the reduced-rank performance-complexity tradeoff for limited secondary data.

Fig. \ref{fig:post_STAP_sionna} repeats the same experiment for the RT-generated mixed-clutter environment. Compared with the stochastic case, the low-Doppler cold-clutter in Fig. \ref{fig:post_STAP_sionna}(a) exhibits less uniform angular support and an overall weaker background level. This is because the cold-clutter is dominated by site-specific building backscatter concentrated over a non-uniform set of azimuth angle regions. Meanwhile, the hot clutter is more prominent and has increased Doppler spread in the RT scene. This is due to the fact that the ray tracing captures multiple indirect propagation paths between the external source and the receiver through environmental interactions (as before, with a maximum interaction depth of three). Despite this stronger and more diffuse interference footprint, STAP in Fig.~\ref{fig:post_STAP_sionna}(b) still provides strong suppression of both the near-zero-Doppler and the hot-clutter leakage. It also mitigates the impact of the dominant UAV in the local angle-Doppler neighborhood, thereby improving the ToI visibility.

\subsection{Knowledge-Aided Methods and Machine Learning}

Despite the effectiveness of the previously discussed temporal, spatial, and space-time clutter suppression methods, practical ISAC scenarios may require more sophisticated processing strategies. Conventional approaches depend heavily on homogeneous snapshots and stationary clutter assumptions, which may not hold in real dynamic environments. To address these limitations, knowledge-aided and machine learning-based approaches have emerged. These ``intelligent'' methods leverage external prior knowledge or data to improve clutter suppression performance under limited data availability and imprecisely understood clutter conditions.

\subsubsection{Knowledge-Aided Methods} Knowledge-aided (KA) methods improve adaptive processing by leveraging externally available prior environmental information \cite{Guerci2010}, such as geometric information derived from digital elevation models (DEMs), building footprints, road networks, and/or communication channel statistics. The key idea is to translate known or learned environmental characteristics into priors for interference modeling, covariance estimation, and adaptive filtering, thereby reducing the dependence on large homogeneous secondary datasets and improving robustness under clutter heterogeneity and nonstationarity.

A common theme across recent studies is to first construct a clutter representation that is informed by available prior knowledge and then incorporate it into the sensing receiver. One representative direction focuses on structural and subspace information. Prior knowledge constrains where clutter energy is expected to concentrate and it therefore reduces the effective adaptive dimension. This perspective is closely related to reduced-dimension and reduced-rank STAP, where the projection basis is guided by the clutter support so that adaptive degrees of freedom are spent on the relevant interference region. In ISAC systems, channel knowledge maps provide a practical mechanism to store offline-learned location-dependent clutter characteristics \cite{Xu-ICCC-2024,ZXu-arxiv-2025,DWu-arxiv-2025}. The stored information may include dominant scattering angles and delay-Doppler signatures. During online operation, the receiver can use these priors to perform spatial projection or zero-forcing to suppress the known clutter directions before parameter estimation. This map assisted pre-nulling is particularly useful for low-velocity or quasi-stationary targets because Doppler-based suppression can be ineffective in that regime and may suppress the desired echo together with low-Doppler clutter.

Another direction emphasizes statistical and covariance information. Prior knowledge is used to regularize covariance estimation through shrinkage, structured modeling, and loading strategies. A typical formulation combines a sample covariance with a predicted or historical covariance model as $\widehat{\mathbf{R}}_I = \alpha \widehat{\mathbf{R}}_\text{s} + (1-\alpha)\mathbf{R}_{\text{prior}}$. These techniques improve numerical conditioning and stabilize adaptive filters in small-sample regimes and under heterogeneous training. In wideband MIMO-OFDM ISAC, the prior can also reflect waveform-induced statistics and frequency-dependent array responses, which are important when effects such as beam-squint are non-negligible. KA sequential angle-range processing detectors provide an example where prior interference covariances are exploited to construct filters that approach SINR optimal performance under practical wideband conditions \cite{UBSarac-TCOM-2025}.

Prior knowledge can also be incorporated throughout the entire processing chain. In this case, environmental information guides upstream operations that affect subsequent clutter suppression. Examples include time-frequency synchronization, sensing parameter association, and hypothesis management. In asynchronous perceptive vehicular networks, stable clutter returns can serve as environment specific fingerprints to support synchronization and association, which then enables more coherent processing and improved clutter mitigation \cite{Wang-JSAC-2024}. In addition, priors derived from channel knowledge maps can provide information about multipath geometry and can reduce ambiguity in non-LoS sensing. This is achieved by narrowing the feasible hypothesis space and separating target related echoes from multipath induced artifacts \cite{DWu-arxiv-2025}.

Knowledge-aided methods provide a physically interpretable and modular route to robust clutter suppression. They complement purely data driven adaptation by incorporating map and geometry driven constraints, prior regularized covariance learning, and knowledge-guided receiver processing. These advantages become more pronounced in heterogeneous ISAC environments, including bistatic and multistatic deployments and wideband operation, where clutter statistics vary with location, waveform, and network configuration.
 
\subsubsection{Machine Learning Methods}
Machine learning (ML) offers a complementary route to KA processing for clutter-aware ISAC, especially when the clutter statistics are heterogeneous, nonstationary, or difficult to capture with a parametric model. Beyond purely data-driven adaptation, model-driven learning provides a practical bridge between optimization and learning by embedding physics- and model-based clutter structure into trainable networks, enabling more reliable performance and reduced computational complexity \cite{PJiang-TCOM-2025}.

Recent advances illustrate that ML can also enhance clutter-aware ISAC over the entire processing chain. In cooperative and cell-free deployments, learning-based representations and learned priors can capture network-coupled interference and enable scalable, geometry-adaptive fusion/inference as clutter statistics vary with the environment and network configuration \cite{Jiang 2025}. At the receiver, learned denoising or interference-aware filtering can suppress clutter and residual self-interference, followed by end-to-end estimation/tracking networks that infer sensing states directly from communication waveforms with reduced reliance on handcrafted models \cite{YWang-ICC-2025}. At the transmitter, reinforcement learning and policy optimization can further support online waveform/beam control to proactively avoid clutter-dominant scatterers, naturally coupling receiver-side suppression with transmitter-side adaptation and sensing--communication trade-offs \cite{HCho-MILCOM-2024,AUmar-arxiv-2025}. Hybrid KA and ML designs that combine interpretable priors with learned components are particularly promising for generalization across different types of deployments.

\section{Clutter Management in ISAC Systems}

The previous section explored receiver-side processing techniques that suppress clutter after it has already contaminated the received signals. In this section we shift focus to \emph{proactive clutter management} through joint transmit and receive design. Modern wireless communication systems routinely exploit transmit-side DoFs such as beamforming and power control to adapt to time-varying channels and satisfy QoS requirements. In ISAC, these same DoFs can be leveraged to shape the radiated field and resulting reflections, which can reduce clutter illumination, improve the conditioning of the interference statistics, and preserve communication reliability under sensing. Beyond spatial beamforming and power allocation, transmit waveform design across slow time (and frequency for wideband ISAC) also provides additional proactive DoFs that control range and Doppler sidelobes and reduce clutter leakage before receiver-side suppression \cite{PLi-TSP-2025}. Building on this motivation, Sections~VI-A and VI-B develop joint spatial and space-time transceiver optimization for clutter mitigation. Sec.~VI-C discusses joint transmit/receive design that selectively exploits resolvable target-related non-line-of-sight (NLoS) components, and Sec.~VI-D summarizes clutter-aware multi-domain optimization that extends the design space to the EM and network domains.

\subsection{Joint Transmit/Receive Spatial Beamforming Optimization}

We first consider joint transceiver design in the spatial domain based on the linear block-level precoding model in \eqref{eq:tx signal xnl}. For each subcarrier $n$, the BS transmits zero-mean vectors $\mathbf{x}_n[\ell]$ whose second-order statistics are characterized by $\mathbf{R}_{X,n} = \mathbb{E}\{\mathbf{x}_n[\ell]\mathbf{x}_n^H[\ell]\}$. Under the model in Sec. II-B, we have $\mathbf{R}_{X,n} = \mathbf{W}_n\mathbf{W}_n^H$ for the linear precoder $\mathbf{W}_n$. This design of $\mathbf{W}_n$ controls the per-subcarrier illumination pattern, but does not impose any slow-time coding across OFDM symbols. As a result, the cold-clutter return on each subcarrier is determined by the contribution of $\mathbf{R}_{X,n}$ towards the clutter directions.
Since this subsection focuses on spatial-only processing, all covariances are $N_\text{r}\times N_\text{r}$ unless otherwise stated, and we drop the superscript $(\cdot)^{\text{sp}}$ for simplicity.

The sensing receiver applies a per-subcarrier spatial combiner $\mathbf{u}_n\in\mathbb{C}^{N_\text{r}}$ to the frequency‑domain snapshot $\mathbf{y}_n[\ell]$ at subcarrier $n$ and OFDM symbol $\ell$ to produce
\begin{equation}
r_n[\ell] = \mathbf{u}_n^H \mathbf{y}_n[\ell].
\end{equation}
The resulting observations are then range-Doppler focused by coherent accumulation across subcarriers and OFDM symbols:
\begin{equation}
\widehat{y}(\tau, f_{\text{D}}) = \sum_{n=0}^{N-1}\sum_{\ell=0}^{L-1} r_n[\ell] e^{-\jmath2\pi f_{\text{D}}\ell T_{\text{sym}}} e^{\jmath2\pi n\Delta_f \tau}.
\end{equation}
For angle $\theta$, we define the transmit and receive beamforming gains on subcarrier $n$ as, respectively,
\begin{align}
g_n(\theta)=\mathbf{a}_n^H(\theta)\mathbf{R}_{X,n}\mathbf{a}_n(\theta),\quad     h_n(\theta)=|\mathbf{u}_n^H\mathbf{b}_n(\theta)|^2.
\end{align}

For spatial-only design, the disturbance covariance can be written as 
$\mathbf{R}_{I,n}(\mathbf{R}_{X,n}) = \mathbf{R}_{\text{cc},n}(\mathbf{R}_{X,n}) + \mathbf{R}_{\bm{\eta},n}$, where $\mathbf{R}_{\text{cc},n}(\mathbf{R}_{X,n})$ is the cold-clutter covariance induced by the transmit waveform and $\mathbf{R}_{\bm{\eta},n}$ captures waveform-independent hot clutter interference, thermal noise, and residual SI. We characterize cold-clutter scattering through the spatial inner-kernel formulation in Sec. IV-B.5, $\text{vec}\{\mathbf{R}_{\text{cc},n}(\mathbf{R}_{X,n})\} = \mathbf{V}^{\text{sp}}_{\text{cc},n}\text{vec}\{\mathbf{R}_{X,n}\}$, where $\mathbf{V}^{\text{sp}}_{\text{cc},n}$ is a waveform-independent environmental signature learned from target-free secondary data, while $\mathbf{R}_{\bm{\eta},n}$ is estimated separately, e.g., during quiet periods as discussed in Sec.~IV-E. In the formulation below, we assume estimates of $\mathbf{V}^{\text{sp}}_{\text{cc},n}$ and $\mathbf{R}_{\bm{\eta},n}$ are available.

The SCNR for target angle $\theta_t$ is
\begin{equation}
\text{SCNR}_{\text{BLP}}(\theta_t) = \frac{\sum_{n=0}^{N-1} \sigma_{t,n}^2 h_n(\theta_t)g_n(\theta_t)}{\displaystyle\sum_{n=0}^{N-1}\mathbf{u}_n^H\big[ \mathbf{R}_{\text{cc},n}(\mathbf{R}_{X,n})+ \mathbf{R}_{\bm{\eta},n}\big]\mathbf{u}_n},
\end{equation}
where $\sigma_{t,n}^2$ is the reflected target power on subcarrier $n$. This RD-cell-based SCNR expression reflects coherent processing over both the slow-time and frequency grids, which differs from non-coherent SCNR accumulation across symbols and subcarriers as considered in \cite{TWei-TCOM-2023}.

We then formulate a clutter-aware joint transmit and receive beamforming problem that maximizes the SCNR at a desired RD cell while satisfying the downlink communication QoS constraints and the total transmit power budget:
\begin{subequations}\label{eq:BLP SCNR max}
\begin{align}
\max_{\mathbf{R}_{X,n},\mathbf{u}_n} & \quad \text{SCNR}_{\text{BLP}}(\theta_t) \\
\text{s.t.}&\quad \text{SINR}_{n,k}\ge\gamma_{n,k},\quad \forall n,k, \\
&\quad \sum_{n=0}^{N-1}\text{tr}(\mathbf{R}_{X,n}) \le P_\text{tot},
\end{align}
\end{subequations}
where $\gamma_{n,k}$ denotes the required communication QoS for user $k$ on subcarrier $n$, and $P_\text{tot}$ represents the total transmit power constraint. Additional constraints including per-subcarrier and per-antenna power limits, as well as hardware-related considerations such as peak-to-average power ratio (PAPR), could also be incorporated into \eqref{eq:BLP SCNR max}. In the following, we use the problem formulation of~\eqref{eq:BLP SCNR max} to illustrate approaches for addressing this class of non-convex optimization problems.

To effectively address the complexity of the joint transmit--receive beamforming optimization, we adopt an alternating optimization (AO) approach that decomposes the original problem into two tractable subproblems: Receive spatial filtering and transmit covariance matrix optimization, which are solved iteratively until convergence. Given fixed transmit covariance matrices $\mathbf{R}_{X,n}$, the optimal receive beamformer $\mathbf{u}_n$ for subcarrier $n$ can be independently obtained using MVDR:
\begin{equation}
\mathbf{u}_n^\star=\frac{\mathbf{R}_{I,n}^{-1}(\mathbf{R}_{X,n})\mathbf{b}_n(\theta_t)}{\mathbf{b}_n^H(\theta_t)\mathbf{R}_{I,n}^{-1}(\mathbf{R}_{X,n})\mathbf{b}_n(\theta_t)}, 
\label{eq:MVDR_receive}
\end{equation}
with $\mathbf{R}_{I,n}(\mathbf{R}_{X,n})=\mathbf{R}_{\text{cc},n}(\mathbf{R}_{X,n})+ \mathbf{R}_{\bm{\eta},n}$. 
Then, with the $\mathbf{u}_n$ fixed, the transmit covariance optimization results in a fractional form. To efficiently tackle the fractional objective, we define $\mathbf{R}_{n,k} = \mathbf{w}_{n,k}\mathbf{w}_{n,k}^H$ and apply Dinkelbach’s method \cite{Dinkelbach1967} with an auxiliary scalar $\eta$:
\begin{subequations}
\label{eq:TX_Dinkelbach}
\begin{align}
&\max_{{\mathbf{R}_{X,n}, \mathbf{R}_{n,k}\succeq\mathbf{0}}} \quad \sum_{n=0}^{N-1}\text{tr}\left((\widetilde{\mathbf{A}}_{t,n}-\eta\widetilde{\mathbf{A}}_{\text{c},n})\mathbf{R}_{X,n}\right)\\
&\text{s.t.}~ (1+\gamma_{n,k}^{-1})\text{tr}(\mathbf{H}_{n,k}\mathbf{R}_{n,k}) \geq \text{tr}(\mathbf{H}_{n,k}\mathbf{R}_{X,n})+\sigma_\text{comm}^2, \forall k,n,\\
&\quad\quad \sum_{n=0}^{N-1}\text{tr}(\mathbf{R}_{X,n})\leq P_\text{tot},\\
&\quad\quad \mathbf{R}_{X,n} \!-\! \sum_{k=1}^K\!\mathbf{R}_{n,k} \succeq\mathbf{0},~ \text{Rank}\{\mathbf{R}_{n,k}\} = 1,~\forall k,n,\!
\end{align}
\end{subequations}
where $\mathbf{H}_{n,k} \triangleq \mathbf{h}_{n,k}\mathbf{h}_{n,k}^H$, 
$\widetilde{\mathbf{A}}_{t,n} \triangleq \sigma^2_{t,n}h_n(\theta_t)\mathbf{a}_n(\theta_t)\mathbf{a}^H_n(\theta_t)$, $\widetilde{\mathbf{A}}_{\text{c},n}\triangleq\text{unvec}\{(\mathbf{V}^{\text{sp}}_{\text{cc},n})^H\text{vec}\{\mathbf{u}_n\mathbf{u}_n^H\}\}_{N_\text{t}\times N_\text{t}}$ and $\sigma_\text{comm}^2$ is the noise power at the communication users. Temporarily relaxing the non-convex rank-one constraints yields a convex semidefinite programming (SDP) problem, which can be efficiently solved via standard numerical optimization tools. After obtaining the SDP solutions $\mathbf{R}_{X,n}^\star$ and $\mathbf{R}_{n,k}^\star$, the optimal rank-one transmit beamforming solutions for the communication users can be recovered using
\begin{align}
    \mathbf{w}_{n,k}^\star = (\mathbf{h}_{n,k}^H\mathbf{R}^\star_{n,k}\mathbf{h}_{n,k})^{-\frac{1}{2}} \mathbf{R}^\star_{n,k} \mathbf{h}_{n,k},~\forall k, n.
\end{align}
The sensing beamformers are subsequently obtained via an eigendecomposition:
\begin{align}
    \mathbf{W}_{\text{s},n}\mathbf{W}_{\text{s},n}^H = \mathbf{R}_{X,n}^\star - \sum_{k=1}^K\mathbf{w}_{n,k}^\star(\mathbf{w}_{n,k}^\star)^H.
\end{align}
After each SDP solution, the auxiliary scalar $\eta$ is updated as
\begin{equation}
\eta=\frac{\sum_{n=0}^{N-1}\text{tr}(\widetilde{\mathbf{A}}_{t,n}\mathbf{R}_{X,n}^\star)}{\sum_{n=0}^{N-1} \big[ \text{tr}(\widetilde{\mathbf{A}}_{\text{c},n}\mathbf{R}_{X,n}^\star)+\mathbf{u}_n^H\mathbf{R}_{\bm{\eta},n}\mathbf{u}_n\big] },  
\label{eq:beta_update}
\end{equation}
and algorithm iterations continue to convergence.

Alternatively, explicitly representing $\mathbf{R}_{n,k}$ as $\mathbf{w}_{n,k}\mathbf{w}_{n,k}^H$, the optimization problem has a difference-of-convex (DC) formulation. Applying Dinkelbach’s transformation, the fractional objective reduces to
\begin{equation} 
\max_{{\mathbf{w}_{n,i}}} \quad \sum_{n=0}^{N-1}\sum_{i=1}^{N_\text{s}}(\mathbf{w}_{n,i}^H\widetilde{\mathbf{A}}_{t,n}\mathbf{w}_{n,i}-\eta\mathbf{w}_{n,i}^H\widetilde{\mathbf{A}}_{\text{c},n}\mathbf{w}_{n,i}). 
\end{equation} 
Leveraging the DC structure, the convex-concave procedure (CCP) is used. Specifically, the convex term is linearized around the beamforming vector $\mathbf{w}_{n,i}^{(j)}$ from the previous iteration as follows: 
\begin{equation} 
\mathbf{w}_{n,i}^H\widetilde{\mathbf{A}}_{t,n}\mathbf{w}_{n,i} \geq 2\Re\{\mathbf{w}_{n,i}^H\widetilde{\mathbf{A}}_{t,n}\mathbf{w}_{n,i}^{(j)}\} + \text{const}. 
\end{equation} 
Hence, the original non-convex DC problem reduces to a convex second-order cone programming (SOCP) problem with SINR and total power constraints expressed as:
\begin{align} 
&|\mathbf{h}_{n,k}^H\mathbf{w}_{n,k}| \geq\! \sqrt{\gamma_{n,k}} \Big\|\Bigl[\mathbf{h}_{n,k}^H\mathbf{W}_{n,-k} \; \; \; \sigma_\text{comm}\Bigr] \Big\|_2,~\forall n,k,\!\! \\
&\sum_{n=0}^{N-1}\sum_{i=1}^{N_\text{s}} \|\mathbf{w}_{n,i}\|_2^2 \leq P,
\end{align} 
where $\mathbf{W}_{n,-k}\triangleq [\mathbf{w}_{n,1},\dots,\mathbf{w}_{n,k-1},\mathbf{w}_{n,k+1},\dots,\mathbf{w}_{n,N_\text{s}}]$. This SOCP formulation enhances computational efficiency and can be solved using existing convex solvers.

Recent studies have explored analogous optimization frameworks with diverse emphases. For instance, \cite{LChen-JSAC-2022} investigates the dual optimization perspectives of maximizing SCNR under sum-rate constraints and vice versa. In \cite{ZHe-JSAC-2023}, a full-duplex ISAC scenario is developed, optimizing either transmit power or sum-rate subject to radar and communication SINR constraints. Robust beamforming optimization considering practical constraints like similarity measures, constant modulus, and per-antenna power is studied in \cite{YNiu-TCCN-2025}. In addition, \cite{JChoi-TWC-2024} incorporates imperfect channel state information due to quantization errors into the optimization. Despite these advances, purely spatial beamforming strategies may not fully resolve all clutter sources, motivating further exploration into STAP and advanced waveform design methods, as discussed next.

\begin{figure}[!t]\centering
\begin{subfigure}[t]{0.9\linewidth}\centering
    \includegraphics[width=\linewidth]{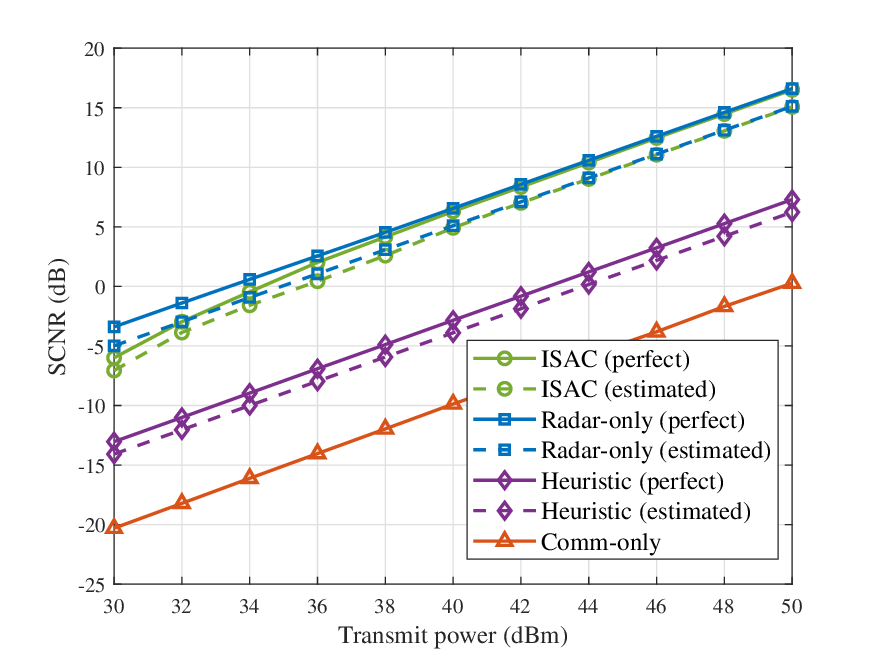}
    \caption{Sensing SCNR versus transmit power $P_\text{tot}$.}\label{fig:SCNR_Ptx}
    \end{subfigure}\hfill
\begin{subfigure}[t]{0.9\linewidth}\centering
    \includegraphics[width=\linewidth]{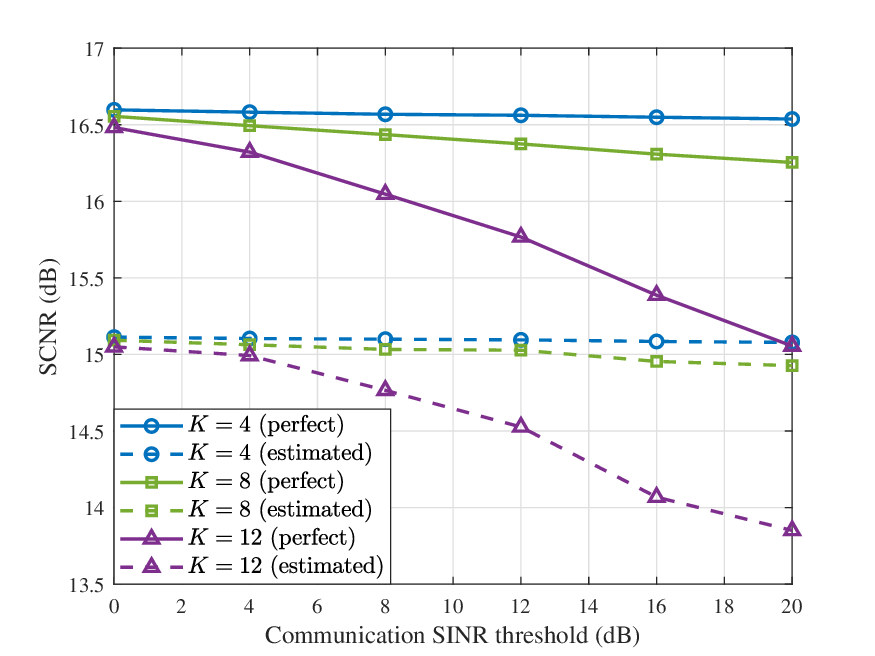}
    \caption{Sensing SCNR versus communication SINR.}\label{fig:SCNR_SINR_BLP}
    \end{subfigure}

    \vspace{1mm}
    \caption{Sensing performance for joint transmit--receive beamforming optimization.}
    \label{fig:BLP}
    \vspace{-4mm}
\end{figure}

Fig.~\ref{fig:BLP} evaluates the sensing performance of the proposed joint transmit–receive beamforming design in a multi-user downlink ISAC scenario. We compare it with three baselines: \textit{(i)} radar-only, which allocates all transmit resources to sensing, \textit{(ii)} communication-only, which optimizes the downlink precoder solely for multi-user communications, and \textit{(iii)} a heuristic scheme that steers beams toward the user channels and an additional sensing stream toward the target direction, followed by an MVDR receive combiner designed from the resulting interference covariance.
Solid curves (``perfect'') assume that the waveform-independent clutter kernel $\mathbf{V}_{\text{cc},n}$ and the remaining disturbance covariance $\mathbf{R}_{\bm{\eta},n}$ are known, whereas dashed curves (``estimated'') use estimates $\widehat{\mathbf{V}}_{\text{cc},n}$ and $\widehat{\mathbf{R}}_{\bm{\eta},n}$ obtained from the covariance-fitting procedure in Sec. IV-B.5. We use $N_\text{tr}=5000$ probing realizations and collect $L=32$ target-free secondary snapshots per realization, where the transmit data $\{\mathbf{X}_n^{(i)}\}$ are generated from $64$-quadrature amplitude modulation (QAM) OFDM signals and are known at the BS. To ensure fairness, SCNR is always evaluated using the true disturbance statistics, and only the design stage uses perfect versus estimated quantities.
Fig.~\ref{fig:BLP}(a) shows the sensing SCNR versus transmit power for $K=6$ users with a $10$dB SINR requirement. The optimized ISAC design consistently approaches the radar-only benchmark and substantially outperforms the communication-only baseline, demonstrating that joint transceiver optimization can maintain sensing performance while meeting multi-user QoS constraints. The heuristic scheme exhibits a noticeable loss, indicating that simple beam steering fails to fully exploit the available sensing DoFs. The gap between solid and dashed curves is modest, reflecting finite-sample errors in estimating $\mathbf{V}_{\text{cc},n}$. Fig.~\ref{fig:BLP}(b) illustrates the sensing--communication tradeoff by showing SCNR versus the SINR threshold for different numbers of users. As the SINR requirement tightens or the number of users increases, more spatial DoFs and power must be allocated to communications, which reduces the achievable sensing SCNR.

\subsection{Joint Transmit/Receive STAP Waveform Optimization}

Sec. VI-A designs a joint spatial transceiver through the per-subcarrier transmit covariance $\mathbf{R}_{X,n}=\mathbb{E}\{\mathbf{x}_n[\ell]\mathbf{x}_n^H[\ell]\}$, which shapes the spatial illumination on each subcarrier but leaves the slow-time waveform across OFDM symbols unconstrained. When cold clutter exhibits Doppler spread \cite{BTang-TSP-2016}--\cite{LWu-TSP-2018}, when hot clutter varies within a CPI, or when clutter and target overlap in angle, spatial-only control is insufficient and the STAP framework in Sec. V-C becomes particularly useful. Since cold clutter is coherent with the probing waveform, the space-time interference covariance in STAP depends on the transmitted waveform through $\mathbf{X}_n$. Thus, here we develop a proactive joint transmit and receive STAP design that optimizes the symbol-level waveforms together with the STAP receive filters under communication QoS constraints.

We assume STAP operates on the stacked snapshot $\mathbf{y}_n\in\mathbb{C}^{N_\text{r}L}$ on each subcarrier $n$. Accordingly, $\mathbf{R}_{I,n}(\mathbf{X}_n)\in\mathbb{C}^{N_\text{r}L\times N_\text{r}L}$ denotes the space-time disturbance covariance. Since we focus solely on space-time processing in what follows, all covariances are of size $N_\text{r}L\times N_\text{r}L$ unless otherwise stated, and we omit the superscript $(\cdot)^{\text{st}}$ for simplicity. 
Following the inner-kernel factorization in Sec.~IV-B.5, the waveform-dependent cold-clutter covariance admits the representation $\mathbf{R}_{I,n}(\mathbf{X}_n) \approx \mathbf{R}_{\text{cc},n}(\mathbf{X}_n)+\mathbf{R}_{\bm{\eta},n} = \mathbf{X}_n\mathbf{V}^{\text{st}}_{\text{cc},n}\mathbf{X}_n^H + \mathbf{R}_{\bm{\eta},n}$, where $\mathbf{V}^{\text{st}}_{\text{cc},n}$ is a waveform-independent space-time clutter kernel that depends only on the scene geometry and angle-Doppler scattering power. The estimates of $\mathbf{V}^{\text{st}}_{\text{cc},n}$ and $\mathbf{R}_{\bm{\eta},n}$ can be found using the methods discussed in Sec. IV.

Extending STAP waveform optimization to ISAC requires enforcing instantaneous multiuser QoS constraints while directly controlling $\mathbf{x}_n[\ell]$ over slow time. In this example we adopt symbol-level precoding as the transmit-side mechanism, since its QoS constraints are imposed directly on the symbol-wise waveform and are compatible with the STAP coupling through $\mathbf{X}_n$ \cite{RLiu-JSAC-2022,BWang-WCL-2022,AAubry-JSTSP-2015}. Let the hypothesized target parameters be $(\theta_t,f_{\text{D},t})$ and define $\mathbf{v}_{t,n}\triangleq \mathbf{v}_n(\theta_t,f_{\text{D},t})$.  
We jointly design the transmit waveforms $\{\mathbf{X}_n\}$ and the corresponding space-time receive filters $\{\mathbf{w}_n\}$. To balance complexity and robustness, we adopt a per-subcarrier STAP strategy and aggregate the sensing performance across the OFDM bandwidth. The radar SCNR for the target range-Doppler cell is expressed as   
\begin{align}\label{eq:SCNR SLP}
\text{SCNR}_{\text{SLP}} =
\frac{\sum_{n=0}^{N-1}\sigma_{t,n}^2\left|\mathbf{w}_n^H\mathbf{X}_n\mathbf{v}_{t,n}\right|^2}{\sum_{n=0}^{N-1}\mathbf{w}_n^H(\mathbf{X}_n\mathbf{V}^{\text{st}}_{\text{cc},n}  \mathbf{X}_n^H+\mathbf{R}_{\bm{\eta},n})\mathbf{w}_n}.
\end{align}
Compared with the block-level spatial SCNR in Sec. VI-A, \eqref{eq:SCNR SLP} exploits space-time DoFs at the receiver and enables slow-time waveform shaping at the transmitter through $\mathbf{X}_n$. 
We then formulate the joint transmit/receive optimization:
\begin{subequations}\label{eq:slp_stap}
\begin{align}
\max_{\{\mathbf{x}_n[\ell]\},\{\mathbf{w}_n\}}\quad&\text{SCNR}_\text{SLP}(\mathbf{X}_n,\mathbf{w}_n)\\
\text{s.t.}\quad&\delta_{n,k}[\ell]\ge\bar{\gamma}_{n,k},~\forall n,k,\ell,\\
&\sum_{n=0}^{N-1}\sum_{\ell=0}^{L-1}\|\mathbf{x}_n[\ell]\|_2^2\le P_{\text{tot}},\\
&\mathbf{x}_n[\ell]\in\mathcal{X},~\forall n,\ell,
\end{align}
\end{subequations}
where $\delta_{n,k}[\ell]$ is the SLP safety margin defined in \eqref{eq:SLP QoS}, and $\bar{\gamma}_{n,k}$ is the required QoS threshold that guarantees satisfactory SER. 
The set $\mathcal{X}$ is used to capture practical waveform constraints such as PAPR limits, spectral masks, similarity to a reference waveform, or constant-envelope signaling.

Problem \eqref{eq:slp_stap} is non-convex due to the fractional objective and the coupling between $\{\mathbf{x}_n[\ell]\}$ and $\{\mathbf{w}_n\}$. We adopt an AO procedure similar to that in the previous section. 
For fixed transmit waveforms, $\mathbf{R}_{I,n}(\mathbf{X}_n)$ is fixed, and the optimal STAP filter for the target bin $(\theta_t,f_{\text{D},t})$ is  
\begin{equation}\label{eq:mvdr_receive} 
\mathbf{w}_n^\star= \frac{\mathbf{R}^{-1}_{I,n}(\mathbf{X}_n)\mathbf{X}_n\mathbf{v}_{t,n}} {\mathbf{v}_{t,n}^H\mathbf{X}_n^H\mathbf{R}^{-1}_{I,n}(\mathbf{X}_n)\mathbf{X}_n\mathbf{v}_{t,n}}. 
\end{equation}
This formulation is consistent with the waveform-aware STAP structure in Sec. V-C, with $\mathbf{R}_{I,n}(\mathbf{X}_n)$ made explicit for transmit-side optimization.
For fixed receive filters, optimizing $\{\mathbf{x}_n[\ell]\}$ shapes $\mathbf{X}_n$ to maximize the SCNR under per-symbol SLP QoS constraints and the waveform feasibility set $\mathcal{X}$. The resulting subproblem is a fractional QCQP with affine QoS constraints in the waveform variables. It can be handled using the same tools as in Sec. VI-A, including Dinkelbach-type fractional programming combined with semidefinite relaxation (SDR) or CCP, with the main difference that the design variable is now the space-time waveform $\{\mathbf{x}_n[\ell]\}$ rather than the covariance-level matrix $\mathbf{R}_{X,n}$. When additional constraints such as constant modulus, PAPR control, or waveform similarity are imposed, the corresponding SLP based waveform design techniques in \cite{RLiu-JSAC-2022} can be incorporated into the transmit update step.

\subsection{Joint Transmit/Receive Design for NLoS Exploitation}

The proactive designs in Sections~VI-A and VI-B shape the transmit waveform and receive filtering to mitigate clutter for improved sensing. ISAC systems also have communication objectives that are enhanced by increased multipath scattering that leads to improved diversity and channel rank. However, inducing such scattering also potentially increases clutter, creating an inherent trade-off for ISAC applications \cite{SLu-TVT-2023}. Rather than eliminating target echoes generated via non-line-of-sight (NLoS) paths, a complementary paradigm is to exploit strong specular NLoS target reflections, since they can provide useful sensing information. This approach is taken in radar multipath exploitation models, where the received signal is categorized into direct target returns, multipath target returns, and clutter-only returns \cite{ZXu-TSP-2021}. In obstructed or sensing-through-wall scenarios, virtual-anchor-based modeling provides a geometry-consistent interpretation of specular reflections, which helps maintain consistent hypotheses and mitigate ghost artifacts when dominant reflectors can be identified \cite{MLeigsnering-ICASSP-2015}.

When multipath components are resolvable in delay, Doppler, or angle, detection and estimation can benefit from coherent fusion across multiple hypotheses. Representative GLRT-type designs in OFDM radar and OFDM-ISAC show that detection performance is governed by a noncentrality parameter, which naturally motivates adaptive waveform selection and multipath-aware combining \cite{SSen-TSP-2011,XLv-Globecom-2025}. When different NLoS paths exhibit heterogeneous reliability and SNR, weighted combining improves robustness by emphasizing stronger or better-tracked components, and the weights can be optimized together with transmit power allocation under ISAC communication constraints \cite{XLv-Globecom-2025}. More broadly, NLoS exploitation is inherently a joint transmit/receive design problem. On the sensing side, the transmitter can selectively illuminate informative reflectors to stabilize target-related NLoS echoes, while the receiver coherently fuses the resulting multipath structure instead of nulling it as interference. On the communication side, the same multipath components contribute to spatial diversity and degrees of freedom, yet they also couple with the probing waveform and the interference environment. As a result, the exploitation-reduction tradeoff must be addressed at the transceiver rather than using receiver-only post-processing \cite{SLu-TVT-2023,ZXu-TSP-2021}.

In practice, NLoS exploitation requires reflector identification and path association across time, which can be supported by geometry-aware models and site-specific priors \cite{MLeigsnering-ICASSP-2015}. Beyond passively using existing reflectors, one may also create strong and controllable NLoS paths via active reconfigurable intelligent surfaces (RIS) to provide additional diversity and improve radar detection performance through joint radar/RIS design \cite{MRihan-SPL-2022}. NLoS exploitation is most beneficial when the dominant reflectors are strong and resolvable so that their parameters can be reliably determined. This hybrid viewpoint motivates clutter-aware multi-domain transceiver optimization that jointly controls the waveform, transceiver processing, and propagation environment, as discussed next.

\subsection{Clutter-Aware Multi-Domain Optimization}
Effective clutter management in ISAC calls for proactive strategies that go beyond the baseband spatial and space-time optimization in Sections~VI-A-VI-C. By exploiting additional DoFs in the electromagnetic (EM) and network domains, ISAC transceivers can better control clutter generation and observation, improving radar detection sensitivity while enhancing communication reliability \cite{Liu WCM 2025}. This subsection outlines a unified clutter-aware optimization viewpoint spanning the EM, baseband, and network domains.

\subsubsection{EM-Domain Clutter Management} 
EM-domain clutter management aims to shape the radiated field and, when possible, the effective propagation environment so that diffuse clutter is weakened \emph{before} it reaches the baseband processor. Two representative EM-domain mechanisms are advanced antennas and RIS.

\textbf{Advanced Antennas for Clutter Control:} Recent advances in antenna technologies provide unprecedented opportunities for clutter-aware radiation control at the EM layer. \textit{Fluid or movable antennas} enable dynamic adjustment of the effective antenna position or geometry, which improves spatial diversity and allows the transmitter to steer energy away from clutter-dominant regions \cite{Zou 2025}--\cite{Chen Zhao 2025}. Likewise, \textit{electronically reconfigurable antennas}, including pixel, parasitic, metasurface-based, and holographic-surface architectures, support rapid reconfiguration of radiation patterns, polarization states, and impedance profiles, enabling fine-grained spatial and polarization-domain control \cite{LiuMengzhen CM 2025}--\cite{SZeng-JSAC-2026}. By actively suppressing transmissions toward known clutter directions or unfavorable polarization states, these antennas reduce the clutter power at its physical origin rather than relying solely on digital post-processing. Polarization agility further helps separate target echoes from clutter by exploiting differences in scattering responses. Key challenges include reliable environment awareness, trade-offs between reconfigurability and radiation efficiency, and implementation complexity in compact and mobile platforms. Recent studies have reported encouraging gains through adaptive radiation control and cross-layer integration, suggesting that EM-domain adaptation is a viable enabler of proactive clutter mitigation.

\textbf{RIS for Propagation Environment Reshaping:} 
Beyond avoiding clutter illumination, RIS offer a mechanism to actively reshape the propagation environment \cite{Liu WCM 2023}. By adjusting their element-wise phase responses, RIS can modify the spatial distribution and intensity of the reflected fields, influencing the scattering from both targets and clutter. Three representative clutter-related strategies have emerged: (\textit{i}) using RIS as clutter sinks to steer clutter reflections away from the receiver or to form destructive-interference nulls \cite{Liu WCM 2023}--\cite{Zhang-TCCN-2025}, (\textit{ii}) repurposing strong reflections as beneficial paths toward intended targets or communication users to enhance SCNR and throughput \cite{RLiu-JSAC-2022}, and (\textit{iii}) creating controllable NLoS paths that introduce additional spatial, delay, and Doppler diversity to improve detection and clutter discrimination \cite{TWei-TCOM-2023,Zhang-TCCN-2025}. Early studies support the potential of RIS-enabled clutter management, with performance gains driven by advances in low-loss materials, efficient control, and integrated architectures.

\subsubsection{Network-Domain Clutter Mitigation}
At the network level, clutter suppression is a cooperative and distributed optimization problem. In cell-free massive MIMO or distributed ISAC networks, geographically separated access points (APs) coordinate illumination and reception to reduce energy delivered to clutter-dominant regions while preserving target visibility through multi-perspective diversity. The key is to exploit inter-node clutter correlation together with the ability to align target returns in delay, angle, and Doppler. With tight synchronization, coherent fusion after target-centric alignment maximizes suppression. With looser synchronization, noncoherent energy fusion still benefits from inter-node clutter de-correlation. Coordinated scheduling and beam/power planning can avoid simultaneous illumination of the same clutter patches, distributed apertures can place joint nulls toward dominant clutter, and heterogeneous range-angle-Doppler views can be fused so that local clutter peaks are attenuated.

Recent related work includes cooperative ISAC architectures that jointly schedule TX/RX roles, beams, and power to suppress illumination of clutter-dominant sectors while preserving sensing coverage and target handover \cite{Meng 2025 A,Li Xiao TWC 2024}. Analytical optimization and derived scaling laws results characterize improvements with increased AP density, coordination granularity, and backhaul capacity \cite{Meng 2025 B}. In cell-free architectures, formulations that couple multi-perspective sensing with clutter-aware AP assignment and beamforming improve robustness under limited or stale CSI \cite{SLiu-WCNC-2024,Galappaththige 2025}. To enhance scalability, graph-learning-based fusion has been proposed to leverage angular diversity and mitigate local clutter outliers \cite{Jiang 2025}. Moreover, exchanging compact clutter maps or sufficient statistics enables distributed beamforming and STAP-like joint filtering with modest backhaul overhead \cite{Liu Zhang 2025}.

Network-level deployments must address {\em (i)} synchronization and calibration for coherent combining and stable nulling, {\em (ii)} fronthaul/backhaul constraints that limit real-time exchange of CSI and maps, {\em (iii)} non-stationary clutter and mobility that render maps stale and require low-latency re-optimization, and {\em (iv)} scalable distributed covariance estimation as the numbers of antennas, subcarriers, and nodes grow. Cross-layer orchestration is also needed, including lightweight MAC/control protocols for clutter-aware scheduling and map dissemination, along with privacy and security safeguards when sharing environmental knowledge.

\subsubsection{Cross-Domain Optimization for Clutter Management}
Joint operation across the EM, baseband, and network domains enables a closed-loop mechanism for proactive clutter management. Knowledge of dominant reflector directions, Doppler signatures, and geographic distributions can be shared across domains to guide coordinated resource allocation and waveform adaptation. In practice, a hierarchical control strategy can be established across distinct timescales. At \textit{coarse timescales} (e.g., per frame or scheduling epoch), the system optimizes slowly varying parameters such as AP activation, antenna topology, and polarization configurations based on long-term clutter statistics. At \textit{fine timescales} (e.g., per OFDM symbol or pulse), dynamic adaptation refines baseband waveform and beamforming, or symbol-level precoding, to track instantaneous variations. Meanwhile, reconfigurable antennas and RIS can adjust phase or polarization states using fast feedback from clutter estimators or site-specific maps.

This cross-domain hierarchy forms an adaptive loop. Long-term EM and network configurations establish a clutter-resilient baseline, while short-term baseband adaptation responds rapidly to local perturbations. Such synergy reduces the effective clutter rank prior to reception, eases post-processing, and strengthens the sensing--communication tradeoff. In summary, integrating EM-domain reconfiguration, baseband waveform optimization, and network-level cooperation establishes a unified paradigm for clutter-aware ISAC, transforming clutter from an unavoidable impairment into a controllable, and in some cases exploitable, environmental parameter.

\section{Future Directions}
The implementation of clutter-aware ISAC is not limited by the availability of suppression algorithms, but rather by the ability to sustain reliable performance under environmental complexity and fast nonstationarity. Practical systems must cope with time-varying clutter statistics, heterogeneous hardware and waveform configurations, and stringent latency and signaling budgets. Against this backdrop, the discussion below highlights open problems that couple physical-scene dynamics, statistical learning, and protocol support, and that may shape next-generation clutter-aware ISAC architectures.

\subsection{Robust Clutter Handling in Dynamic Environments}
Highly dynamic clutter challenges ISAC receivers because the interference statistics can drift within a CPI and across CPIs, directly undermining covariance estimation and adaptive filtering. Classical tracking algorithms such as Kalman filtering are effective when the clutter evolution is well captured by low-order state space models, but their performance degrades under abrupt regime changes, multi-mechanism dynamics, and heterogeneous or limited secondary data. A key research direction is to develop robust online processing that can track clutter statistics with limited snapshots while remaining stable under distribution shifts and model mismatch.

A promising approach in this direction is to couple model-based estimators with learned components that assist adaptation rather than replace physical modeling. For example, neural sequence models can predict or regularize the evolution of covariance parameters, enabling fast updates and reducing sensitivity to nonstationary conditions. Foundation models trained on diverse radar and ISAC datasets may further provide transferable priors for clutter structure and help recognize recurring scene-dependent patterns \cite{Hu 2025}. Realizing these gains requires lightweight architectures for real-time inference, data-efficient training and adaptation under limited labels, and evaluation methodologies that certify performance for heterogeneous environments. In addition, self-supervised adaptation can enable continual refinement without labels, physics-informed learning can impose Doppler and geometry consistency to improve sample efficiency and interpretability, and predictive digital twins can provide short-horizon forecasts of clutter evolution to support proactive tracking and filtering.

\subsection{Clutter and Multipath Exploitation}
 
Building on the multipath/NLoS exploitation paradigm reviewed in Sec. VI‑C, a possible future direction is to reliably identify, associate, and coherently combine informative multipaths in cluttered, time‑varying scenes. Such scattering can produce ``ghost'' target images that can be exploited for target detection and classification \cite{Smith TIP 2014}. Learning-based approaches, such as deep neural networks trained on extensive multipath-rich datasets, are possible solutions \cite{Guendel TGRS 2024}. Environment-aware training, and leveraging ray-tracing simulations or large-scale multipath datasets can improve robustness in complex propagation scenarios. Beyond passive exploitation, intentional multipath exploitation methods, including radio-simultaneous localization and mapping (radio-SLAM), use reflectors as virtual transmitters or receivers. Coherent multipath combining achieves diversity gain by carefully aligning informative multipath signals across multiple nodes, requiring precise path labeling and synchronization. By strategically using multipath reflections, future ISAC systems can convert clutter into a beneficial sensing asset, enabling improved NLoS target detection and enhanced situational awareness. 

\subsection{Digital Twins and Clutter Maps}
Integrating ISAC with digital twin technology is a promising approach to enhance environmental awareness and clutter prediction. A distributed network of ISAC nodes can cooperatively construct real-time clutter maps, enabling predictive identification of clutter effects before they degrade system performance. For example, the distributed intelligent ISAC (DISAC) framework of \cite{Strinati 6G Summit 2024} demonstrates how sharing sensor observations and semantic information among nodes yields network-level situational awareness. When linked to digital twin models, these distributed maps may enable forecasting of clutter dynamics and proactive mitigation. Key challenges include synchronizing and fusing high-rate data streams from heterogeneous sensors in real time, maintaining map consistency across nodes, and preserving data privacy and security.

AI-enhanced multimodal environmental modeling is also gaining traction as a way to create more accurate digital replicas of the physical world. Modern approaches fuse radar, LiDAR, cameras, and even communication signals to improve map fidelity and robustness \cite{Peng 2025}--\cite{Mao 2025}. For example, decomposing a scene into static structures and dynamic objects enables sensing approaches tailored to each component, as demonstrated by methods that combine point-cloud reconstruction with real-time object tracking. Future research should focus on generalizations to diverse environments, cross-modal calibration and uncertainty handling, and efficient real-time inference on edge devices. With these advances, digital-twin-driven clutter prediction could facilitate highly adaptive clutter suppression strategies for ISAC systems.

\subsection{Emerging Waveform Designs}
Unlike conventional OFDM, which can be fragile in doubly dispersive and clutter-rich channels, emerging waveforms such as orthogonal time frequency space (OTFS) and affine frequency division multiplexing (AFDM) exhibit inherent robustness against mobility and multipath. OTFS operates in the delay--Doppler domain and can coherently aggregate energy from multiple propagation paths, allowing indirect or clutter-induced components to be modeled explicitly rather than absorbed into unstructured noise \cite{Yuan JSTSP 2021}--\cite{Shtaiwi}. This can improve clutter characterization and facilitate detection of weak targets. AFDM employs chirp-based modulation that also offers fine delay--Doppler resolution and resilience to Doppler shifts \cite{Rou 2024}--\cite{Ni TWC 2024}, which helps separate moving targets from near-static clutter. In essence, both OTFS and AFDM make clutter a more structured part of the signal model, which can be isolated or even exploited through suitable processing.

Ideally, waveform designs should adapt to environment-dependent clutter characteristics. For example, an AFDM transmitter could dynamically adjust chirp rate, or an OTFS system could modify symbol mapping and pilot placement, to increase target--clutter separability. Learning-assisted controllers could select or synthesize waveforms from a library under spectral regulations and hardware constraints. 
Integrating heterogeneous waveforms within a single ISAC system would leverage their complementary strengths for various mobility and clutter regimes, but it introduces new challenges in synchronization, cross-waveform interference management, and adaptive switching policies.

\subsection{Standardization and Protocol Evolution}
Next-generation wireless standards should embrace clutter awareness as a key design objective. One immediate need is a unified channel model that explicitly incorporates clutter and target scattering, bridging the gap in current 5G models that lack dedicated sensing representations. Ongoing 6G research is extending geometry-based stochastic models to integrate target RCS and environmental clutter within a common framework while maintaining compatibility with legacy assumptions. In parallel, standards should introduce clutter-centric performance metrics (e.g., suppression ratio or sensing reliability under interference) and define representative scenarios, from dense urban canyons to industrial factory floors, for consistent evaluation of ISAC performance. Adopting such unified models and metrics, as advocated in \cite{Gong 2024}, will provide a common foundation for clutter-aware ISAC evaluation and guide future standardization.

Another crucial aspect is the development of MAC-layer protocols and signaling mechanisms that enable network-wide clutter awareness. New protocol primitives are needed for sharing sensing information and coordinating clutter mitigation among distributed nodes. Emerging network-level ISAC architectures envision a centralized sensing server that aggregates environmental feedback such as clutter maps and dynamically schedules transmissions across base stations for cooperative tracking and interference suppression. To support interoperability, standardized feedback and control messages for clutter awareness should be defined, potentially including simple clutter-status indicators and lightweight sensing broadcast channels for timely dissemination of environment updates, while keeping signaling overhead minimal. Establishing these interfaces will allow nodes to coordinate beamforming, power control, and other adaptations as clutter conditions evolve, in line with distributed signaling strategies highlighted in \cite{Han 2025}.

Equally important is deeper cross-layer integration so that real-time environmental context permeates the entire network stack. Physical-layer sensing outputs such as obstacle positions or clutter power levels should directly inform MAC scheduling, resource allocation, and beam management. This enables adaptive policies whereby the radio access network adjusts communication and sensing resources based on clutter level and mobility. Environment-agnostic scheduling can waste resources in stable scenes and degrade sensing under rapid dynamics; predictive or cognitive MAC designs that leverage short-term clutter forecasts could mitigate these issues. Realizing fully clutter-aware ISAC in 6G will require not only standard interfaces (potentially open APIs enabling external systems such as V2X to contribute sensing data) but also rigorous privacy and security mechanisms for shared environmental information. This end-to-end integration remains an open challenge and a key direction for future ISAC standardization.

\section{Conclusions}
This paper has presented a clutter-aware ISAC framework that unifies clutter modeling, estimation, suppression, and proactive management. Starting from a wideband MIMO-OFDM signal model that captures both cold and hot clutter across the space, time, and frequency domains, we connected amplitude distributions, robust SIRV statistics, and structured covariance estimation under limited-snapshot conditions. We systematically reviewed receiver-side suppression methods, spanning slow-time processing, spatial filtering, and STAP and SFTAP extensions, highlighting how joint multi-domain processing is essential when clutter is dynamic, structured, or overlaps with the target. We further discussed clutter-aware transceiver co-design, where beamforming and waveform optimization are integrated with communication QoS constraints to proactively control illumination and mitigate interference. Looking ahead, key research priorities include dynamic clutter handling via learning-enhanced filtering, leveraging clutter/multipath as sensing assets, intelligent digital twins and clutter maps, delay-Doppler robust waveform designs, and standardization of clutter-centric channel models, performance metrics, and protocols. These advances will be instrumental in enabling truly environment-adaptive and clutter-resilient ISAC systems for next-generation networks.

\end{document}